\documentclass[aps,twocolumn,amsmath,amssymb,notitlepage,groupedaddress,floatfix]{revtex4-2}
\usepackage{graphicx}
\usepackage{dcolumn}
\usepackage{bm}
\usepackage{upgreek}
\usepackage{siunitx}
\usepackage{colonequals}
\usepackage[caption=false]{subfig}
\usepackage[dvipsnames]{xcolor}
\usepackage{bbold}

\usepackage{enumitem}

\usepackage{xr-hyper}
\definecolor{darkblue}{rgb}{0,0,0.6}
\definecolor{darkred}{rgb}{0.6,0,0}
\usepackage[colorlinks=true,urlcolor=darkblue, citecolor=darkblue, linkcolor=black, hyperfootnotes=false]{hyperref}

\usepackage{mathrsfs}

\usepackage{tikz}
\usepackage[kerning=true]{microtype}
\usetikzlibrary{patterns}

\newcommand{\dd}{\mathrm{d}}

\DeclareMathOperator{\erfc}{erfc}
\DeclareMathOperator{\sign}{sign}
\def \equi#1{\mathrel{\mathop{\kern 0pt\sim}\limits_{#1}}} 

\def\XXint#1#2#3{{\setbox0=\hbox{$#1{#2#3}{\int}$}
    \vcenter{\hbox{$#2#3$}}\kern-.5\wd0}}

\newcommand{\D}{\mathcal{D}}

\def\e{\mathrm{e}}
\def\I{\mathrm{i}}
\def\O{\mathcal{O}}

\def\Nsc{\Lambda}
\def\fm{f_{\mathrm{M}}}

\newcommand{\abs}[1]{\ensuremath{\left| #1 \right|}}
\newcommand{\moy}[1]{\ensuremath{\left\langle #1 \right\rangle}}

\newcommand{\V}[1]{\boldsymbol{#1}}
\def \z {^{(0)}}
\def \o {^{(1)}}

\def\Vnn{V_{\mathrm{NN}}}
\def\Vtnn{\tilde{V}_{\mathrm{NN}}}

\begin{document}

\title{Macroscopic fluctuation theory of interacting Brownian particles}

\author{Aur\'elien Grabsch}
\affiliation{Sorbonne Universit\'e, CNRS, Laboratoire de Physique Th\'eorique de la Mati\`ere Condens\'ee (LPTMC), 4 Place Jussieu, 75005 Paris, France}

\author{Davide Venturelli}
\affiliation{Sorbonne Universit\'e, CNRS, Laboratoire de Physique Th\'eorique de la Mati\`ere Condens\'ee (LPTMC), 4 Place Jussieu, 75005 Paris, France}

\author{Olivier Bénichou}
\affiliation{Sorbonne Universit\'e, CNRS, Laboratoire de Physique Th\'eorique de la Mati\`ere Condens\'ee (LPTMC), 4 Place Jussieu, 75005 Paris, France}

\date{\today}

\begin{abstract}
We apply the macroscopic fluctuation theory (MFT) to study the large-scale dynamical properties of Brownian particles with arbitrary pairwise interaction.
By combining it with standard results of equilibrium statistical mechanics for the collective diffusion coefficient, the MFT gives access to the exact large-scale dynamical properties of the system, both in- and out-of-equilibrium.
In particular, we obtain exact results for dynamical correlations between the density and the current of particles. For one-dimensional systems, this allows us to obtain a precise description of these correlations for emblematic models, such as the Calogero and Riesz gases, and for systems with nearest-neighbor interactions such as the Rouse chain of hardcore particles or the recently introduced model of tethered particles. 
Tracer diffusion with the single-file constraint (but for arbitrary pairwise interaction) are also studied.
For higher-dimensional systems, we quantitatively characterize these dynamical correlations by relying on standard methods such as the virial expansion.
\end{abstract}

\maketitle

\tableofcontents

\hypersetup{linkcolor=darkred}

\section{Introduction}

The dynamical behavior of interacting particle systems, in both equilibrium and non-equilibrium regimes, has attracted a lot of attention over the past decades~\cite{Spohn:1991,Evans:2005,Derrida:2007,Campa:2009,Chou:2011,Bertini:2015,Derrida:2025a}. 
For instance, many studies have been devoted to lattice gas models, such as the simple exclusion process (SEP), for which various exact results have been obtained~\cite{Derrida:2009,Chou:2011,Mallick:2015,Mallick:2022}.

In this article, we consider the paradigmatic model of pairwise interacting Brownian particles, which can represent various systems such as colloidal suspensions, supercooled liquids, polymers, or ions in solution~\cite{Rouse:1953,Dufreche:2005,Liu:2015}. The positions $\boldsymbol{x}_i$ of the particles in $d$ dimensions evolve according to a set of overdamped Langevin equations,
\begin{equation}
    \label{eq:EqBrownianPart}
    \frac{\dd \boldsymbol{x}_i}{\dd t} = 
    - \mu_0 \sum_{j\neq i} \boldsymbol{\nabla} V_0(\boldsymbol{x}_i - \boldsymbol{x}_j)
    + \sqrt{2 D_0} \: \boldsymbol{\eta}_i
    \:,
\end{equation}
where $\mu_0$ is the bare mobility of a particle, $V_0$ the interaction potential between two particles, $\boldsymbol{\eta}_i$ a Gaussian white noise with unit variance, and $D_0 = \mu_0 k_{\mathrm{B}} T$ the bare diffusion coefficient. Note that the Langevin equations~\eqref{eq:EqBrownianPart} constitute an effective description of the particles after integration of the bath degrees of freedom. We will nevertheless refer to this description as ``microscopic'', since it constitutes our starting point.
In the following, we will mostly consider the case of hardcore particles, with $V_0(0) = +\infty$, but we can also address the case of soft particles for which $V_0(0)$ is finite (except for 1D tracer diffusion where the single-file constraint is required).
We will focus on the case where the interaction potential $V_0(\V{x})$ decays faster than $||\V{x}||^{-d}$, so that the dynamics of the system is diffusive at large scales (i.e.~the mean density obeys a diffusion equation). By contrast, if $V_0(\V{x})$ decays slower than $||\V{x}||^{-d}$, then the energy of the system is no longer extensive and the dynamics is not diffusive (see for instance Ref.~\cite{Dandekar:2023}).
Although the evolution equations~\eqref{eq:EqBrownianPart} are well suited for Brownian dynamics simulations, the analytical treatment of these many-body systems remains challenging. Indeed, very few results have been obtained from a direct investigation of these equations, with the notable exception of the Calogero potential $V_0(x) \propto 1/x^2$ in a one-dimensional system at low temperature~\cite{Touzo:2024}.
See also Refs.~\cite{Spohn:1987,Touzo:2024a,Dandekar:2024} for explicit results in cases where the potential decays slower than $1/x$ in $d=1$.

An alternative path consists in considering the microscopic density of particles instead of the set of their positions $\{ \boldsymbol{x}_i \}$. Since these positions are random, the density is also random, and can be shown to obey a stochastic partial differential equation, known as the Dean--Kawasaki equation~\cite{Kawasaki:1994,Dean:1996}. This equation is formally equivalent to the set of Langevin equations~\eqref{eq:EqBrownianPart} and has served as a starting point to many works over the last decades, see for instance Ref.~\cite{Illien:2024} for a recent review.
However, the Dean--Kawasaki equation cannot be solved analytically and one must resort to approximations that hold in limiting situations such as weak interactions~\cite{Demery:2014,Dean:2014,Demery:2016,Poncet:2017,Kruger:2017,Kruger:2018,Mahdisoltani:2021,Venturelli:2024,Muzzeddu:2024}. 

In parallel, another formalism has been developed to describe the \textit{macroscopic} density of particles, which describes the system of particles~\eqref{eq:EqBrownianPart} at large scales (long time and large distances). The fluctuating hydrodynamics~\cite{Spohn:1983} and the macroscopic fluctuation theory (MFT, not to be confused with mean-field theory)~\cite{Bertini:2015} indeed provide a systematic approach to analyse the large-scale dynamical properties of diffusive systems (which obey Fick's or Fourier's law). 
The main advantage of this approach is that it is analytically 
amenable.
The MFT has indeed permitted to obtain various exact results concerning, for instance, the displacements of tracer particles or integrated currents for various models ranging from stochastic lattice gases, such as the SEP, to heat transfer models like the Kipnis-Marchioro-Presutti model~\cite{Derrida:2009a,Krapivsky:2012,Krapivsky:2014,Krapivsky:2015,Krapivsky:2015a,Sadhu:2015,Poncet:2021,Grabsch:2022,Bettelheim:2022,Mallick:2022,Krajenbrink:2022,Rizkallah:2022,Grabsch:2023,Grabsch:2024,Grabsch:2024b,Bodineau:2025,Saha:2025}. Recently, we have shown how to apply the MFT to describe the large-scale properties of systems of interacting Brownian particles described by~\eqref{eq:EqBrownianPart}~\cite{Grabsch:2025b}.

In this article, we provide full details on this result, and show explicit applications to 
specific examples of Brownian particles with pairwise interaction, in one dimension and beyond. For completeness, we recall the case of the Calogero potential studied in Ref.~\cite{Grabsch:2025b}, as well as the virial expansion used to apply the MFT formalism to an arbitrary interaction potential~\cite{Grabsch:2025b}.
We additionally consider here one-dimensional systems of Brownian particles with nearest-neighbour interactions, such as the Rouse chain~\cite{Rouse:1953}, the model of sticky hard rods~\cite{Baxter:1968,Percus:1982}, and the model of tethered particles~\cite{Yuste:2025}.
We recall from~\cite{Grabsch:2025b} that this formalism gives access to the displacement of a tagged particle (a tracer) in one dimension. We additionally apply here this formalism to obtain exact results on the statistics of the current through a given surface in a $d$-dimensional channel. We also characterize the large-scale density and current fluctuations that emerge in the system in any dimension (while only the density-density fluctuations were considered in~\cite{Grabsch:2025b}).

The article is organized as follows. We first recall the general framework of MFT in Section~\ref{sec:MFT}, and in particular the transport coefficients $D(\rho)$ and $\sigma(\rho)$ that play a central role in this formalism. We then show in Section~\ref{sec:TrCoefs} that $D$ and $\sigma$ take a simple form for the model of interacting Brownian particles described by~\eqref{eq:EqBrownianPart}. We also determine these coefficients explicitly, either exactly or relying on approximations, for various interaction potentials in any dimension.
In Section~\ref{sec:OneDimensionMFT},
we combine these expressions with recent results obtained in the MFT literature
to characterise the position of a tracer and the current in one-dimensional systems of interacting Brownian particles.
In higher dimensions, we obtain new results by applying the MFT to $d$-dimensional channels in Section~\ref{sec:MFTchannels}, and in infinite $d$-dimensional systems in Section~\ref{sec:MFTdDimension}.

\section{Macroscopic fluctuation theory}
\label{sec:MFT}

Let us first introduce the \textit{microscopic} density of particles at positions $\V{x}_n(t)$, defined as
\begin{equation}
    \label{eq:DensMicro}
    \rho_0(\V{x},t) = \sum_n \delta(\V{x} - \V{x}_n(t))
    \:.
\end{equation}
The macroscopic fluctuation theory focuses on a coarse-grained version of this density, which describes the system at the \textit{macroscopic} scale,
\begin{equation}
    \label{eq:RescalingDensity}
    \rho(\V{x},t) = \rho_0(\Nsc \V{x}, \Nsc^2 t)
    \:,
\end{equation}
where $\Nsc \gg 1$ is a dimensionless rescaling parameter, which can be for instance proportional to the size of the system. 
Note that we perform here a diffusive rescaling of space and time because of the diffusive nature of the system we consider.
We will see in Section~\ref{sec:MFTdDimension} below that the precise choice of $\Lambda$ is irrelevant, as long as $\Lambda \gg 1$. Since the number of particles is conserved, the density obeys a conservation equation
\begin{equation}
    \label{eq:ConsEqDens}
    \partial_t \rho +
   \boldsymbol{\nabla} \cdot \V{j} = 0
   \:,
\end{equation}
where $\V{j}(\V{x},t)$ is the current of particles at position $\V{x}$ and time $t$. The key idea of MFT is to write the current in closed form, in terms of the collective diffusion coefficient $D(\rho)$ and the mobility $\sigma(\rho)$, as~\cite{Bertini:2015}
\begin{equation}
   \label{eq:jflucHydro}
    \V{j} = -
    D(\rho) \boldsymbol{\nabla} \rho
    - \sqrt{\frac{\sigma(\rho)}{\Nsc^d} } \: \boldsymbol{\eta}
    \:,
\end{equation}
where $\V{\eta}$ is a Gaussian white noise in space and time,
\begin{equation}
    \label{eq:DefNoiseCorrel}
    \moy{ \boldsymbol{\eta}_i(\boldsymbol{x},t) \boldsymbol{\eta}_j(\boldsymbol{y},t') }
    = \delta_{i,j} \delta( \boldsymbol{x}- \boldsymbol{y})
    \delta(t-t')
    \:,
\end{equation}
with $\delta_{i,j}$ the Kronecker delta function.
A few comments are in order. 
(i)~The first term in the current~\eqref{eq:jflucHydro} corresponds to Fick's law, so that Eq.~\eqref{eq:jflucHydro} can be seen as a stochastic version of Fick's law.
(ii)~The noise term is small due to the factor $\Nsc^{-d/2}$, which comes from the coarse graining to the macroscopic scale~\cite{Bertini:2015}.
The noise can thus be treated perturbatively for $\Lambda \gg 1$. This is the key property that makes the MFT applicable in practice, as we will discuss below.
(iii)~Equations~(\ref{eq:ConsEqDens},\ref{eq:jflucHydro}) fully determine the (stochastic) dynamics of the system at large scale, including fluctuations and large deviations (which characterize rare events), even in out-of-equilibrium settings.
(iv)~All the microscopic details, such as temperature $T$, bare mobility $\mu_0$ or interaction potential $V_0$, are encapsulated in the effective diffusion coefficient $D(\rho)$ and mobility $\sigma(\rho)$, which, as we will see below, can be determined from equilibrium properties. The determination of these transport coefficients is therefore essential to apply the MFT to the system of interacting Brownian particles described by~\eqref{eq:EqBrownianPart}.
(v) The fact that the macroscopic current~\eqref{eq:jflucHydro} involves transport coefficients computed close to equilibrium is a consequence of the local equilibrium property. At large scales, even if the full system is far from equilibrium, 
a small subsystem can be considered to be locally at equilibrium, with a density $\rho(\V{x},t)$. Based on this property, the MFT formalism~(\ref{eq:jflucHydro},\ref{eq:DefNoiseCorrel}) is able to describe the large-scale dynamics of systems far from equilibrium, for instance the relaxation of an initial step of density~\cite{Derrida:2009a,Bertini:2015}.

\subsection{Comparison with the Dean--Kawasaki equation}

The Dean--Kawasaki equation~\cite{Kawasaki:1994,Dean:1996} provides an alternative description of the \textit{microscopic} density $\rho_0$, in terms of the evolution equations
\begin{gather}
    \label{eq:ConsEqDensMicro}
    \partial_t \rho_0 +
   \boldsymbol{\nabla} \cdot \V{j}_0 = 0
   \:, \\
    \label{eq:MicroCurrent}
    \V{j}_0 = - D_0 \V{\nabla} \rho_0 
    - \mu_0 \rho_0 (\V{\nabla} V_0 \star \rho_0) 
    + \sqrt{2 D_0 \rho_0} \: \V{\eta}
    \:,
\end{gather}
where $(\V{\nabla} V \star \rho_0)(\V{x},t) = \int \V{\nabla} V(\V{x} - \V{y}) \rho_0(\V{y},t) \dd^d \V{y}$ and $\V{\eta}$ is a Gaussian white noise with delta correlations~\eqref{eq:DefNoiseCorrel}.
This equation is formally exact and equivalent to the set of Langevin equations~\eqref{eq:EqBrownianPart}.

Although the microscopic~\eqref{eq:MicroCurrent} and macroscopic currents~\eqref{eq:jflucHydro} present some similarities, 
there are important differences between the two approaches:
\begin{enumerate}[label={(\roman*)},nosep,leftmargin=*]
    \item First, the microscopic current~\eqref{eq:MicroCurrent} has the advantage of being expressed in terms of the interaction potential $V_0$ explicitly. However, it is involved in a convolution that yields a nonlocal structure. On the other hand, the macroscopic current~\eqref{eq:jflucHydro} is local, thanks to the coarse graining~\eqref{eq:RescalingDensity}, but the interaction potential appears through the transport coefficients $D(\rho)$ and $\sigma(\rho)$ that must be determined.
    \item \label{pt:weaknoise} The noise in the macroscopic current~\eqref{eq:jflucHydro} is intrinsically weak, due to the rescaling~\eqref{eq:RescalingDensity} with the large parameter $\Lambda \gg 1$. This makes the macroscopic formalism fully operational by relying on well-controlled weak-noise approaches (see Section~\ref{sec:Action} below).
    This is \textit{not} the case in the microscopic equation~\eqref{eq:MicroCurrent}, for which such approaches are limited to low-temperature or high-density limits~\cite{Illien:2025}. As a result, the microscopic equation~\eqref{eq:MicroCurrent} remains mostly formal and has essentially been applied to limiting situations such as weak potentials, for which a linearised equation has been derived~\cite{Demery:2014,Dean:2014,Demery:2016,Poncet:2017,Kruger:2017,Kruger:2018,Mahdisoltani:2021,Venturelli:2024,Muzzeddu:2024}. 
\end{enumerate}
Here, we will use only the macroscopic approach which, thanks to point~\ref{pt:weaknoise} above, will allow us to obtain \textit{exact} results for the large-scale properties of interacting Brownian particles described by~\eqref{eq:EqBrownianPart}. Before addressing the determination of the transport coefficients in Section~\ref{sec:TrCoefs} below, we first reformulate the macroscopic equations~(\ref{eq:ConsEqDens}-\ref{eq:DefNoiseCorrel}) in terms of an action formalism.

\subsection{An action formalism}
\label{sec:Action}

The MFT is a reformulation of the equations~(\ref{eq:ConsEqDens}-\ref{eq:DefNoiseCorrel})
for the current $\V{j}$ and density $\rho$ as an action formalism~\cite{Bertini:2015}. The probability of observing a given evolution of $\rho$ and $\V{j}$ is
\begin{equation}
    \label{eq:MFTprob0}
    P[\{ \rho, \V{j} \}] \propto
    \e^{-\Lambda^d \int \dd \V{x} \int \dd t \frac{(\V{j} + D(\rho) \V{\nabla} \rho)^2}{2 \sigma(\rho)}}
    \delta(\partial_t \rho + \V{\nabla} \cdot \V{j})
    \:.
\end{equation}
The first term corresponds to the probability density of the Gaussian noise $\V{\eta}$, expressed in terms of $\rho$ and $\V{j}$ using~\eqref{eq:jflucHydro}, while the delta function imposes the conservation law~\eqref{eq:ConsEqDens} at all points in space and time. The delta function can be represented as an integral over an auxilliary field $H$,
\begin{equation}
    \label{eq:MFTprob1}
    P[\{ \rho, \V{j} \}] \propto
    \int \D H \:
    \e^{-\Lambda^d \int \dd \V{x} \int \dd t  \left[ \frac{(\V{j} + D(\rho) \V{\nabla} \rho)^2}{2 \sigma(\rho)} 
    + H(\partial_t \rho + \V{\nabla} \cdot \V{j}) \right]}
    \:.
\end{equation}
If one is only interested in the evolution of the density $\rho$, the Gaussian integral over $\V{j}$ can be performed explicitly to yield the probability of observing a given evolution $\rho(\V{x},t)$,
\begin{equation}
    \label{eq:MFTprob}
    P[\{ \rho(\V{x},t) \}] \propto
    \int \D H \:
    \e^{-\Lambda^d S[\rho,H]}
    \:,
\end{equation}
where
\begin{multline}
    \label{eq:MFTaction}
    S[\rho,H] = \int \dd^d \V{x} \int \dd t \Big[
    H \partial_t \rho 
    \\
    + D(\rho) \V{\nabla} \rho \cdot \V{\nabla} H
    - \frac{\sigma(\rho)}{2} (\V{\nabla} H)^2
    \Big]
\end{multline}
is the MFT action~\cite{Derrida:2009a,Bertini:2015}. Because the action in~\eqref{eq:MFTprob} is multiplied by the large parameter $\Lambda^d \gg 1$, the integral is dominated by the minimum of the action. This means that the probability of evolving from a given density $\rho(\V{x},0)$ to $\rho(\V{x},t)$ for a fixed $t>0$ is entirely controlled by the solution of the Euler-Lagrange equations~\cite{Derrida:2009a,Bertini:2015}
\begin{subequations}
\label{eq:MFTeqs}
\begin{align}
    \partial_\tau q &= \V{\nabla} \cdot \left[
    D(q) \V{\nabla} q - \sigma(q) \V{\nabla} p
    \right]
    \:,
    \\
    \partial_\tau p &= 
    - D(q) \Delta p - \frac{\sigma'(q)}{2} (\V{\nabla} p)^2
    \:,
\end{align}
\end{subequations}
where $q(\V{x},\tau)$ is the optimal evolution of $\rho(\V{x},\tau)$, from $q(\V{x},\tau = 0) = \rho(\V{x},0)$ to $q(\V{x},\tau = t) = \rho(\V{x},t)$, and $p$ is the optimal evolution of the conjugate field $H$. The drastic simplification from the integral over all possible evolutions~\eqref{eq:MFTprob} to only the most probable one, described by the MFT equations~\eqref{eq:MFTeqs}, is what makes the MFT tractable in practice. It all boils down to the presence of the large rescaling parameter $\Lambda$, which comes from the fact that the MFT is a \textit{macroscopic} description in which the noise is \textit{intrinsically} weak. To apply this powerful formalism to the system of interacting Brownian particles described by~\eqref{eq:EqBrownianPart}, there only remains to determine the transport coefficients $D(\rho)$ and $\sigma(\rho)$.

\section{Transport coefficients}
\label{sec:TrCoefs}

The classic problem of determining the diffusion coefficient $D(\rho)$ for a system of pairwise interacting Brownian particles has been widely explored, often relying on accurate approximation schemes~\cite{Lekkerkerker:1981,Cichocki:1991,Butta:1999,Felderhof:2009}.
In turn, the mobility $\sigma(\rho)$ can be deduced from it using a fluctuation-dissipation relation~\cite{Bertini:2015}. But surprisingly, these results have never been used in the MFT literature until very recently~\cite{Grabsch:2025b}. In this Section, we first provide for the sake of consistency a simple derivation of these transport coefficients. We then obtain either exact or approximate expressions for these coefficients for various interaction potentials $V_0$, mostly in the one-dimensional case, but also for $d>1$.

\subsection{Derivation of the transport coefficients}
\label{sec:DerivTransCoefs}

From the constitutive equation~\eqref{eq:jflucHydro}, the collective diffusion coefficient $D(\rho)$ quantifies the linear response of the mean current to the application of a density gradient, while the mobility $\sigma(\rho)$ quantifies the amplitude of current fluctuations in equilibrium. These two quantities are related by the fluctuation-dissipation relation~\cite{Bertini:2015,Derrida:2025a}
\begin{equation}
    \label{eq:FlucDissDSigma}
    \frac{2 k_{\mathrm{B}} T D(\rho)}{\sigma(\rho)} = f''(\rho)
    \:,
\end{equation}
where $f(\rho)$ is the equilibrium free energy density, and $f''(\rho)$ is its second derivative with respect to the density $\rho$. The free energy is defined for a finite system of volume $V$ with $N$ particles as~\cite{Hill:1986}
\begin{equation}
    \label{eq:DefFreeEnerDens}
    \beta f(\rho) V
    \underset{N \to \infty}{\simeq}
    -\ln Z_{N,V}(\beta)
    \:,
    \text{ with }
    \rho = \frac{N}{V} \text{ fixed,}
\end{equation}
and $Z_{N,V}(\beta)$ is the usual partition function for a system of particles of mass $m$,
\begin{multline}
    \label{eq:DefFullPartFct}
    Z_{N,V}(\beta) = \frac{1}{h^{dN} N!} \int_{-\infty}^\infty \dd^d \V{p}_1 \cdots \dd^d \V{p}_N
    \\
    \int_V \dd^d \V{x}_1 \cdots \dd^d \V{x}_N \: \e^{- \sum_{i=1}^N \frac{\beta p_i^2}{2m}
    - \frac{\beta}{2} \sum_{i \neq j} V_0(x_i - x_j)}
    \:.
\end{multline}
Performing the integrals over the momenta, we have
\begin{align}
    \label{eq:DefPartFct}
    Z_{N,V}(\beta) = \frac{1}{N! \ell_0^{dN}}
    \int_V \dd^d \V{x}_1 \cdots \dd^d \V{x}_N \: 
    \e^{- \frac{\beta}{2} \sum_{i \neq j} V_0(x_i - x_j)}
    \:,
\end{align}
with $\ell_0 = \sqrt{\beta h^2/(2m\pi)}$ the thermal de Broglie wavelength, which is needed for $Z_{N,\beta}$ to be dimensionless. Note that the partition function, and therefore the free energy $f(\rho)$, involve the microscopic parameter $\ell_0$. This quantity is however irrelevant for the system of interacting Brownian particles, because it does not appear in the evolution equation~\eqref{eq:EqBrownianPart}. Remarkably, this can be shown from the definition of the free energy~\eqref{eq:DefFreeEnerDens}, which indeed implies from the partition function~\eqref{eq:DefPartFct}, $f(\rho) =  \frac{d}{\beta} \ln \ell_0 + \text{(independent of $\ell_0$)}$, so that $\ell_0$ does not appear in $f''(\rho)$. Hence, the fluctuation-dissipation relation~\eqref{eq:FlucDissDSigma} involves only the microscopic quantities present in Eq.~\eqref{eq:EqBrownianPart}. This will also be explicit in the examples below.

Since the fluctuation-dissipation relation~\eqref{eq:FlucDissDSigma} relates $D(\rho)$ and $\sigma(\rho)$, we only need to determine one of these coefficients to deduce the other. For models of lattice gases, it is usually simpler to determine the diffusion coefficient $D(\rho)$ and then deduce the mobility $\sigma(\rho)$ from~\eqref{eq:FlucDissDSigma}~\cite{Arita:2017,Arita:2018,Derrida:2025a}. Here, it turns out that it is simpler to first determine the mobility. Indeed, the mobility $\sigma$ can also be defined as the linear response of the mean current to the application of a small constant external force $\V{F}_0$ on all the particles (if there is no phase transition in the system)~\cite{Bertini:2015},
\begin{equation}
    \label{eq:DefSigmaForce}
    \moy{\V{j}} = 
    \frac{\sigma(\rho)}{2 k_{\mathrm{B}} T} 
    \V{F}_0
    \:,
\end{equation}
where the dynamics is now given by
\begin{equation}
    \frac{\dd \boldsymbol{x}_i}{\dd t} = 
    - \mu_0 \sum_{j\neq i} \boldsymbol{\nabla} V(\boldsymbol{x}_i - \boldsymbol{x}_j)
    + \mu_0 \boldsymbol{F}_0
    + \sqrt{2 D_0} \boldsymbol{\eta}_i
\end{equation}
instead of~\eqref{eq:EqBrownianPart}. The force $\V{F}_0$ can be absorbed by a change of reference frame, $\V{x}_i = \V{y}_i + \V{v}_0 t$ with $\V{v}_0 = \mu_0 \V{F}_0$, with the $\V{y}_i$ obeying the original evolution equation~\eqref{eq:EqBrownianPart}. By symmetry $\V{y}_i \to - \V{y}_i$, there is no average current in the moving frame, therefore in the fixed frame
\begin{equation}
    \moy{\V{j}} = \rho \V{v}_0 = \rho \mu_0 \V{F}_0
    \:.
\end{equation}
Comparing with the definition of the mobility~\eqref{eq:DefSigmaForce}, we deduce
\begin{equation}
    \label{eq:Mobility}
    \sigma(\rho) = 
    2 \mu_0 k_{\mathrm{B}} T \rho
    \:.
\end{equation}
This expression of the mobility is independent of the interaction potential $V_0$, and results directly from the Brownian dynamics~\eqref{eq:EqBrownianPart}~\footnote{Note that the same argument also holds for an underdamped dynamics.}.

Having determined the mobility~\eqref{eq:Mobility}, we straightforwardly deduce the diffusion coefficient from the fluctuation-dissipation relation~\eqref{eq:FlucDissDSigma},
\begin{equation}
    \label{eq:DiffCoef0}
    D(\rho) = \mu_0 \rho f''(\rho)
    \:.
\end{equation}
Although this expression fully determines the diffusion coefficient, it is often more practical to write it in a different form. Indeed, using the standard definition of 
pressure in equilibrium statistical mechanics, one finds
\begin{align}
    P(\rho) 
    &= - \left( \frac{\partial F}{\partial V} \right)_{T,N}
 = -  \left( \frac{\partial [V f(N/V)]}{\partial V} \right)_{T,N}
 \nonumber
 \\
 &= \rho f'(\rho) - f(\rho)
 \label{eq:LinkPf}
 \:,
\end{align}
where $F = V f(\rho = N/V)$ is the free energy. Taking the derivative of~\eqref{eq:LinkPf} with respect to $\rho$, we deduce
\begin{equation}
    \label{eq:RelDpFpp}
    P'(\rho) = \rho f''(\rho)
    \:.
\end{equation}
This allows to rewrite the diffusion coefficient~\eqref{eq:DiffCoef0} in terms of the equilibrium pressure,
\begin{equation}
    \label{eq:DiffCoef}
    D(\rho) = \mu_0 \: P'(\rho)
    \:.
\end{equation}
This is a well-known relation for colloidal systems, see for instance~\cite{Lekkerkerker:1981,Cichocki:1991,Butta:1999,Felderhof:2009}, but which has only very recently been used in MFT~\cite{Grabsch:2025b}. The diffusion coefficient~\eqref{eq:DiffCoef} can be written in different forms by using classical thermodynamic identities. For instance, we can express it in terms of the chemical potential
\begin{align}
    \mu(\rho) 
    &\equiv \left( \frac{\partial F}{\partial N} \right)_{T,V}
    = \left( \frac{\partial (V f(N/V))}{\partial N} \right)_{T,V}
    \nonumber
    \\
    &= f'(\rho)
    \label{eq:DefChemPot}
    \:,
\end{align}
so that the diffusion coefficient~\eqref{eq:DiffCoef0} takes the equivalent form
\begin{equation}
    \label{eq:DiffCoefMu}
    D(\rho) = \mu_0 \rho \: \mu'(\rho)
    \:.
\end{equation}
In practice, it can be simpler to use one of the equivalent expressions~(\ref{eq:DiffCoef0},\ref{eq:DiffCoef},\ref{eq:DiffCoefMu}), depending on the situation.

To summarize, the large-scale dynamical properties of systems of interacting Brownian particles are entirely controlled by the universal mobility $\sigma(\rho)$~\eqref{eq:Mobility}, which does not depend on the interaction potential $V_0$, and by the collective diffusion coefficient $D(\rho)$~\eqref{eq:DiffCoef}, which depends on the interaction potential $V_0$ through the equilibrium pressure $P(\rho)$. In practice, this means that one needs to know the equation of state of the system. For several one-dimensional systems, the equation of state can be obtained explicitly, but in general one must resort to standard approximations. We discuss both cases below by considering different examples.

\subsection{One-dimensional systems}
\label{sec:1d}

For one-dimensional systems, one can obtain explicit expressions for the equation of state of several models of interacting Brownian particles, and thus for the diffusion coefficient $D(\rho)$. We consider a few cases here.

\subsubsection{Independent particles}

We first consider the simplest and well-known case, which is the one of independent particles, i.e.~$V_0(\V{x}) = 0$,
to illustrate the method. In this case the partition function~\eqref{eq:DefPartFct} becomes
\begin{equation}
    Z_{N,V}(\beta) = \frac{1}{N!} \left( \frac{V}{\ell_0} \right)^N
    \:,
\end{equation}
where 
the volume $V$ is the length of the system. Using Stirling's formula, we obtain that the free energy density~\eqref{eq:DefFreeEnerDens} reads
\begin{equation}
    f(\rho) = k_{\mathrm{B}} T 
    \left[
    \rho \ln (\rho \ell_0) - \rho
    \right]
    \:.
    \label{eq:free-energy-noninteracting}
\end{equation}
As discussed in Section~\ref{sec:DerivTransCoefs} above, this expression depends on the thermal de Broglie wavelength $\ell_0$, which is not relevant for the model considered here. However, $\ell_0$ cancels out when computing the pressure from 
Eqs.~\eqref{eq:LinkPf}~and~\eqref{eq:free-energy-noninteracting},
\begin{equation}
    \label{eq:PerfGas}
    P(\rho) = k_{\mathrm{B}} T \rho 
    \:.
\end{equation}
This is exactly the equation of state of 
an ideal gas, which involves only macroscopic quantities, as it should. We therefore recover from~\eqref{eq:DiffCoef} the collective diffusion coefficient
\begin{equation}
    \label{eq:Dfree}
    D(\rho) = \mu_0 k_{\mathrm{B}} T = D_0
    \:,
\end{equation}
which is indeed identical to the diffusion coefficient of a single particle because there is no interaction.

\subsubsection{The Calogero potential}

We now consider particles interacting pairwise via the Calogero potential
\begin{equation}
\label{eq:CaloPot}
    V_0(x) = \frac{g}{x^2}
    \:,
\end{equation}
which has been widely studied~\cite{Calogero:1975,Moser:1975,Polychronakos:2006,Touzo:2024,Touzo:2024a,Lewin:2022}.
For this model, the grand potential has been computed explicitly~\cite{Choquard:2000}. We can thus use this result to compute the diffusion coefficient. Explicitly, we introduce the grand canonical partition function for a system of size $V$ and chemical potential $\mu$,
\begin{equation}
    \mathcal{Z}_V(\beta,\mu)
    = \sum_{N=0}^\infty \e^{\beta \mu N} Z_{N,V}(\beta)
    \:.
\end{equation}
The grand potential density is defined in the thermodynamic limit as
\begin{equation}
    \phi_{\mathrm{G}}(\beta, \mu) \equiv
    - \lim_{V \to \infty} \frac{k_{\mathrm{B}} T}{V}
    \ln \mathcal{Z}_V(\beta,\mu)
    \:.
\end{equation}
It is explicitly given by~\cite{Choquard:2000}
\begin{equation}
    \label{eq:GrandPotCalo}
    \phi_{\mathrm{G}}(\beta, \mu) = 
    -\frac{1}{\pi \beta \sqrt{2 \beta g}} \int_{-\infty}^\infty 
    W \left(  \frac{\sqrt{2 \pi \beta g }}{\ell_0} \e^{\beta \mu-k^2} \right) \dd k
    \:,
\end{equation}
with $W(x)$ the Lambert-W function, solution of $W(x) \e^{W(x)} = x$. This expression being written in a grand-canonical formalism, it involves the chemical potential instead of the density $\rho$. We can reintroduce the density and obtain the free energy by performing a Legendre transform
\begin{equation}
    f(\rho) = \min_{\mu} \: 
    [\phi_{\mathrm{G}}(\beta,\mu) + \mu \rho]
    \:.
\end{equation}
We thus have that
\begin{equation}
    f(\rho) = \phi_{\mathrm{G}}(\beta,\mu^\star(\rho))
    + \mu^\star(\rho) \rho
    \:,
\end{equation}
where $\mu^\star(\rho)$ is solution of
\begin{equation}
    \label{eq:EqforMuCalo}
    \rho + \partial_\mu \phi_{\mathrm{G}}(\beta,\mu) = 0
    \:.
\end{equation}
Since this equation directly determines the chemical potential $\mu(\rho)$, it is here simpler to express the diffusion coefficient using Eq.~\eqref{eq:DiffCoefMu}. Taking a derivative of~\eqref{eq:EqforMuCalo} with respect to $\rho$, we obtain $\mu'(\rho) = -1/\partial_\mu^2 \phi_{\mathrm{G}}(\beta,\mu(\rho))$, and thus we get the parametric representation
\begin{equation}
    \label{eq:DCalo}
    D(\rho) = - \frac{\mu_0 \rho}{\partial_{\mu}^2 \phi_{\mathrm{G}}(\beta, \mu(\rho))}
    \:,
\end{equation}
with $\mu(\rho)$ determined by~\eqref{eq:EqforMuCalo}, which explicitly reads
\begin{equation}
    \label{eq:MuCalo}
   \rho = \frac{\e^{\beta \mu(\rho)}}{\ell_0 \sqrt{\pi}} \int_{-\infty}^\infty 
    \e^{-k^2}
    W' \left(  \frac{\sqrt{2 \pi \beta g }}{\ell_0} \e^{\beta \mu(\rho)-k^2} \right) \dd k
    \:.
\end{equation}
These two equations give a parametric representation of the diffusion coefficient for the Calogero gas of Brownian particles. Importantly, the thermal de Broglie wavelength $\ell_0$ appears in both equations, but actually cancels out in $D(\rho)$. Indeed, Eq.~\eqref{eq:MuCalo} actually determines $\e^{\beta \mu(\rho)}/\ell_0$, which is exactly the combination that appears in $\phi_{\mathrm{G}}$~\eqref{eq:GrandPotCalo}. This is a general feature, that holds for any interaction potential $V_0$, as discussed in Appendix~\ref{App:DeBroglie}.

\begin{figure*}
    \centering
    \subfloat[]{
        \includegraphics[width=0.32\textwidth]{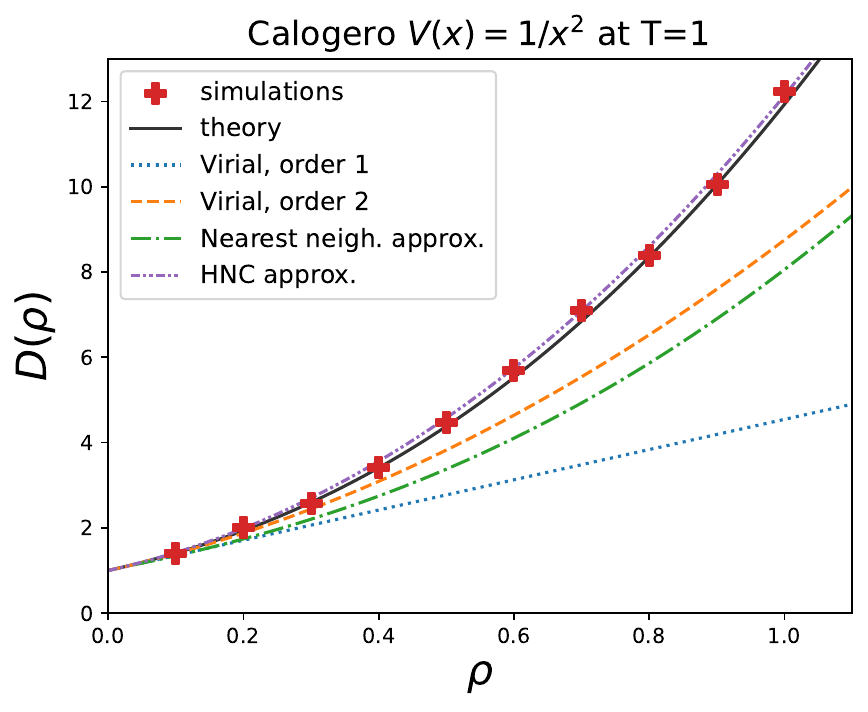}
        \label{fig:DCalo}
        \put(-168,10){(a)}
    }
    \subfloat[]{
        \includegraphics[width=0.325\textwidth]{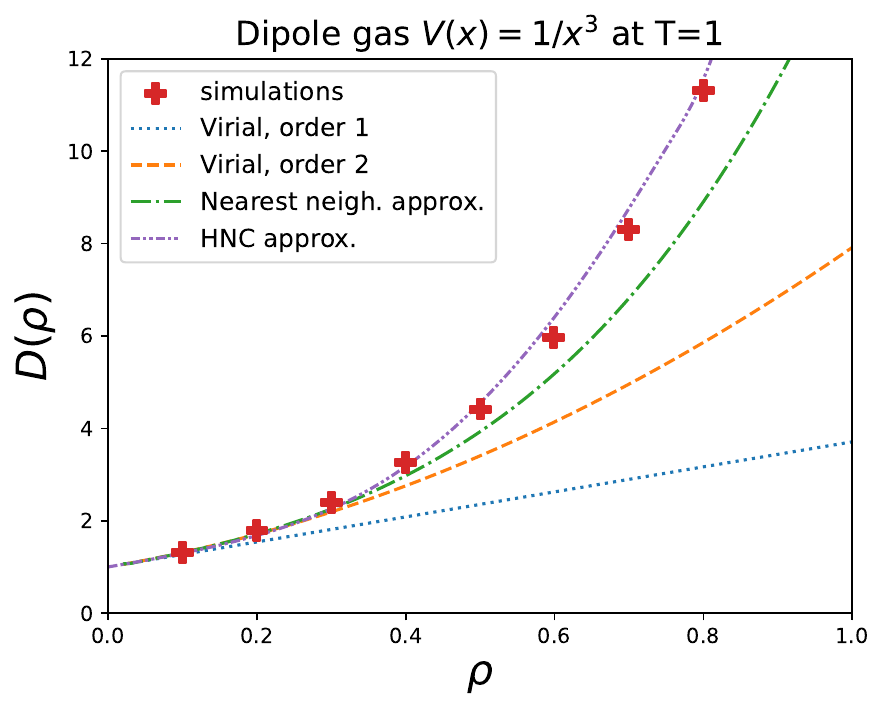}
        \label{fig:DDip}
        \put(-168,10){(b)}
    }
    \subfloat[]{
        \includegraphics[width=0.32\textwidth]{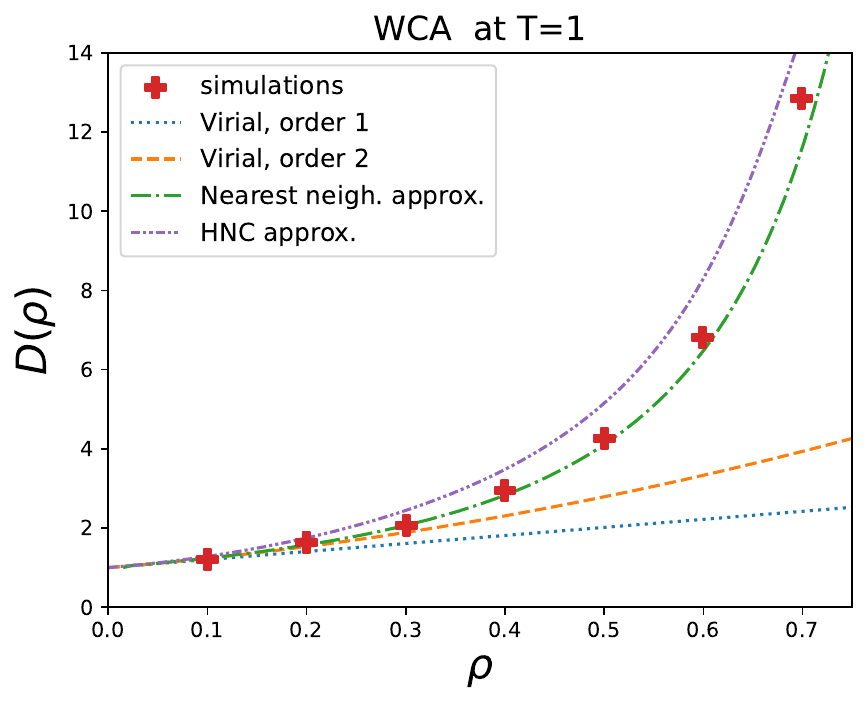}
        \label{fig:DWCA}
        \put(-168,10){(c)}
    }
    \vspace{-10pt}
    \caption{
    Diffusion coefficient $D(\rho)$ for some of the one-dimensional models presented in Sec.~\ref{sec:1d}. 
    (a) The Calogero gas with potential~\eqref{eq:CaloPot}; the solid line is the analytical expression~(\ref{eq:DCalo},\ref{eq:MuCalo}). 
    (b) The dipole gas with potential~\eqref{eq:DipPot}. (c) The WCA gas with potential~\eqref{eq:WCAPot}, with $l = 1$.
    In all panels, the red symbols are obtained from numerical simulations, with $g=1$, $T=1$, $\mu_0 = 1$ and $k_{\mathrm{B}} = 1$ (see Appendix~\ref{app:NumDSigma}). 
    The dashed lines are different approximations: the virial expansion (see Section~\ref{sec:Virial}), the nearest-neighbor approximation~\eqref{eq:DiffNN} (see Section~\ref{sec:NearNeigh}), and the hypernetted-chain (HNC) approximation (see Section~\ref{sec:PfromGr}).
    }
    \label{fig:D-first3}
\end{figure*}

The parametric representation~(\ref{eq:DCalo},\ref{eq:MuCalo}) can be easily implemented to evaluate numerically $D(\rho)$, which is shown in Fig.~\ref{fig:DCalo}.
It is also well suited to extract asymptotic behaviors in the low/high density or temperature regimes. For instance, solving~\eqref{eq:MuCalo} for $\rho \to \infty$ or $T \to 0$, we get
\begin{equation}
    \mu(\rho) \simeq \frac{g \pi^2}{2} \rho^2
    \quad \text{for } T \to 0 \text{ or } \rho \to \infty 
    \:.
\end{equation}
Inserting this expression into~\eqref{eq:DiffCoefMu}, we deduce
\begin{equation}
    \label{eq:DCaloHighDens}
    D(\rho) \simeq \mu_0 g \pi^2 \rho^2
    \quad \text{for } T \to 0 \text{ or } \rho \to \infty 
    \:.
\end{equation}
Similarly, in the low-density or high-temperature limit, we obtain
\begin{equation}
    \mu(\rho) = \frac{1}{\beta} \ln (\rho \ell_0)
    + \rho \sqrt{\frac{\pi g}{\beta}} + 3 (2-\sqrt{3}) \pi g \rho^2
    + O(\rho^3)
    \:.
\end{equation}
Note that, again, $\mu(\rho)$ involves the parameter $\ell_0$ through an additive constant, so that the diffusion coefficient~\eqref{eq:DiffCoefMu}, which involves only the derivative of $\mu(\rho)$, is independent of $\ell_0$, as it should:
\begin{multline}
    \label{eq:DCaloLowDens}
    D(\rho) = \mu_0 k_{\mathrm{B}} T 
    + \mu_0 \rho \sqrt{\pi g k_{\mathrm{B}} T}
    \\
    + 6 (2-\sqrt{3}) \mu_0 \pi g \rho^2
    + O(\rho^3)
    \:.
\end{multline}
Note that this $\rho \to 0$ expansion also holds for $T \to \infty$. This expansion is also shown in Fig.~\ref{fig:DCalo}.

\subsubsection{Arbitrary nearest-neighbor interactions}
\label{sec:NearNeigh}

We now consider a variation of the model~\eqref{eq:EqBrownianPart} in which a particle interacts only with its two nearest neighbors (one on each side), so that the evolution of the positions is now given by
\begin{equation}
    \label{eq:BrownianNN}
    \frac{\dd x_i}{\dd t}
    = - \mu_0 \Vnn'(x_{i} - x_{i-1}) - \mu_0 \Vnn'(x_i - x_{i+1})
    + \sqrt{2 D_0} \: \eta_i
    \:,
\end{equation}
where we now denote $\Vnn$ the interaction potential between nearest neighbours.
This model can be seen as an approximation of the original model~\eqref{eq:EqBrownianPart} at small interaction or low density, when the neighbouring particles dominate the interaction (see below). Equation~\eqref{eq:BrownianNN} can also represent other classes of systems like polymers for instance (see Section~\ref{sec:Rouse}).
The arguments of Section~\ref{sec:DerivTransCoefs} still hold, so the transport coefficients are again given by~(\ref{eq:Mobility},\ref{eq:DiffCoef}).
The grand potential, and thus the diffusion coefficient $D(\rho)$~\eqref{eq:DiffCoef}, for this model can be computed explicitly~\cite{Percus:1982,Santos:2016}. We reproduce here the derivation of Ref.~\cite{Percus:1982} for completeness.
We define a ``truncated'' partition function describing $N$ particles between $0$ and $x$, with a $(N+1)^{\mathrm{th}}$ particle at position $x$,
\begin{align}
    \label{eq:ZNnn}
    Z_{N}(\beta; x) =&
    \frac{1}{\ell_0^N}
    \int_0^x \dd x_1\, \e^{-\beta \Vnn(x-x_1)} \int_0^{x_1} \dd x_2\,
    \nonumber\\
    &\cdots \int_0^{x_{N-1}} \dd x_N
    \:
    \e^{- \beta \sum_{i=1}^{N-1} \Vnn(x_{i+1} - x_{i})}
    \:.
\end{align}
This partition function satisfies a recursion relation, which reads
\begin{equation}
  Z_{0}(\beta;x) = 1
  \:,
  \quad
  Z_{N+1}(\beta;x) =  \mathscr{L}[Z_{N}](x)
  \:,
\end{equation}
where $\mathscr{L}$ is the integral operator
\begin{equation}
  \mathscr{L}[f](x) = \frac{1}{\ell_0} \int_0^x \e^{-\beta \Vnn(x-x')} f(x') \dd x'
  \:.
\end{equation}
We can thus obtain explicitly the grand canonical partition function
\begin{equation}
    \Xi(\beta,\mu;x) = \sum_{N=0}^\infty \e^{N \beta \mu} Z_N(\beta;x)
    = (1 - \e^{\beta \mu} \mathscr{L})^{-1}[1](x)
    \:.
\end{equation}
Applying the operator $(1 - \e^{\beta \mu} \mathscr{L})$ on each side, we obtain an integral equation satisfied by $\Xi$,
\begin{equation}
    \label{eq:IntegEqGrandPartNN}
    \Xi(\beta,\mu;x) = 1 
    + \frac{\e^{\beta \mu}}{\ell_0} \int_0^x \e^{-\beta \Vnn(x-x')}\Xi(\beta,\mu;x') \dd x'
    \:.
\end{equation}
This equation can be solved by a Laplace transform. Indeed, we can obtain an equation for the Laplace transform of $\Xi$,
\begin{equation}
    \hat{\Xi}(\beta,\mu;s) = \int_0^\infty \e^{-s x} \Xi(\beta,\mu;x) \dd x
    \:,
\end{equation}
by multiplying each side of~\eqref{eq:IntegEqGrandPartNN}
by $\e^{-s x}$ and integrating over $x$; we thus obtain
\begin{equation}
    \label{eq:LaplaceEqGrandPartNN}
    \hat{\Xi}(\beta,\mu;s) = \frac{1}{s} 
    + \frac{\e^{\beta\mu}}{\ell_0} \hat{\Xi}(\beta,\mu;s) \hat{\mathcal{V}}(\beta;s)
    \:,
\end{equation}
where we have defined
\begin{equation}
    \label{eq:LaplaceExpV}
    \hat{\mathcal{V}}(\beta;s) = \int_0^\infty \e^{-s x - \beta \Vnn(x)} \dd x
    \:.
\end{equation}
Solving~\eqref{eq:LaplaceEqGrandPartNN} we obtain
\begin{equation}
    \label{eq:ResLapGrandPart}
    \hat{\Xi}(\beta,\mu;s) = \frac{1}{s( 1 - \frac{\e^{\beta \mu}}{\ell_0} \hat{\mathcal{V}}(\beta;s))}
    \:.
\end{equation}
The grand partition function can be obtained by an inverse Laplace transform of~\eqref{eq:ResLapGrandPart}. In the thermodynamic limit of a large system size $x$, $\Xi(\beta,\mu;x)$ will grow exponentially as $\e^{x s_\star}$, with $s_\star$ the rightmost pole of~\eqref{eq:ResLapGrandPart}. We therefore obtain that the grand potential density $\phi_G$ is given by
\begin{equation}
    \label{eq:DefSforNN}
    \beta \phi_G(\beta,\mu) \equiv -\lim_{x \to \infty} \frac{1}{x} \ln \Xi(\beta,\mu;x) = - s_\star(\beta,\mu)
    \:,
\end{equation}
where $s_\star(\beta,\mu)$ is solution of
\begin{equation}
    \label{eq:EqForSNN}
    \hat{\mathcal{V}}(\beta;s_\star(\beta,\mu)) = \ell_0 \e^{-\beta \mu}
    \:.
\end{equation}
As for the Calogero gas above, we perform a Legendre transform to go from a grand-canonical picture (with the chemical potential $\mu$ as a parameter) to a canonical description (with $\rho$ as parameter). The chemical potential $\mu(\rho)$ is again solution of~\eqref{eq:EqforMuCalo}, which gives here
\begin{equation}
  \label{eq:EqMuNN0}
    \beta \rho = \partial_\mu s_\star(\beta, \mu(\rho))
    \:.
\end{equation}
This equation involves a derivative with respect to $\mu$ of $s_\star$, which can be computed by differentiating~\eqref{eq:EqForSNN}, which gives
\begin{equation}
    \partial_\mu s_\star(\beta,\mu) =
    - \frac{\beta \ell_0 \e^{-\beta \mu}}{\partial_s \hat{\mathcal{V}}(\beta,s_\star(\beta,\mu))}
    \:.
\end{equation}
Evaluating this expression at $\mu = \mu(\rho)$ and using~\eqref{eq:EqMuNN0} gives a new equation for the chemical potential $\mu(\rho)$, which is more convenient for numerical evaluations,
\begin{equation}
    \label{eq:EqMuNN}
    \rho \: \partial_s \hat{\mathcal{V}}(\beta,s_\star(\beta,\mu(\rho))) 
    + \ell_0 \e^{-\beta \mu(\rho)} = 0
    \:.
\end{equation}
Computing the derivative with respect to $\rho$ of this equation, we deduce the diffusion coefficient from~\eqref{eq:DiffCoefMu},
\begin{equation}
  \label{eq:DiffNN}
  D(\rho) = \frac{\mu_0 k_{\mathrm{B}} T}
  {\rho^2 \e^{\beta\mu(\rho)} \partial_s^2 \hat{\mathcal{V}}(\beta,s_\star(\beta,\mu(\rho)))/\ell_0 - 1}
  \:.
\end{equation}
Finally, using Eq.~(\ref{eq:EqForSNN}), we can eliminate the chemical potential
from~(\ref{eq:EqMuNN},\ref{eq:DiffNN}) to obtain a parametric
representation:
\begin{subequations}
  \label{eq:RepDNNwithS}
  \begin{align}
    \label{eq:RhoFromS}
    \rho &=
           -\frac{\hat{\mathcal{V}}(\beta,s)}{\partial_s \hat{\mathcal{V}}(\beta,s)}
           \:,
    \\
    D(\rho)
         &= \frac{\mu_0 k_{\mathrm{B}} T}{\rho^2 \partial_s^2 \hat{\mathcal{V}}(\beta,s)/\hat{\mathcal{V}}(\beta,s) - 1 }
           \:,
  \end{align}
\end{subequations}
with $s$ now acting as a parameter.
Note that these equations no longer involve the thermal de Broglie wavelength $\ell_0$, as expected.

The result~\eqref{eq:RepDNNwithS} can be directly applied to obtain the exact diffusion coefficient in a system of nearest-neighbor interacting Brownian particles for paradigmatic models --- this will be done in the next sections. Alternatively, it can be used as an approximation of the original system described by~\eqref{eq:EqBrownianPart} (in which all the particles interact with each other) if the interaction potential decays sufficiently fast, or if the density is low enough so that the interaction becomes effectively nearest neighbor. To test this approximation, we show in Fig.~\ref{fig:DCalo} the result for the Calogero gas, in Fig.~\ref{fig:DDip} for the Dipole gas with potential
\begin{equation}
    \label{eq:DipPot}
    \Vnn(x) = \frac{g}{x^3}
    \:,
\end{equation}
and in Fig.~\ref{fig:DWCA} for the Weeks-Chandler-Andersen potential
\begin{equation}
    \label{eq:WCAPot}
    \Vnn(x) = g \left\lbrace
    \begin{array}{cc}
    \displaystyle
    4 \left[ \left( \frac{l}{x} \right)^{12} - \left( \frac{l}{x} \right)^{6} \right] + 1 \:,
    & \abs{x} < l \: 2^{1/6}
    \:,
    \\[0.3cm]
    0 \:, & \abs{x} > l \: 2^{1/6}
    \:,
    \end{array}
    \right.
\end{equation}
which are involved in experimental realisations of colloids, such as~\cite{Wei:2000}.
We observe that the approximation is always correct for $\rho \to 0$, with a range of validity that increases as the decay of the potential becomes faster. In particular, both for the dipole and the WCA potential, the nearest-neighbour approximation is systematically better than the $O(\rho^2)$ expansion of the diffusion coefficient that can be computed from the virial expansion (see Section~\ref{sec:Virial} below).

\subsubsection{The gas of hard rods}
\label{sec:HardRods}

We first consider, as a direct illustration of the result~\eqref{eq:RepDNNwithS}, the case of the gas of hard rods of length $\ell$. This corresponds to a nearest-neighbor interaction~\eqref{eq:BrownianNN}, with the potential
\begin{equation}
\label{eq:PotHR}
    \Vnn(x) = 
    \left\lbrace
    \begin{array}{ll}
         + \infty & \text{for } |x| < \ell \\
        0 & \text{for } |x| > \ell 
    \end{array}
    \right.
    \:.
\end{equation}
Applying the formalism of Section~\ref{sec:NearNeigh}, the diffusion
coefficient~(\ref{eq:RepDNNwithS}) is expressed in terms of
\begin{equation}
  \hat{\mathcal{V}}(\beta,s) = \int_0^\infty \e^{-s x - \beta \Vnn(x)} \dd x
  = \frac{\e^{-s \ell}}{s}
  \:.
\end{equation}
Plugging this expression into~(\ref{eq:RepDNNwithS}) gives the
parametric representation
\begin{equation}
  \rho = \frac{s}{1+\ell s}
  \:,
  \quad
  D(\rho) = \frac{\mu_0 k_{\mathrm{B}}T s^2}{2\rho^2(1 +  \ell s) - s^2(1-\rho^2 \ell^2) }
  \:.
\end{equation}
Solving the first equation for $s$ and inserting the result into the second equation yields
\begin{equation}
  \label{eq:DiffHR}
  D(\rho) = \frac{\mu_0 k_{\mathrm{B}}T}{(1-\rho \ell)^2}
  \:.
\end{equation}
This is indeed the well-known diffusion coefficient for a gas of hard rods~\cite{Lin:2005}. For $\ell = 0$, we recover the diffusion coefficient of pointlike particles~\eqref{eq:Dfree}. For $\ell > 0$, the diffusion coefficient diverges at the maximal density $1/\ell$ of the system.

\subsubsection{The gas of sticky hard rods}

Another classical model of nearest-neighbour interacting particles in one dimension is given by the sticky hard rods, corresponding to the interaction potential $V_0$ defined by~\cite{Baxter:1968,Percus:1982}
\begin{equation}
    \e^{- \beta \Vnn(x)} = \Theta(\abs{x} - \ell) + \gamma \delta(\abs{x} - \ell)
    \:,
\end{equation}
where $\ell$ is the size of the particles and $\gamma$ a parameter which controls the adhesiveness of the particles. This potential can be seen as the limit of a piecewise constant potential
\begin{equation}
    \Vnn(x) = \left\lbrace
    \begin{array}{ll}
        + \infty & \abs{x} < \ell \:, \\
        kT \ln[\varepsilon/\gamma] & \ell < \abs{x} < \ell + \varepsilon \:,\\
        0 & \abs{x} > \ell + \varepsilon \:,
    \end{array}
    \right.
\end{equation}
when $\varepsilon \to 0$~\cite{Baxter:1968}.
The diffusion coefficient of this model has recently been computed in~\cite{Schweers:2023}, relying on a previous result on the pair correlation function~\cite{Yuste:1993}. We can easily recover this result by using the parametrisation~\eqref{eq:RepDNNwithS}, which becomes
\begin{subequations}
  \begin{align}
      \rho &= \frac{s(1+\gamma s)}{1+ \ell s (1+\gamma s)}
      \:,
      \\
      D(\rho) &=
      \frac{\mu_0 k_{\mathrm{B}}T  \: s^2(1+\gamma s)}
      {2 \rho^2 (1 + 2 \ell s) - s^2 (1 + s\gamma) (1-\ell^2 \rho^2)}
      \:.
  \end{align}
\end{subequations}
Solving the first equation for $s$ and inserting it into the second equation yields and explicit form for the diffusion coefficient,
\begin{equation}
    D(\rho) = \frac{\mu_0 k_{\mathrm{B}} T}{(1 - \rho \ell)^2
    \sqrt{1 + \frac{4 \gamma \rho}{1 - \rho \ell}}} 
    \:,
\end{equation}
which coincides with the result of~\cite{Schweers:2023}, as it should. This diffusion coefficient is similar to the one of the hard-rods gas~\eqref{eq:DiffHR}, but divided by the factor $\sqrt{1 + 4 \gamma \rho/(1 - \rho \ell)}$ which encodes the effect of the stickiness of the particles. In particular, we see that the adhesiveness decreases the diffusion coefficient, and this effect becomes stronger as the density increases, as could be expected.

\bigskip

Up to now, we have recovered known expressions of $D(\rho)$ from our approach. We now turn to situations in which the diffusion coefficient is not known, but can be obtained from~\eqref{eq:RepDNNwithS}.

\subsubsection{The Rouse chain with hardcore repulsion}
\label{sec:Rouse}

We now consider the important model of the Rouse chain, here with additional and physically relevant hardcore repulsion. The standard Rouse model (without hardcore interaction) has been introduced to model a polymer, in which the monomers are attached by harmonic springs~\cite{Rouse:1953}. We derive here the collective diffusion coefficient of the Rouse model with hardcore interaction in one dimension, based on~\eqref{eq:RepDNNwithS}.
To this end, we consider
a chain of particles of length $\ell$ attached by harmonic springs. This is indeed a model in which a particle interacts only with its nearest neighbors~\eqref{eq:BrownianNN}, so that we can apply the formalism of Section~\ref{sec:NearNeigh}, with the interaction potential
\begin{equation}
\label{eq:PotRouse}
    \Vnn(x) = 
    \left\lbrace
    \begin{array}{ll}
         + \infty & \text{for } |x| < \ell \\
        g (x-\ell)^2 & \text{for } |x| > \ell 
    \end{array}
    \right.
    \:.
\end{equation}
Note that, in this model, there is a maximal density of particles: $\rho < \frac{1}{\ell}$.

\begin{figure*}
    \centering
    \subfloat[]{
        \includegraphics[width=0.32\textwidth]{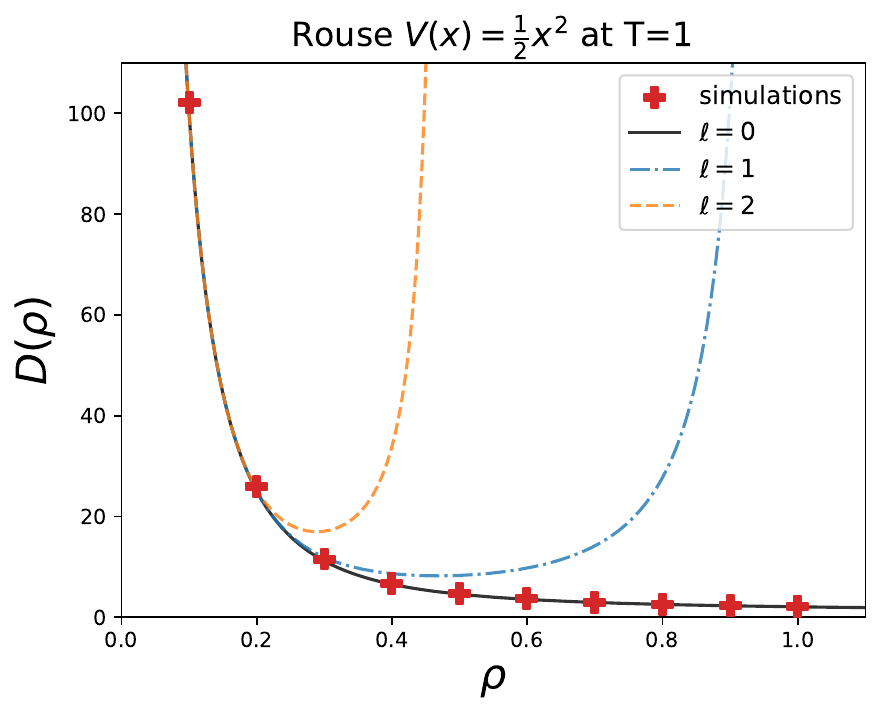}
        \label{fig:DiffRouse}
        \put(-168,10){(a)}
    }
    \subfloat[]{
        \includegraphics[width=0.32\textwidth]{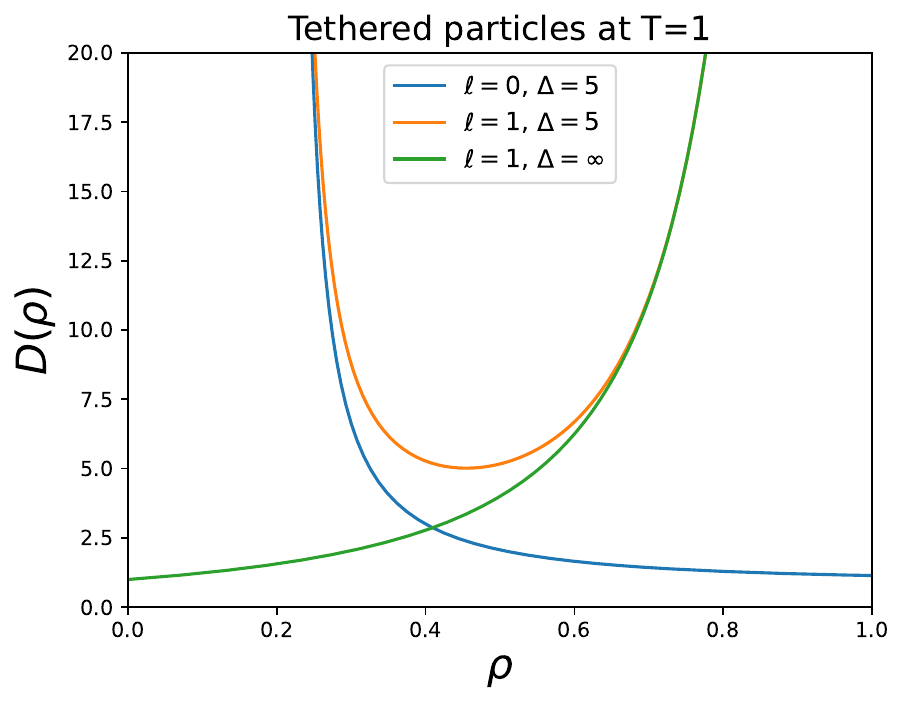}
        \label{fig:DiffTethered}
        \put(-168,10){(b)}
    }
    \subfloat[]{
        \includegraphics[width=0.32\textwidth]{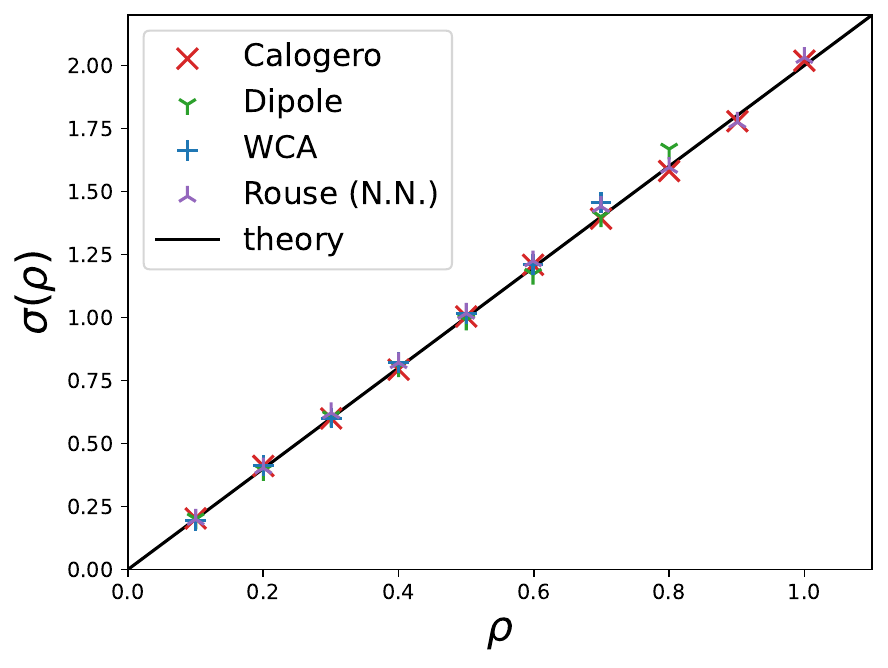}
        \label{fig:Sigma}
        \put(-168,10){(c)}
    }
    \vspace{-10pt}
    \caption{(a) Diffusion coefficient for the Rouse chain with nearest-neighbour interaction potential~(\ref{eq:PotRouse}) with $g=\frac{1}{2}$, for different
    values of the particle size $\ell$.
    We have performed simulations for the case $\ell=0$ only, since the arbitrary size can be deduced from it, see Section~\ref{sec:ParticSize}.
    (b) Diffusion coefficient for the chain of tethered particles with nearest-neighbour interaction potential~(\ref{eq:PotTethered}), for different
    values of the particle size $\ell$ and length $\Delta$ of the tether. We have not performed numerical simulations in this case, because the sharp potential~\eqref{eq:PotTethered} is tricky to implement numerically.
    (c) Mobility $\sigma(\rho)$ for the different interaction potentials considered in this article. The points are obtained from numerical simulations.
    }
\end{figure*}

The diffusion coefficient~\eqref{eq:DiffNN} is fully determined by the Laplace transform of $\e^{-\beta \Vnn}$~\eqref{eq:LaplaceExpV}, which is here explicitly given by
\begin{equation}
  \hat{\mathcal{V}}(\beta;s) = 
  \sqrt{\frac{\pi}{4\beta g}} \: 
  \e^{\frac{s^2}{4\beta g}}
  \erfc \left(
    \frac{s + 2 \beta g \ell}{2 \sqrt{\beta g}}
  \right)
  \:.
\end{equation}
Inserting this expression into the parametric representation~(\ref{eq:RepDNNwithS}) allows to compute numerically the diffusion coefficient. The result obtained is shown in Fig.~\ref{fig:DiffRouse} for different sizes $\ell$ of the particles.
For $\ell \neq 0$, the diffusion coefficient is non monotonic, and diverges both at low and high density. The behaviour near these divergences can be computed explicitly from the parametric representation~(\ref{eq:RepDNNwithS}). For $s \to \infty$, we obtain
\begin{subequations}
  \begin{align}
    \rho
    &=
      \frac{1}{\ell}
      - \frac{1}{\ell^2 s }
      + O(s^{-2})
      \:,
    \\
    D(\rho) &= \frac{\mu_0 \ell^2 s^2}{\beta}
              \:.
  \end{align}
\end{subequations}
Combining these expressions, we get
\begin{equation}
  \label{eq:DRouseDense}
  D(\rho)
  \underset{\rho \to 1/\ell}{\simeq}
  \frac{\mu_0 k_{\mathrm{B}}T}{(1-\rho \ell)^2}
  \:.
\end{equation}
This way we recovered exactly the diffusion coefficient of a gas of hard
rods~(\ref{eq:DiffHR}). This is expected, since for $\rho \to 1/\ell$
the particles are very close to each other, so their interaction is
negligible.
Note that, in the case $\ell = 0$, the same procedure yields
\begin{equation}
  \label{eq:DRouseDenseInfinity}
  D(\rho)  \underset{\rho \to \infty}{\simeq}
  \mu_0 k_{\mathrm{B}}T
  + \frac{4 g \mu_0}{\rho^2}
  + \cdots
  \:.
\end{equation}
Performing the same analysis in the opposite limit
$s \to - \infty$, we obtain
\begin{equation}
  \label{eq:DRouseDilute}
  D(\rho) \underset{\rho \to 0}{\simeq}
  \frac{2 \mu_0 g}{\rho^2}
  \:.
\end{equation}
In particular, in the dilute limit $\rho \to 0$ the diffusion coefficient does not
depend on the size $\ell$ of the particles, which is expected since
the size $\ell$ is negligible compared to the interparticle distance
$1/\rho$.  The diffusion coefficient~(\ref{eq:DRouseDilute}) diverges
as $\rho \to 0$, due to the large harmonic forces at play in this
case.

Remarkably, here the parameter $s$ covers the full real axis when the density varies between its minimal and maximal values. This was not the case, for instance, for the 
hard rods discussed in Section~\ref{sec:HardRods}, where $s$ was restricted to the positive axis. Going back to the definition of $s$~\eqref{eq:DefSforNN}, $s<0$ implies positive grand potential and thus negative pressure. This happens for densities $\rho < \rho_c$, where $\rho_c$ is obtained from~\eqref{eq:RhoFromS} with $s=0$,
\begin{equation}
    \rho_c =  \sqrt{\pi \beta g} \: \e^{\beta g \ell^2} \erfc \left( \ell \sqrt{\beta g} \right)
    \:.
\end{equation}
The origin of this change of sign of the pressure can be traced back to the fact that the interaction potential $V_0(x)$ does not decay to zero at infinity. This means that at long distances the system is dominated by attractive forces, leading to negative pressures. We checked in numerical simulations, see Fig.~\ref{fig:DiffRouse}, that the diffusion coefficient obtained from the derivative of the pressure~\eqref{eq:DiffCoef} is indeed correct for all densities $\rho$.

\subsubsection{The model of tethered particles with hardcore repulsion}

We consider now the recently introduced model of tethered random
walkers~\cite{Yuste:2025}, corresponding to hardcore particles of size
$\ell$, attached to the next particle with a cable of length
$\Delta - \ell > 0$, so that the centers of two neighbouring particles cannot be separated by a distance larger than $\Delta$. This
corresponds to the nearest-neighbor model~\eqref{eq:BrownianNN} with the potential
\begin{equation}
  \label{eq:PotTethered}
  \Vnn(x) = 
  \left\lbrace
    \begin{array}{ll}
      + \infty & \text{for } |x| < \ell \:, \\
      0 & \text{for } \Delta > |x| > \ell \:, \\
      + \infty & \text{for } |x| >  \Delta \:.
    \end{array}
  \right.
\end{equation}
In this model, there is both a maximal density (when all particles touch) and a minimal density (when all the particles are spaced by $\Delta$),
\begin{equation}
  \frac{1}{\Delta} < \rho < \frac{1}{\ell}
  \:.
\end{equation}
The diffusion coefficient~(\ref{eq:RepDNNwithS}) is controlled by the
Laplace transform of $\e^{-\beta \Vnn}$, which is given
in this case
by
\begin{equation}
  \hat{\mathcal{V}}(\beta, s) = \frac{\e^{-\ell s} - \e^{-\Delta s}}{s}
  \:.
\end{equation}
Combining this expression with Eq.~(\ref{eq:RepDNNwithS}), we can
compute numerically the diffusion coefficient. The result is shown in
Fig.~\ref{fig:DiffTethered}. For $\Delta \to \infty$, we recover the diffusion coefficient of the hard rods, as expected since the tethers play no role in this case. The limiting behaviors of the diffusion coefficient can be obtained by
considering the limits $s \to \pm \infty$. For $s \to \infty$, we get
\begin{equation}
  \label{eq:DTethDense}
  D(\rho) \underset{\rho \to 1/\ell}{\simeq}
  \frac{\mu_0 k_{\mathrm{B}}T}{(1-\rho \ell)^2}
  \:,
\end{equation}
which is again identical to the diffusion coefficient of a gas of hard rods~\eqref{eq:DiffHR}, since at high density only the hardcore interaction between the particles is relevant, and not the presence of the tethers. In the opposite limit $s \to -\infty$, we obtain
\begin{equation}
  \label{eq:DTethDilute}
  D(\rho) \underset{\rho \to 1/\Delta}{\simeq}
  \frac{\mu_0 k_{\mathrm{B}}T}{(1-\rho \Delta)^2}
  \:.
\end{equation}
Note that, as for the case of the Rouse chain, the parameter $s$ takes both positive and negative values, indicating that the system exhibits again a negative pressure due to the effective attractive forces caused by the tethers.

\subsubsection{A comment on the effect of the size of the particles}
\label{sec:ParticSize}

Let us consider the case of particles of length $\ell > 0$ interacting via nearest-neighbour interaction, with an arbitrary interaction potential
\begin{equation}
  \Vnn(x) = 
  \left\lbrace
    \begin{array}{ll}
      + \infty & \text{for } |x| < \ell \:, \\
      \Vtnn(x - \ell) & \text{for } |x| >  \ell \:.
    \end{array}
  \right.
\end{equation}
We now show that the effect of the finite size $\ell$ can be absorbed by a change of function, so that it is sufficient to study the case $\ell = 0$. Indeed, the partition function~\eqref{eq:ZNnn} takes the form
\begin{multline}
    Z_{N}(\beta; x) =
    \frac{1}{\ell_0^N}
    \int_0^{x-\ell} \dd x_1 \e^{-\beta \Vtnn(x-x_1 - \ell)} 
    \int_0^{x_1 - \ell}
    \dd x_2
    \\
    \cdots 
    \int_0^{x_{N-1} - \ell} \dd x_N
    \:
    \e^{- \beta \sum_{i=1}^{N-1} \Vtnn(x_{i+1} - x_{i} - \ell)}
    \:.
\end{multline}
Performing the change of variables $x_i = y_i +(N-i) \ell$, this becomes
\begin{multline}
    Z_{N}(\beta; x) =
    \frac{1}{\ell_0^N}
    \int_0^{x-N\ell} \dd y_1 \e^{-\beta \Vtnn(x - N\ell-y_1)} 
    \int_0^{y_1}
    \dd y_2
    \\
    \cdots 
    \int_0^{y_{N-1}} \dd y_N
    \:
    \e^{- \beta \sum_{i=1}^{N-1} \Vtnn(y_{i+1} - y_{i})}
    \:.
\end{multline}
This is exactly the partition function of $N$ pointlike particles in a region of size $x - N \ell$. Therefore, for large $x$ with $N/x = \rho$ constant, we have
\begin{equation}
    -\frac{1}{\beta} \ln Z_{N}(\beta;x) 
    \simeq  x f(\rho) = (x - N \ell) \tilde{f} \left( \frac{N}{x - N \ell} \right)
    \:,
\end{equation}
where $\tilde{f}(\rho)$ is the free energy density for the system of pointlike Brownian particles interacting with the nearest-neighbour potential $\Vtnn$. Therefore, we have
\begin{equation}
    f(\rho) = (1 - \rho \ell) \tilde{f} \left( \frac{\rho}{1 - \rho \ell} \right)
    \:.
\end{equation}
Using the expression of the diffusion coefficient in terms of the free energy~\eqref{eq:DiffCoef0}, we obtain
\begin{equation}
    D(\rho) = \mu_0 \frac{\rho}{(1-\rho \ell)^3} 
    \tilde{f}'' \left( \frac{\rho}{1 - \rho \ell} \right)
    \:.
\end{equation}
Introducing the diffusion coefficient $\tilde{D}(\rho) = \mu_0 \rho \tilde{f}''(\rho)$ for the system of pointlike particles, we get the relation
\begin{equation}
    \label{eq:TransfDFiniteSizePart}
    D(\rho) = \frac{1}{(1-\rho \ell)^2} 
    \tilde{D} \left( \frac{\rho}{1 - \rho \ell} \right)
    \:.
\end{equation}
It is thus sufficient to study the system with pointlike particles $\ell = 0$ to determine the diffusion coefficient of the system with finite-size particles $\ell > 0$. In particular, this yields the diffusion coefficient of the hard rod gas~\eqref{eq:DiffHR} from the one of reflecting Brownian particles $\tilde{D}(\rho) = \mu_0 k_{\mathrm{B}} T$. For the case of the Rouse chain~\eqref{eq:PotRouse} or the tethered particles~\eqref{eq:PotTethered}, this implies that one can focus on the case $\ell = 0$ only, 
while also obtaining
results for $\ell > 0$ from~\eqref{eq:TransfDFiniteSizePart}.

\subsection{Arbitrary dimensions}
\label{sec:d-coefficients}

In arbitrary spatial dimension, there are few exact results for the pressure in systems of interacting particles, besides the case of noninteracting particles described by the equation of state of 
ideal gases~\eqref{eq:PerfGas}. Different approximations have been developed to obtain corrections to this equation of state in the presence of interactions, the most famous being the virial expansion~\cite{Hill:1986}, which we now recall.
We also discuss how the pressure, and thus the diffusion coefficient, can be derived from classical approximations, such as the Percus–Yevick or Hypernetted-chain, for the pair correlation function~\cite{Hansen:2005}.

\subsubsection{Virial expansion}
\label{sec:Virial}

The virial expansion gives the equation of state $P(\rho)$ as a power series in the density $\rho$~\cite{Hill:1986},
\begin{equation}
    \label{eq:Pvirial}
    \frac{P(\rho)}{ k_{\mathrm{B}} T } = \rho + 
    B_2(T) \rho^2 + B_3(T) \rho^3 + \O(\rho^4)
    \:,
\end{equation}
where $B_2$ and $B_3$ are the virial coefficients, which are explicitly given by
\begin{align}
    \label{eq:VirialB2}
    B_2(T) 
    &= - \frac{1}{2V} \int_V \dd^d \boldsymbol{x}_1 \int_V \dd^d \boldsymbol{x}_2
    \left( \e^{-\beta V_0(\boldsymbol{x}_1 - \boldsymbol{x}_2)} - 1\right)
    \nonumber
    \\
    &\simeq
    - \frac{1}{2} \int \dd^d \boldsymbol{x}_1 
    \left( \e^{-\beta V_0(\boldsymbol{x}_1)} - 1\right)
    \:,
\end{align}
and
\begin{align}
    B_3(T) 
    =& - \frac{1}{3V} \int_V \dd^d \boldsymbol{x}_1 
    \int_V \dd^d \boldsymbol{x}_2
    \int_V \dd^d \boldsymbol{x}_3
    \fm(\boldsymbol{x}_1  - \boldsymbol{x}_2) 
    \nonumber
    \\
   & \times 
    \fm(\boldsymbol{x}_1  - \boldsymbol{x}_3) 
    \fm(\boldsymbol{x}_2  - \boldsymbol{x}_3)
    \label{eq:VirialB3}
    \\
    \simeq&
    - \frac{1}{3} \int \dd^d \boldsymbol{r}_1 
    \int \dd^d \boldsymbol{r}_2
    \fm(\boldsymbol{r}_1) 
    \fm(\boldsymbol{r}_2) 
    \fm(\boldsymbol{r}_2  - \boldsymbol{r}_1)
    \nonumber
    \:,
\end{align}
in the thermodynamic limit of infinite volume $V \to \infty$ and where
\begin{equation}
    \fm(\V{x}) = \e^{- \beta V_0(\V{x})} - 1
\end{equation}
is the Mayer function. Inserting these expressions into the relation~\eqref{eq:DiffCoef}, we obtain the exact low-density behavior of the diffusion coefficient for any interaction potential $V_0$,
\begin{align}
    \label{eq:DfromVirial}
    &\frac{D(\rho)}{\mu_0 k_{\mathrm{B}} T} \simeq 1 
    - \rho \int \fm(\V{x}) \dd^d \V{x}
    \\
    &- \rho^2 \int \dd^d \V{x} 
    \int \dd^d \V{y} \fm(\V{x}) \fm(\V{y}) \fm(\V{x}-\V{y})
    + O( \rho^3 )
    \:. \nonumber
\end{align}
This expansion, up to the first two orders, is compared to numerical simulations in Fig.~\ref{fig:D-first3} for different interaction potentials $V_0$. Even if the virial expansion quickly deviates from the true value as the density increases, it provides a good approximation of the diffusion coefficient at low density, as it should.

\subsubsection{Virial expansion for the Riesz gas}

The Riesz gas is a well-studied model of interacting particles (see the review~\cite{Lewin:2022}), corresponding to the pairwise potential~\cite{Riesz:1988}
\begin{equation}
    V(\V{x}) = \frac{g}{|| \V{x} ||^s}
    \:.
\end{equation}
Except for the specific case $s=2$ in $d=1$, corresponding to the Calogero gas discussed above, the equation of state for the Riesz gas is not known. We can nevertheless apply the virial expansion~\eqref{eq:DfromVirial} to obtain the first orders of the density dependence of the diffusion coefficient $D(\rho)$. The second virial coefficient~\eqref{eq:VirialB2} is given by
\begin{equation}
    B_2(T) = - \frac{1}{2} \int 
    \left( \e^{- \beta g || \V{x} ||^{-s}} - 1\right) \dd^d \V{x} 
    \:.
\end{equation}
Going to spherical coordinates and integrating over the angular variables, we get
\begin{equation}
    B_2(T) =- \frac{\pi^{d/2}}{\Gamma(\frac{d}{2})}
    \int_0^\infty r^{d-1} \left( \e^{-\beta g/r^s} - 1\right) \dd r
    \:.
\end{equation}
This integral converges for $s > d$, which corresponds to the ``short-range'' case of the Riesz gas~\cite{Lewin:2022}. In the ``long-range'' situation $s \leq d$, the energy of the gas is no longer extensive and non-diffusive behavior arises~\cite{Dandekar:2023,Touzo:2024,Touzo:2024a}. Here we restrict ourselves to the diffusive regime, for which
\begin{equation}
    B_2(T) = \frac{\pi^{d/2}}{\Gamma(\frac{d}{2})} \frac{(\beta g)^{d/s}}{s}
    \Gamma \left( -\frac{d}{s} \right)
    \:.
\end{equation}
Therefore, combining the expression for the pressure~\eqref{eq:Pvirial} with~\eqref{eq:DiffCoef}, we obtain for the Riesz gas
\begin{equation}
    \label{eq:DRieszVirial}
    D(\rho) = \mu_0 k_{\mathrm{B}} T \left[
    1 + \rho \frac{2 \pi^{\frac{d}{2}} (\beta g)^{\frac{d}{s}}}
    {s \Gamma(\frac{d}{2})}
    \Gamma \left( \frac{-d}{s} \right)
    + O(\rho^2)
    \right]
    \:.
\end{equation}
In the case of the Calogero gas (i.e.~$d=1$ and $s=2$), this reduces to the leading order of~\eqref{eq:DCaloLowDens}, as expected. We have also checked numerically that the third virial coefficient~\eqref{eq:VirialB3} gives the correct next term of~\eqref{eq:DCaloLowDens}, although we have not been able to perform the integrals analytically.

Note that the Riesz gas was also studied in~\cite{Dandekar:2023}, both in the short-range and the long-range cases, relying on the Dean--Kawasaki equation~\eqref{eq:MicroCurrent}. However, in the short-range situation considered here, the method of Ref.~\cite{Dandekar:2023} applies only in the high-density limit, so it does not reproduce~\eqref{eq:DRieszVirial} (see the Supplemental Material of Ref.~\cite{Grabsch:2025b} for more details), unlike our approach which gives access to the full density dependence of the diffusion coefficient.

\subsubsection{Pressure from the pair correlation function}
\label{sec:PfromGr}

Another possibility to determine the pressure is to use the relation with the pair correlation function $g(r)$,
\begin{equation}
    \label{eq:Pfromgr}
    P(\rho) = \rho k_{\mathrm{B}} T 
    - \rho^2 \frac{\pi^{d/2}}{d \: \Gamma (\frac{d}{2})}
    \int_0^\infty r^d V_0'(r) g(r) \dd r
    \:,
\end{equation}
where $d$ is again the dimension.
In this expression, $g(r)$ depends on the density $\rho$, although it is not written explicitly. Since 
various approximations to determine the pair correlation function exist~\cite{Hansen:2005}, the pressure~\eqref{eq:Pfromgr} and thus the diffusion coefficient~\eqref{eq:DiffCoef} can be determined approximately in any dimension. For instance, $g(r)$ can be obtained from the hypernetted-chain approximation, which gives the integral equation~\cite{Hansen:2005}
\begin{multline}
    \label{eq:HNC}
    \ln g(\V{r}) + \beta V_0(\V{r})
    =
    \rho \int
    \left[ g(\V{r}') - 1 - \ln g(\V{r}') - \beta V_0 (\V{r}')\right]
    \\
    \times
    \left[ g(\V{r} - \V{r}') - 1 \right] 
    \dd^d \V{r}'
    \:.
\end{multline}
This equation can be solved numerically (see Appendix~\ref{app:SolHNC}) and gives an approximation of the diffusion coefficient $D(\rho)$. The result is shown in Fig.~\ref{fig:D-first3} for different interaction potentials $V_0$. In the case of the Calogero~\eqref{eq:CaloPot} and dipole~\eqref{eq:DipPot} potentials, it is the best approximation, in perfect agreement with the numerical simulations (and the exact result~\eqref{eq:DCalo} for the Calogero case). For the WCA potential~\eqref{eq:WCAPot}, which has a finite range, it still gives a better approximation than the virial expansion to second order, but the nearest-neighbour approximation is better in this case. This is 
presumably due to the fact that the potential has a finite range in this case, so that the nearest-neighbour approximation is almost exact in a 
wide range of densities.

\bigskip

To summarise, in this Section, we have provided (i) exact calculations for the collective diffusion coefficient $D(\rho)$ of interacting Brownian particles in one dimension, in particular for paradigmatic models, and (ii) approximations that give access to $D(\rho)$ for any interaction potential, in dimension one and beyond.

\section{Statistics of currents and tracers in one-dimensional systems}
\label{sec:OneDimensionMFT}

\begin{figure*}
    \centering
    \subfloat[]{
        \includegraphics[width=0.4\textwidth]{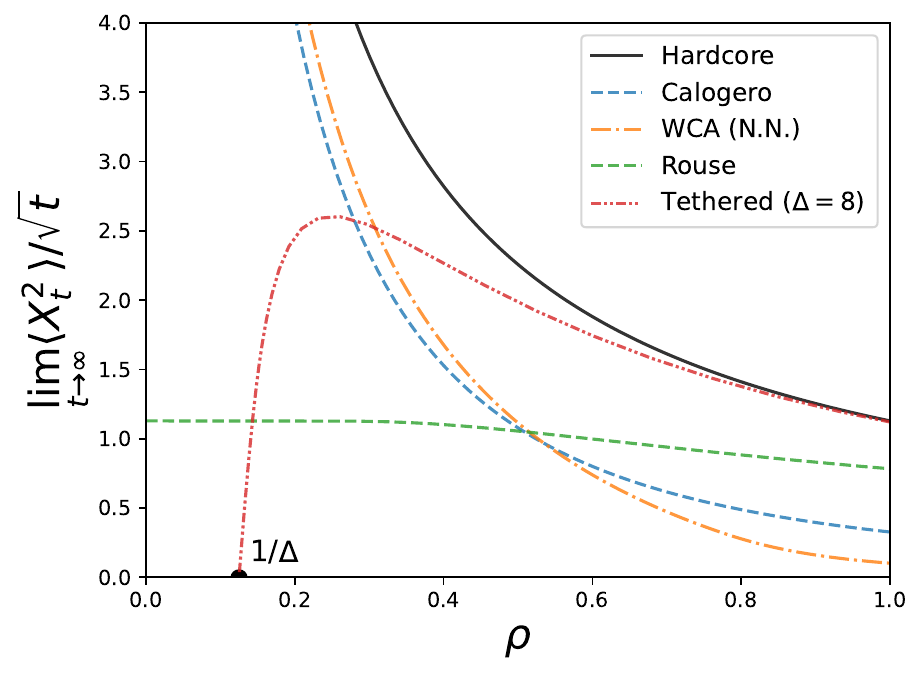}
        \label{fig:Xt2}
        \put(-205,7){(a)}
    }
    \hspace{20pt}
    \subfloat[]{
        \includegraphics[width=0.4\textwidth]{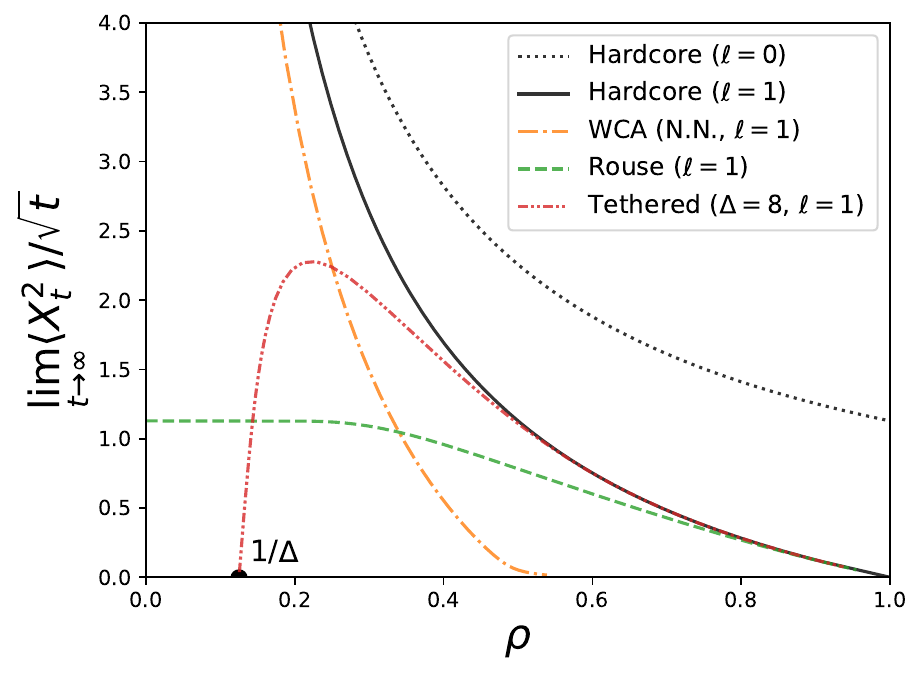}
        \label{fig:Xt2l1}
        \put(-205,7){(b)}
    }
    \vspace{-10pt}
    \caption{Prefactor of the asymptotic behaviour of $\moy{X_t^2}_c \sim \sqrt{t}$ as a function of the mean density $\bar\rho$, given in~\eqref{eq:Xt2} at $T=1$, in the case of (a) pointlike particles $\ell=0$, or (b) extended particles $\ell = 1$,
    for different interaction potentials considered in Section~\ref{sec:TrCoefs}: 
    (i) the Calogero potential~\eqref{eq:CaloPot} with $g=1$; 
    (ii) the WCA potential~\eqref{eq:WCAPot} with $g=1$ and $l = 1$ for nearest-neighbour interaction (which is a good approximation, see Fig.~\ref{fig:DWCA}); 
    (iii) the Rouse chain~\eqref{eq:PotRouse} with $g=\frac{1}{2}$; (iv) the gas of tethered particles~\eqref{eq:PotTethered} with $\Delta = 8$. 
    In (a), the black solid line represents the case of hardcore Brownian particles, for which $D(\rho)=\mu_0 k_{\mathrm{B}} T$, see Eq.~\eqref{eq:Dfree}.
    In (b), the black solid line represents the case of hardcore Brownian rods of length $\ell = 1$, for which $D(\rho)$ is given by~\eqref{eq:DiffHR}, while the dotted line corresponds to pointlike ($\ell=0$) hardcore Brownian particles.
    }
\end{figure*}%

In this Section, we combine various results previously obtained from the MFT formalism in one dimension~\cite{Krapivsky:2014,Krapivsky:2015a,Poncet:2021,Grabsch:2024b,Berlioz:2025}, and which are valid for any $D(\rho)$ and $\sigma(\rho)$, with the explicit expressions of these transport coefficients obtained above for different models of interacting Brownian particles. This allows us to obtain exact results for two important observables in these systems:
the displacement of a tracer and the integrated current through a given point.
Note that the study of a tracer in one dimension relies on the non-crossing of the particles. It is the only observable that requires the physically relevant hardcore particles. Concerning the integrated current, our discussion applies to hardcore or soft particles equivalently, although we will mostly consider hardcore particles.

In the remaining of the article, we consider the case of ``annealed'' initial conditions, corresponding to a system initially at equilibrium with a mean density $\bar\rho$. The MFT formalism can also be applied to ``quenched'' initial conditions, in which the system is initially prepared in a given configuration, see for instance~\cite{Krapivsky:2015a}, but we do not discuss this case here. Importantly, since both the displacement of a tracer and the integrated current are dynamical quantities, their determination remains a challenging task, even if the system is at equilibrium.

\subsection{Mean squared displacement of a tracer}

We consider a tracer, i.e.~we choose one particle in the system, and follow its displacement, denoted $X_t$ at time $t$. Initially, we have $X_0=0$ and we assume that the system is at equilibrium, with mean density $\bar\rho$. We consider that the interaction potential is repulsive enough so that the particles cannot cross (alternatively, we can add a hardcore interaction to prevent overlapping). 
In one dimension, this imposes the single-file constraint, which leads to a subdiffusive behavior of the displacement
$\moy{X_t^2} \sim \sqrt{t}$~\cite{Harris:1965}. Kollmann~\cite{Kollmann:2003} obtained the exact expression for this mean squared displacement, which was later expressed in terms of $D(\rho)$ and $\sigma(\rho)$ in
the MFT context~\cite{Krapivsky:2014,Krapivsky:2015a},
\begin{equation}
    \label{eq:Xt2}
  \moy{X_t^2}
  \underset{t \to \infty}{\simeq}
  \frac{\sigma(\bar\rho)}{\bar\rho^2 \sqrt{\pi D(\bar\rho)}} \sqrt{t}
  = \frac{2 \mu_0 k_{\mathrm{B}} T}{\bar\rho \sqrt{\pi D(\bar\rho)}} \sqrt{t}
  \:,
\end{equation}
where in the second equality we have used the expression of $\sigma$~\eqref{eq:Mobility}.
The notation $\moy{X_t^n}_c$ denotes the $n^{\mathrm{th}}$ cumulant of $X_t$, in particular $\moy{X_t^2}_c = \moy{X_t^2} - \moy{X_t}^2$ is the variance.
Combining this result with the expressions of $D(\rho)$ derived in Section~\ref{sec:TrCoefs}, we obtain the exact long-time behaviour of the mean squared displacement of a tracer in a single file of interacting Brownian particles.

The prefactor of the $\sqrt{t}$ behaviour given in Eq.~\eqref{eq:Xt2} is shown in Fig.~\ref{fig:Xt2} (pointlike particles, $\ell = 0$) and in Fig.~\ref{fig:Xt2l1} (extended particles, $\ell = 1$) as a function of the mean density $\bar\rho$ for different interaction potentials.
Let us make a few comments.
(i)
In most cases, the prefactor in~\eqref{eq:Xt2} decreases monotonically with the density, as expected since the subdiffusive behaviour of $X_t$ originates from the presence of the other particles around it, which hinder its displacement.
In the case of tethered particles, since at low density the particles cannot move because the tethers between them have reached their maximal length $\Delta$, this leads to a vanishing of the displacement of $X_t$ at the finite density $1/\Delta$, due to the divergence of the diffusion coefficient~\eqref{eq:DTethDilute}. For the Rouse chain, the divergence of the diffusion coefficient at zero density~\eqref{eq:DRouseDilute} is compensated by the $\bar\rho$ in the denominator of~\eqref{eq:Xt2}, yielding a finite value of the prefactor at zero density. 
(ii)
Both for the Rouse chain and the tethered particles, the high-density behaviour is identical to the case of purely hardcore repulsion, see Figs.~\ref{fig:Xt2loglog} and~\ref{fig:Xt2l1loglog}. This is expected, since the potentials can be neglected at short distances.
This is not the case for repulsive potentials, such as the Calogero and WCA potentials (the latter is considered within the nearest-neighbour approximation, which gives the best estimate for $D(\rho)$, see Fig.~\ref{fig:DWCA}). In this case, the low-density behaviour is identical to the noninteracting case, since the potential is small at long distances and can thus be neglected, see Figs.~\ref{fig:Xt2loglog} and~\ref{fig:Xt2l1loglog}. However, the prefactor decays faster than in the hardcore case when the density increases.

\begin{figure*}
    \centering
    \subfloat[]{
        \includegraphics[width=0.4\textwidth]{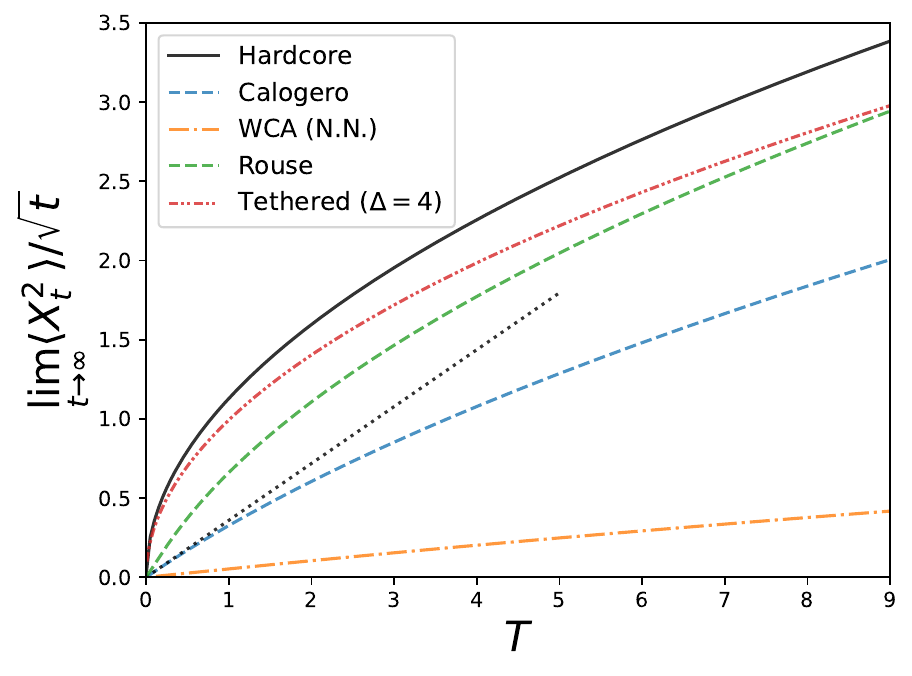}
        \label{fig:Xt2fctT}
        \put(-205,7){(a)}
    }
    \hspace{20pt}
    \subfloat[]{
        \includegraphics[width=0.4\textwidth]{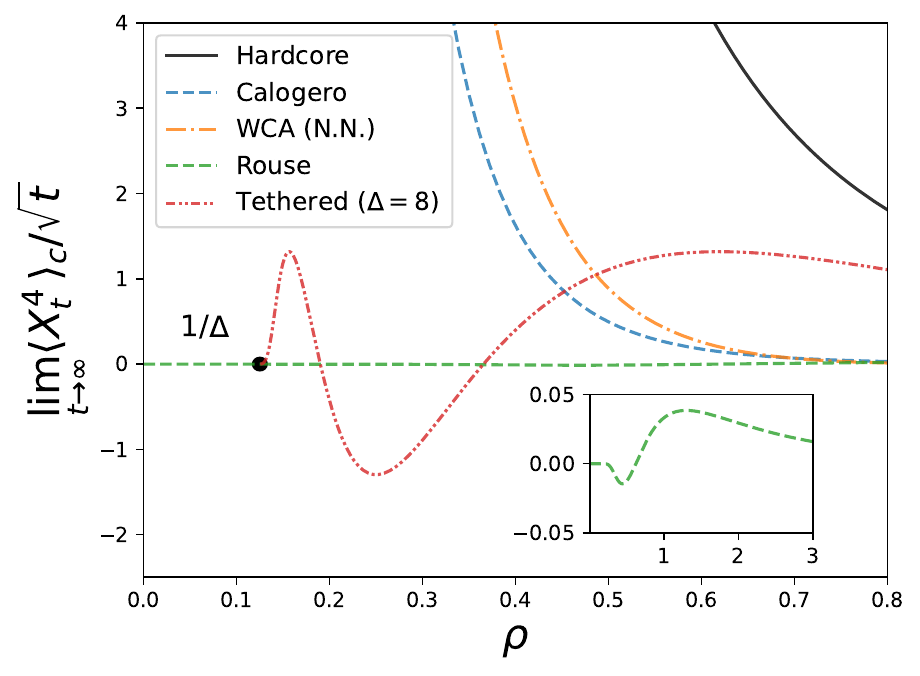}
        \label{fig:Xt4}
        \put(-205,7){(b)}
    }
    \vspace{-10pt}
    \caption{Prefactor of the asymptotic behaviour of (a) $\moy{X_t^2}_c \sim \sqrt{t}$ as a function of the temperature $T$, given in~\eqref{eq:Xt2}, and 
    (b) $\moy{X_t^4}_c \sim \sqrt{t}$ as a function of the mean density $\bar\rho$, given in~\eqref{eq:Xt4}. 
    In both panels we considered 
    the case of pointlike particles $\ell = 0$ at $\bar\rho=1$, for different interaction potentials considered in Section~\ref{sec:TrCoefs}: (i) the Calogero potential~\eqref{eq:CaloPot} with $g=1$; (ii) the WCA potential~\eqref{eq:WCAPot} with $g=1$ and $l = 1$ for nearest-neighbour interaction (which is a good approximation, see Fig.~\ref{fig:DWCA}); (iii) the Rouse chain~\eqref{eq:PotRouse} with $g=\frac{1}{2}$; (iv) the gas of tethered particles~\eqref{eq:PotTethered} with $\Delta = 4$ or $\Delta = 8$ (see the legend). In (a), the dotted line represents the low-temperature behaviour of the Calogero gas computed in~\cite{Touzo:2024,Touzo:2024a}.
    In (b), the inset is a zoom to show the variations of the prefactor for the Rouse chain, which is orders of magnitude below the other models. In both panels, the black solid line represents the case of hardcore Brownian particles.
    }
\end{figure*}%
\begin{figure*}
        \centering
    \subfloat[]{
        \includegraphics[width=0.3\textwidth]{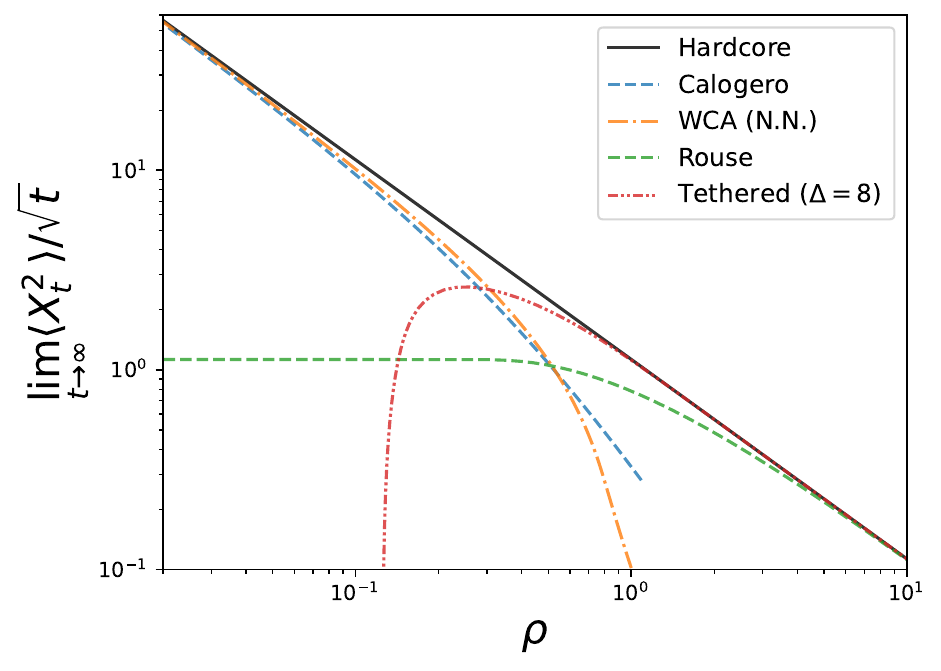}
        \label{fig:Xt2loglog}
        \put(-150,7){(a)}
    }
    \hspace{10pt}
    \subfloat[]{
        \includegraphics[width=0.3\textwidth]{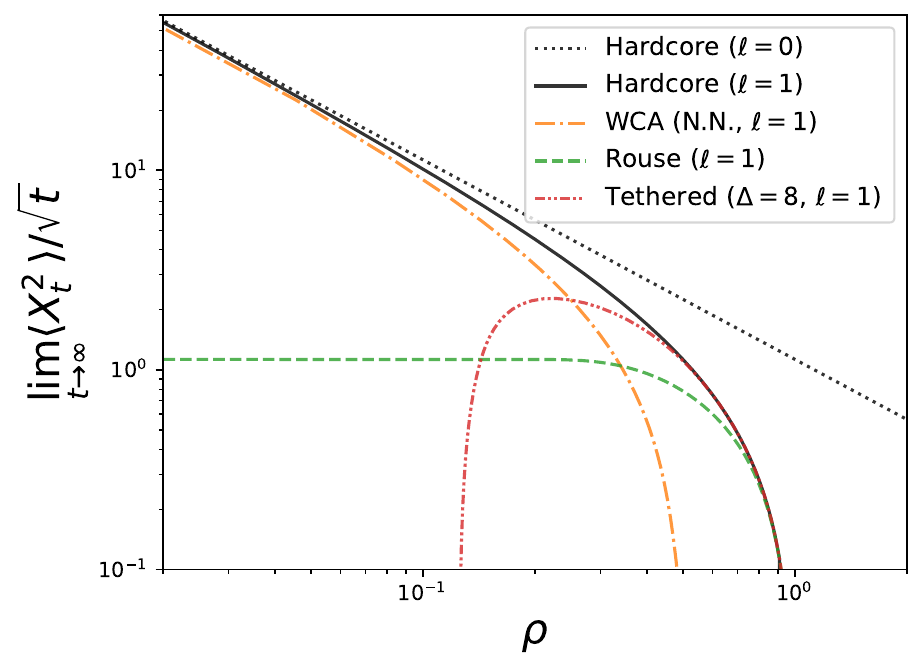}
        \label{fig:Xt2l1loglog}
        \put(-150,7){(b)}
    }
    \hspace{10pt}
    \subfloat[]{
        \includegraphics[width=0.3\textwidth]{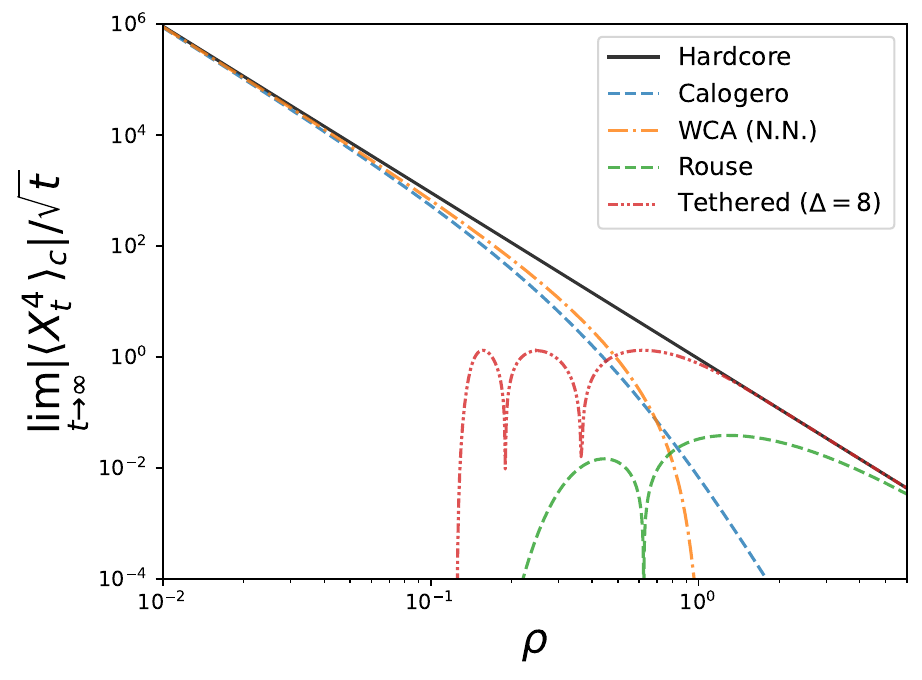}
        \label{fig:Xt4loglog}
        \put(-150,7){(c)}
    }
    \vspace{-10pt}
    \caption{ Same as Figs.~\ref{fig:Xt2},~\ref{fig:Xt2l1} and~\ref{fig:Xt4}, but in log-log scale, to better show the asymptotic behaviors of $\moy{X_t^2}_c$ and $\moy{X_t^4}_c$ at low and large densities.}
    \label{fig:Xt2Xt4LogLog}
\end{figure*}

We show in Fig.~\ref{fig:Xt2fctT} the prefactor of the mean squared displacement~\eqref{eq:Xt2} as a function of the temperature $T$. For all interaction potentials, the prefactor vanishes at $T=0$ due to the numerator of~\eqref{eq:Xt2}, and increases with the temperature. In particular, in the case of the Calogero potential, using the low-temperature behaviour of the diffusion coefficient~\eqref{eq:DCaloHighDens} gives
\begin{equation}
    \lim_{t \to \infty} \frac{\moy{X_t^2}_c}{\sqrt{t}}
    \underset{T \to 0}{\simeq}
    \frac{2 \mu_0 k_{\mathrm{B}} T}{\bar\rho^2 \sqrt{\mu_0 g \pi^3}}
    \:.
\end{equation}
This way we recover the result of~\cite{Touzo:2024,Touzo:2024a}, which was obtained from a microscopic computation at low temperature. This low-temperature prediction is also represented in Fig.~\ref{fig:Xt2fctT}.

\subsection{Fourth cumulant of the position of the tracer}

The distribution of the position $X_t$ of a tracer is known to be non-Gaussian. All the cumulants of the distribution have been obtained for reflecting Brownian particles (with no other interaction)~\cite{Hegde:2014,Krapivsky:2014,Krapivsky:2015a,Sadhu:2015}, and for the SEP~\cite{Illien:2013,Imamura:2017,Imamura:2021}. For a general system with arbitrary $D(\rho)$ and $\sigma(\rho)$, the fourth cumulant of $X_t$, which probes the deviation from the Gaussian distribution, has been computed recently~\cite{Grabsch:2024b} using the MFT formalism. For the specific choice of $\sigma(\rho)$~\eqref{eq:Mobility} corresponding to interacting Brownian particles the result of~\cite{Grabsch:2024b} becomes
\begin{widetext}
\begin{multline}
    \label{eq:Xt4}
    \frac{1}{(\mu_0 k_{\mathrm{B}} T)^3}
    \frac{\moy{X_t^4}_c}{\sqrt{t}} \underset{t \to \infty}{\simeq}
    \frac{24 \left(\bar\rho  D'(\bar\rho )+D(\bar\rho )\right)}
    {\pi ^{3/2} \bar\rho^3 D(\bar\rho )^{7/2}}
   -\frac{6 \left(
   \bar\rho D(\bar\rho )
   + \bar\rho^2 D'(\bar\rho )
   \right)}
   {\sqrt{\pi} \bar\rho^4 D(\bar\rho )^{7/2}}
   +\frac{3  \left(D'(\bar\rho )^2-D(\bar\rho ) D''(\bar\rho )\right)}
   {\sqrt{\pi } \bar\rho D(\bar\rho)^{9/2}}
   \\
   +\frac{3 \left(2 D(\bar\rho ) D''(\bar\rho )-D'(\bar\rho )^2\right)}
   {\pi ^{3/2} \bar\rho D(\bar\rho )^{9/2}}
   + \frac{3\left(\sqrt{2} \pi -2 \sqrt{3}\right) \left(2 D(\bar\rho ) D''(\bar\rho )-3 D'(\bar\rho )^2\right)}
   {2 \pi^{3/2} \bar\rho D(\bar\rho)^{9/2}}
    \:.
\end{multline}
\end{widetext}
Using the expressions of $D(\rho)$ obtained in Section~\ref{sec:TrCoefs}, we obtain the fourth cumulant for various models, see Fig.~\ref{fig:Xt4}. For the Calogero and WCA cases (approximated with only nearest-neighbor interaction, see Fig.~\ref{fig:DWCA}), the behaviour of the fourth cumulant of $X_t$ is qualitatively identical to the hardcore case, with a divergence as $1/\bar\rho^{3}$ at low density, and a monotonic decay to zero at high density. The decay is however much faster for the Calogero and WCA gases than in the hardcore one. For instance, the high-density behaviour of $D(\rho)$ for the Calogero gas~\eqref{eq:DCaloHighDens} implies a decay as $1/\bar\rho^{8}$ at high density (compared to $1/\bar\rho^{3}$ for the hardcore particles).
Conversely, the case of the Rouse chain or the tethered particles yield a bounded prefactor for the fourth cumulant, which changes sign as the density is varied. In particular, in the case of the Rouse chain, the fourth cumulant is orders of magnitude smaller than for the tethered gas (which was not the case for the second cumulant, see Figs.~\ref{fig:Xt2}-\ref{fig:Xt2fctT}), indicating that the distribution of $X_t$ is almost Gaussian in this case.

\subsection{Correlations between two tracers}

Let us now consider the joint displacement of two tracers, $X_t \equiv x_0(t) - x_0(0)$ and $Y_t \equiv x_k(t) - x_k(0)$ for a given label $k$.
Even without long-range interaction, the displacements of these particles are correlated, due to the presence of the surrounding particles.
The covariance of the positions of these two tracers has been computed using the MFT formalism in~\cite{Berlioz:2025}. 
For a large observation time $t$, we introduce $\xi = k/\sqrt{4 \bar\rho^2 D(\bar\rho) t}$, i.e.~the rescaled number of particles between the tracers, corresponding to an initial average distance $k/\bar\rho$ between the two tracers. The covariance is then given by~\footnote{There is a typo in the formula given in Ref.~\cite{Berlioz:2025}. The correct expression is~\eqref{eq:CovXtYt}.}
\begin{equation}
\label{eq:CovXtYt}
   \frac{\moy{X_t Y_t}_c}{\sqrt{t}}
   \underset{t \to +\infty}{\simeq}
   \frac{\sigma (\bar\rho )}{\bar\rho^2 \sqrt{D(\bar\rho )}}
    \mathcal{G}( \xi )
    =  \frac{2 \mu_0 k_{\mathrm{B}} T}{\bar\rho \sqrt{D(\bar\rho )}}
    \mathcal{G}( \xi )
    \:,
\end{equation}
where in the second equality we have used~\eqref{eq:Mobility}, and we have denoted
\begin{equation}
    \mathcal{G}(z) = \frac{\e^{-z^2}}{\sqrt{\pi}} - |z| \erfc |z|
    \:.
\end{equation}
For $\xi = 0$, we recover the variance of the position of a tracer~\eqref{eq:Xt2}. For $\xi \neq 0$, the covariance decays to zero as the initial distance between the tracers is increased.

Note that, for the Calogero gas at low temperature, using the diffusion coefficient~\eqref{eq:DCaloHighDens} we obtain
\begin{equation}
   \frac{\moy{x_0(t) x_k(t)}_c}{\sqrt{t}}
   \underset{t \to +\infty}{\simeq}
    \frac{2 k_{\mathrm{B}} T}{\pi \bar\rho^2} \sqrt{\frac{\mu_0}{g}}
    \mathcal{G} \left( \frac{k}{2 \bar\rho^2 \pi \sqrt{\mu_0 g t} } \right)
    \:.
\end{equation}
We checked that this expression coincides with the low-temperature covariance of tracers in the Calogero gas found in Ref.~\cite{Touzo:2024a} from a microscopic calculation.

\subsection{Bath-tracer correlation profiles}

Qualitatively, the subdiffusive behaviour of the tracer~\eqref{eq:Xt2} can be traced back to the strong correlations present in the system, due to the order of the particles being conserved at all times. Quantitatively, these correlations are encoded in the generating function of bath-tracer correlation profiles~\cite{Poncet:2021,Grabsch:2022}
\begin{equation}
    \frac{\moy{ \rho_0(X_t + x, t) \e^{\lambda X_t} }}{\moy{ \e^{\lambda X_t} }}
    \equiv \sum_{n=0}^\infty \frac{\lambda^n}{n!} \moy{ \rho_0(X_t + x) X_t^n }_c
    \:.
\end{equation}
At large scales, it behaves as~\cite{Poncet:2021}
\begin{equation}
    \frac{\moy{ \rho_0(X_t + x, t) \e^{\lambda X_t} }}{\moy{ \e^{\lambda X_t} }}
    \underset{t \to \infty}{\simeq}
    \Phi\left(\lambda ; z = \frac{x}{\sqrt{t}} \right)
    \:,
\end{equation}
indicating that these correlations are non stationary and spread as $\sqrt{t}$. 
Furthermore, $\Phi(\lambda;z)$ fully determines the cumulants of $X_t$ via the relation~\cite{Grabsch:2024b,Berlioz:2025,Berlioz:2025b}
\begin{equation}
    \label{eq:CumulFromPhi}
    \frac{1}{\sqrt{t}} \ln \moy{\e^{\lambda X_t}}
    \underset{t \to \infty}{\simeq}
    -2 \frac{\partial_z P(\Phi)}{k_{\mathrm{B}} T \Phi} \Bigg|_{0}
    \int_{\Phi(\lambda;0^-)}^{\Phi(\lambda;0^+)} D(r) \dd r
    \:.
\end{equation}
The correlation profile $\Phi$ is therefore both a key physical quantity, which measures the coupling between the displacement of the tracer and the bath of surrounding particles, and a key technical tool which fully determines the cumulants of $X_t$, in particular since they have been shown to obey a closed equation in the case of the SEP~\cite{Grabsch:2022}.

\begin{figure}
    \centering
    \includegraphics[width=0.9\columnwidth]{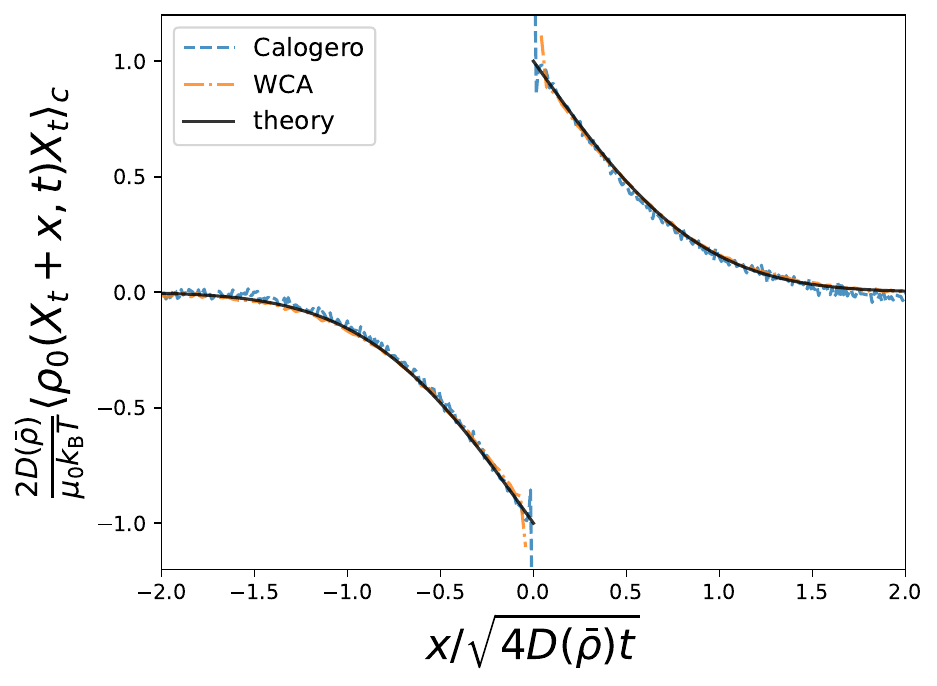}
    \caption{Scaled correlation profile $\moy{\rho_0(X_t + x, t)X_t}_c$, given by~\eqref{eq:XtRho}, as a function of the scaled variable $x/\sqrt{4 D(\bar\rho) t}$, at $T=1$, compared to numerical simulations for: (i) the Calogero potential~\eqref{eq:CaloPot} with $g=1$, at $t=200$; (ii) the WCA potential~\eqref{eq:WCAPot} with $g=1$ and $l = 1$, at $t=200$ (the diffusion coefficient is obtained from the nearest-neighbour approximation, see Fig.~\ref{fig:DWCA}).}
    \label{fig:XtRho}
\end{figure}

The first three orders in $\lambda$ of $\Phi$, which provide the large-scale behaviour of the correlation functions $\moy{\rho_0(X_t + x ,t )X_t^n}_c$ for $n \leq 3$, have been computed exactly using the MFT formalism for any $D(\rho)$ and $\sigma(\rho)$ in~\cite{Poncet:2021,Grabsch:2024b}. In particular, for the case of interacting Brownian particles considered here, the first correlation profile gives the covariance~\cite{Poncet:2021}
\begin{multline}
    \label{eq:XtRho}
    \moy{\rho_0(X_t + x, t)X_t}_c
    \underset{t \to \infty}{\simeq}
    \\
    \sign(x)
    \frac{\mu_0 k_{\mathrm{B}} T}{2 D(\bar\rho)} 
    \erfc \left( \frac{\abs{x}}{\sqrt{4 D(\bar\rho) t}} \right)
    \:.
\end{multline}
This analytical prediction is compared with numerical simulations in Fig.~\ref{fig:XtRho} for different interaction potentials. The small discrepancies observed near $x = 0$ are due to microscopic effects, which are not captured by the macroscopic formalism. These effects appear on a scale that does not depend on time, and are thus squeezed to the origin in the scaling variable $x/\sqrt{t}$ for large $t$.

\subsection{Integrated current}

Another quantity that has attracted a lot of attention~\cite{Derrida:2009,Derrida:2009,Krapivsky:2012} is the integrated current through a given point, say for instance the origin, which counts how many particles have crossed this point (counted positively from left to right and negatively from right to left)
up to a given time $t$. In terms of the microscopic current $j_0(x,t)$, it is simply given by
\begin{equation}
    Q_t = \int_0^t j_0(0,t')\dd t'
    \:.
\end{equation}
Since $j_0$ satisfies the conservation law $\partial_t \rho_0 + \partial_x j_0 = 0$, the integrated current can be written as the variation of the number of particles to the right of the origin
\begin{equation}
    Q_t = \int_0^\infty [\rho_0(x,t) - \rho_0(x,0)] \dd x
    \:.
    \label{eq:current-1}
\end{equation}
The statistical properties of the current $Q_t$ in a one-dimensional diffusive system have been shown to be equal to (the opposite of) those of the position $X_t$ of a tracer in a dual diffusive system~\cite{Rizkallah:2022}, with the dual transport coefficients
\begin{equation}
    \label{eq:DualityMapping}
    \tilde{D}(\tilde\rho) = \frac{1}{\tilde\rho^2} D \left( \frac{1}{\tilde\rho} \right)
    \:,
    \quad
    \tilde{\sigma}(\tilde\rho) = \tilde\rho \: \sigma \left( \frac{1}{\tilde\rho} \right)
    \:.
\end{equation}
The dual ``density'' $\tilde{\rho}(k,t)$ represents the distance between particle $k$ and $k+1$ in the original system. Thus, $\tilde{\rho}$ has the dimension of an inverse density, and its ``spatial'' argument $k$ 
is dimensionless.

In particular, replacing $D$ and $\sigma$ in~\eqref{eq:Xt2} to obtain the mean squared displacement of the dual tracer and then applying~\eqref{eq:DualityMapping}, we obtain the variance of the integrated current in the original system~\cite{Krapivsky:2012},
\begin{equation}
    \label{eq:Qt2}
    \moy{Q_t^2} \underset{t \to \infty}{\simeq}
    \frac{\sigma(\bar\rho)}{\sqrt{\pi D(\bar\rho)}} \sqrt{t}
    = \frac{2 \mu_0 k_{\mathrm{B}} T \bar\rho}{\sqrt{\pi D(\bar\rho)}} \sqrt{t}
    \:,
\end{equation}
where in the second equality we have used the expression of the mobility for interacting Brownian particles~\eqref{eq:Mobility}. The striking similarity between~\eqref{eq:Qt2} and~\eqref{eq:Xt2} is due to the fact that, at the level of the fluctuations, $Q_t = \rho X_t$. This is however not the case for higher-order cumulants. The fourth cumulant of $Q_t$ can nevertheless be obtained from~\eqref{eq:Xt4} by applying the transformation~\eqref{eq:DualityMapping}~\cite{Grabsch:2024b}.

Additionally, a fluctuation of $Q_t$ is correlated with a fluctuation of the density around the origin, which is quantified by the covariance
\begin{multline}
    \label{eq:QtRho}
    \moy{\rho_0(x, t)Q_t}_c
    \underset{t \to \infty}{\simeq}
    \\
    \sign(x)
    \frac{\mu_0 k_{\mathrm{B}} T \bar\rho}{2 D(\bar\rho)} 
    \erfc \left( \frac{\abs{x}}{\sqrt{4 D(\bar\rho) t}} \right)
    \:,
\end{multline}
as for the case of the tracer. An increase of $Q_t$ is therefore correlated with an increase of the density for $x>0$, and a decrease for $x<0$, as expected (the particles that contribute to increase $Q_t$ must come from the left and end up to the right). As for the case of $X_t$, this correlation profile is non stationary and grows diffusively through the system.

\section{Integrated current in \texorpdfstring{$d$}{d}-dimensional channels}
\label{sec:MFTchannels}

\begin{figure}
\centering
\begin{tikzpicture}[line cap=round,line join=round,scale=0.7]

\def\xL{-5.5}  
\def\xl{-5}  
\def\xr{ 4}  
\def\xR{ 4.5}  

\def\yA{ 1.6}  
\def\yB{ 0.4}  
\def\yC{-0.4}  
\def\yD{-1.6}  

\foreach \y in {\yA,\yB,\yD}{
  \draw[dashed] (\xL,\y) -- (\xl,\y);
  \draw          (\xl,\y) -- (\xr,\y);
  \draw[dashed] (\xr,\y) -- (\xR,\y);
}
\draw[dashed,opacity=.45] (\xL,\yC) -- (\xl,\yC);
\draw[opacity=.45]         (\xl,\yC) -- (\xr,\yC);
\draw[dashed,opacity=.45] (\xr,\yC) -- (\xR,\yC);

\coordinate (C) at (-1.2,{-0.6}); 
\coordinate (T) at (-2.2,\yA);
\coordinate (B) at (-1.2,\yD);
\coordinate (L) at (-2.2,\yC);
\coordinate (R) at (-1.2,\yB);

\fill[gray!25,pattern=north east lines,pattern color=black!50]
    (T)--(R)--(B)--(L)--cycle;

\draw[opacity=0.4] (T) -- (L);
\draw[opacity=0.4] (L) -- (B);
\draw (T) -- (R);
\draw (R) -- (B);

\draw[<->] (\xR+0.5,\yB) -- node[right] {$2L$} (\xR+0.5,\yD);

\foreach \p in {(-3.6,1.05), (-3.3,0.15), (-3.7,-0.9)} \draw \p circle (1.8pt);
\foreach \p in {(-0.2,0.95), (0.7,-0.55), (1.7,-0.05), (2.3,-1.05), (3.0,0.85)}
  \draw \p circle (1.8pt);

\end{tikzpicture}
\caption{Schematic representation of interacting Brownian particles evolving in a 3-dimensional channel (see Sec.~\ref{sec:MFTchannels}). 
The length of the channel is assumed to be infinite along the horizontal direction $\hat{\V{e}}_1$, 
and $2L$ along the other directions, with reflective boundary conditions.
The integrated particle current defined in Eq.~\eqref{eq:current-d-def} counts the net number of particles that cross the shaded square cross section (located at $x_1=0$) from left to right.}
\label{fig:sketch-channel}
\end{figure}
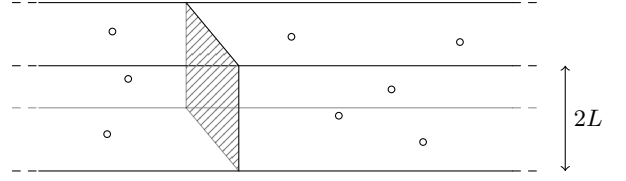

As recalled above, in the last decades the MFT has been key to obtaining several exact results for one-dimensional interacting particle systems~\cite{Bodineau:2004,Bertini:2005a,Bodineau:2005,Derrida:2009a,Krapivsky:2012,Krapivsky:2014,Krapivsky:2015a,Sadhu:2015,Sadhu:2016,Mallick:2022,Bettelheim:2022,Bettelheim:2022a,Krajenbrink:2022,Saha:2023,Grabsch:2024,Grabsch:2024d,Sharma:2024,Bodineau:2025,Saha:2025}. However, its validity is clearly not limited to this class of systems, and extends naturally to higher spatial dimensions. For instance, this framework has proved instrumental to the determination of dynamical laws and symmetries in $d$-dimensional systems far from equilibrium~\cite{Espigares:2016,Escamilla:2017,Escamilla:2017a}. Equipped with the dynamical coefficients $D(\rho)$ and $\sigma(\rho)$, for which we provided constructive approximation schemes in Sec.~\ref{sec:d-coefficients}, in this section we apply the MFT to compute explicitly the integrated current fluctuations in the important case of a $d$-dimensional channel, as illustrated in Fig.~\ref{fig:sketch-channel}.

In this setting, reflective boundary conditions are assumed to act on the particles along all directions but the horizontal one, namely $x_1 = \V x \cdot \hat{\V{e}}_1$. We thus generalize the definition~\eqref{eq:current-1} of the integrated particle current as
\begin{equation}
    Q_{\mathcal T} = \int_0^\infty \dd{x_1} \int_{[-L,L]^{d-1}} \dd^{d-1}{\V x_\perp} [\rho_0(\V x, \mathcal T)-\rho_0(\V x, 0)],
    \label{eq:current-d-def}
\end{equation}
where we denoted by $\V x_\perp$ the component of $\V x$ orthogonal to $\hat{\V{e}}_1$.
We identify $\Lambda=\mathcal T^{1/2}$ as the large macroscopic scale by which the microscopic density $\rho_0$ is rescaled in Eq.~\eqref{eq:RescalingDensity}, so as to rewrite $Q_{\mathcal T} = \Lambda^d \hat Q[\rho]$, with
\begin{equation}
    \hat Q[\rho] = \int_0^\infty \dd{x_1} \int_{[-\tilde L,\tilde L]^{d-1}} \dd^{d-1}{\V x_\perp} [\rho(\V x, 1)-\rho(\V x, 0)],
    \label{eq:Qhat-def}
\end{equation}
where we called $\tilde L = L/\Lambda$.
Its moment generating function can be formally expressed as
\begin{align}
    \moy{\mathrm e^{\lambda Q_{\mathcal T}}} =  \int & \mathcal D \rho(\V x,t) \, \mathcal D H(\V x,t)\, \mathcal D \rho(\V x,0) \, \nonumber \\
    &\mathrm e^{-\Lambda^d \left\lbrace \mathcal S[\rho,H] +F[\rho(\V{x},0)]- \lambda \hat Q[\rho] \right\rbrace},
    \label{eq:moments-Q}
\end{align}
where $\mathcal S$ is the dynamical action introduced in Eq.~\eqref{eq:MFTaction}, but with the spatial integration restricted to the domain $\mathcal V \equiv \mathbb R \cup [-\tilde L,\tilde L]^{d-1}$ (i.e.~infinite along the direction $\hat{\V e}_1$, and bounded along the other orthogonal directions). Note that for a system in which the particles' momentum does not play a role --- which is the case for discrete models such as the SEP, but also for the overdamped Brownian particles considered here --- limiting the spatial domain of integration in the dynamical action actually corresponds to imposing reflective boundary conditions~\footnote{See for instance Ref.~\cite{Saha:2023}, where reflective or absorbing boundary conditions for the one-dimensional SEP in a semi-infinite line are recovered as limiting situations of a system coupled to a particle reservoir.}.
Similarly, we limit the temporal integration in Eq.~\eqref{eq:MFTaction} 
within the domain $t\in [0,1]$.
Finally, we assume that at $t=0$ the system is at equilibrium at mean density $\bar\rho$: this means that the initial density $\rho(\V{x},0)$ is distributed according to $P[\rho(\V{x},0)] \propto
    \mathrm e^{- \Lambda^{d} F[\rho(\V{x},0)]}$, with~\cite{Derrida:2025a}
\begin{equation}
    \label{eq:DistrInitDens}
    F[\rho(\V{x},0)] = \int_{\mathcal V} \dd^d \V{x} \int_{\bar\rho}^{\rho(\V{x},0)} (\rho(\V{x},0)-r) \frac{2D(r)}{\sigma(r)} \dd r
    \:.
\end{equation}

The evaluation of the moments of $Q_{\mathcal T}$ passes through the minimization of the total action in Eq.~\eqref{eq:moments-Q}, which turns out to lead to the Euler-Lagrange equations~\eqref{eq:MFTeqs} endowed with the Neumann boundary conditions
\begin{equation}
   \nabla q\Big|_{\partial \mathcal V}=0, \qquad \nabla p\Big|_{\partial \mathcal V}=0,
    \label{eq:BC}
\end{equation}
where $\partial \mathcal V$ indicates the boundary of the region $\mathcal V$, and with the initial and final conditions
\begin{equation}
    p(\V x,0) = \lambda \Theta (x_1) + \int_{\bar\rho}^{q(\V{x},0)} \frac{2 D(r)}{\sigma(r)} \dd r, \quad
    p(\V x,1) = \lambda \Theta (x_1).
    \label{eq:initial-cond}
\end{equation}
Remarkably, the set of differential equations~\eqref{eq:MFTeqs} with the boundary conditions~\eqref{eq:BC} is separable, meaning that we can construct a solution of the form
\begin{equation}
    q(\V x,t) = q_1(x_1,t)q_\perp(\V x_\perp), \quad p(\V x,t) = p_1(x_1,t)p_\perp(\V x_\perp).
    \label{eq:factorization}
\end{equation}
The problem is then reduced to that of evaluating the current fluctuations in a one-dimensional system, whose solution is reported e.g.~in Ref.~\cite{Krapivsky:2012,Berlioz:2025} --- we adapt it here, for the sake of completeness, to the problem at hand.

First, we note that a spatially uniform choice of $q_\perp(\V x_\perp)$ and 
$p_\perp(\V x_\perp)$ trivially satisfies the Neumann boundary conditions~\eqref{eq:BC}. We identify this solution as the relevant one assumed in the stationary limit, and thus we set both $q_\perp(\V x_\perp)$ and 
$p_\perp(\V x_\perp)$ to unity without loss of generality (i.e.~we reabsorb any multiplicative constant into $ q_1(x_1,t)$ and $ p_1(x_1,t)$). 
Second, we note that in the large-$\Lambda$ limit one gets from Eq.~\eqref{eq:moments-Q}
\begin{equation}
    \ln  \moy{\mathrm e^{\lambda Q_{\mathcal T}}} \simeq \Lambda^d \hat \psi(\lambda) , \quad \hat \psi(\lambda) \equiv \lambda \hat Q[q] -  S[q,p] -F[q(\V{x},0)] ,
    \label{eq:saddle-approx}
\end{equation}
in terms of the saddle-point solution $(q,p)$. 
By construction,
\begin{equation}
    \hat \psi(\lambda) \simeq \Lambda^{-d} \ln  \moy{\mathrm e^{\lambda Q_{\mathcal T}}} =  \Lambda^{-d} \sum_{n=1}^\infty \frac{\lambda^n}{n!}\kappa_n,
    \label{eq:psi-expansion}
\end{equation}
where we called
$ \kappa_n$ the cumulants of $Q_{\mathcal T}$. Besides, note that $\delta \hat \psi/\delta q$ and $\delta \hat \psi/\delta p$ vanish 
at the saddle point, whence
\begin{equation}
    \frac{\mathrm d \hat \psi(\lambda)}{\mathrm d \lambda } = \hat Q[q],
    \label{eq:dpsi_dlambda}
\end{equation}
so that using Eq.~\eqref{eq:psi-expansion} we deduce
\begin{equation}
    \sum_{n=0}^\infty \frac{\lambda^n}{n!}\kappa_{n+1} = \Lambda^{d}\hat Q[q].
\end{equation}
This suggests 
looking for a solution of the Euler-Lagrange equations~\eqref{eq:MFTeqs} in the form of a power series in $\lambda$, 
\begin{equation}
    q_1= \sum_{n=0}^\infty \lambda^n q_1^{(n)}, \quad p_1= \sum_{n=0}^\infty \lambda^n p_1^{(n)},
    \label{eq:series}
\end{equation}
and 
identifying the various cumulants $\kappa_n$ using the expression~\eqref{eq:Qhat-def} of $\hat Q[q]$: 
\begin{align}
    \frac{\kappa_{n+1}}{n!} 
    =\Lambda \cdot (2 L)^{d-1}  \int_0^\infty \dd{x_1}  [q_1^{(n)}(x_1, 1)-q_1^{(n)}(x_1, 0)],
    \label{eq:kappa-explicit}
\end{align}
where we used the factorization property~\eqref{eq:factorization}, and 
we recalled that 
$\tilde L = L/\Lambda$.
Interestingly, 
by using $\Lambda=\mathcal{T}^{1/2}$, one finds that all cumulants $\kappa_n$ of the integrated particle current through a cross section (i.e.~spanning the entire $d$-dimensional channel) display the scaling $\propto \sqrt{\mathcal T}$ typical of one-dimensional systems. This has to be compared with the integrated particle current through a finite slit of length $l\ll L$ in a system of spatial dimension $d>1$, whose variance is known to scale linearly with $\mathcal T$~\cite{Bodineau:2008,Berlioz:2024}.

Since in this section we are interested in the variance $\kappa_2$ of the particle current (indeed, $\kappa_1=0$ by symmetry), it is sufficient to consider the expansion~\eqref{eq:series} up to $\mathcal{O}(\lambda)$ only. First, from the initial conditions~\eqref{eq:initial-cond} we immediately conclude that $p_1\z \equiv 0$ and $q_1\z \equiv \bar \rho$. At the leading order in $\lambda$, the Euler-Lagrange equations~\eqref{eq:MFTeqs} read
\begin{align}
    \partial_t q_1\o &=
    D(\bar \rho) \partial_{x_1}^2 q_1\o - \sigma(\bar \rho)  \partial_{x_1}^2  p_1\o,
    \label{eq:EL-q1}
    \\
    \partial_t p_1\o &= 
    - D(\bar \rho) \partial_{x_1}^2 p_1\o ,
    \label{eq:EL-p1}
\end{align}
where we used Eq.~\eqref{eq:factorization}, whereas the initial conditions~\eqref{eq:initial-cond} reduce to
\begin{align}
    p_1\o(x_1,1) &= \Theta (x_1) , \\
    q_1\o(x_1,0) &= \frac{\sigma(\bar \rho)}{2D(\bar \rho)}
    [p_1\o(x_1,0) -  \Theta (x_1)],
    \label{eq:initial-cond2}
\end{align}
where we noted that $\int_{\bar\rho}^{q_1(x_1,0)} f(r) \dd r = \lambda f(\bar \rho)q_1\o(x_1,0) +\mathcal O(\lambda^2) $. We can immediately solve~\eqref{eq:EL-p1} to obtain
\begin{equation}
    p_1\o(x_1,t) = \frac12 \mathrm{erfc}\left(  -\frac{x_1}{2\sqrt{D(\bar \rho)(1-t)}}\right),
\end{equation}
which can be inserted back into Eqs.~\eqref{eq:EL-q1} and~\eqref{eq:initial-cond2} to find~\cite{Berlioz:2025}
\begin{align}
    q_1\o(x_1,t) = \frac{\sigma(\bar \rho)}{4D(\bar \rho)} \Big[& \mathrm{erfc}\Big(  -\frac{x_1}{2\sqrt{D(\bar \rho)(1-t)}}\Big)\nonumber \\
    &-\mathrm{erfc}\Big(  -\frac{x_1}{2\sqrt{D(\bar \rho)t}}\Big) \Big].
\end{align}
In particular, 
\begin{align}
    q_1\o(x_1,1) &= \mathrm{sign}(x_1) \frac{\sigma(\bar \rho)}{4D(\bar \rho)}  \mathrm{erfc}\left(  \frac{|x_1|}{2\sqrt{D(\bar \rho)}}\right) ,\\
    q_1\o(x_1,0) &= -\mathrm{sign}(x_1) \frac{\sigma(\bar \rho)}{4D(\bar \rho)}  \mathrm{erfc}\left(  \frac{|x_1|}{2\sqrt{D(\bar \rho)}}\right) .
\end{align}
We conclude the calculation by plugging this result into Eq.~\eqref{eq:kappa-explicit}, which gives the variance $\kappa_2 = \moy{Q_{\mathcal T}^2}_c$ of the integrated particle current as
\begin{align}
    \kappa_2 &=  \Lambda \cdot (2 L)^{d-1}  \int_0^\infty \dd{x_1}  [q_1^{(1)}(x_1, 1)-q_1^{(1)}(x_1, 0)] \nonumber \\
    &=\sqrt{\mathcal T} \cdot (2 L)^{d-1} \frac{\sigma(\bar \rho)}{\sqrt{\pi D(\bar \rho)}},
\end{align}
where in the second line we recalled that 
$\Lambda=\mathcal{T}^{1/2}$, and we performed the integration over $x_1$ explicitly. 
Note that this expression correctly reduces to~\eqref{eq:Qt2} upon choosing $d=1$.

\section{Current and density correlations in arbitrary dimension}
\label{sec:MFTdDimension}

The fluctuating hydrodynamics equations~(\ref{eq:ConsEqDens},\ref{eq:jflucHydro}) fully determine the current $\V{j}$ and the density $\rho$, and in particular their correlations, in any dimension.
We consider that at $t=0$ the system is at equilibrium at mean density $\bar\rho$.
As recalled in Sec.~\ref{sec:MFTchannels}, this
means that the initial density $\rho(\V{x},0)$ is distributed according to $P[\rho(\V{x},0)] \propto
    \e^{- \Lambda^{d} F[\rho(\V{x},0)]}$,
with $F[\rho(\V{x},0)]$ given in Eq.~\eqref{eq:DistrInitDens}.
We first determine the two-point correlations of the initial density, and then use the evolution equations~(\ref{eq:ConsEqDens},\ref{eq:jflucHydro}) to determine these correlations at all times.

\subsection{Equilibrium density-density correlations}

The spatial correlations of the initial density can be computed from the generating function
\begin{equation}
    \label{eq:GenCorrelInitDistr}
    G[\kappa] \equiv \ln \int  \e^{- \Lambda^d \left[ F[\rho(\V{x},0)] - \int \kappa(\V{x})  \rho(\V{x},0) \dd^d \V{x} \right]} \D \rho(\V{x},0)
    \:.
\end{equation}
In particular, we have that
\begin{equation}
    \label{eq:CorrelInitDistr}
    \moy{\rho(\V{x},0) \rho(\V{y},0)}_c =
    \Lambda^{-2d}
    \frac{\delta G[\kappa]}{\delta \kappa(\V{x}) \delta \kappa(\V{y})} \Bigg|_{\kappa = 0}
    \:.
\end{equation}
The integral in~\eqref{eq:GenCorrelInitDistr} can be computed from a saddle point for $\Lambda \to \infty$, so that
\begin{equation}
    \label{eq:GenCorrelInitSol}
    G[\kappa] \underset{\Lambda \to \infty}{\simeq} - \Lambda^d \left[ F[\rho_\star] 
    - \int \kappa(\V{x}) \rho_\star(\V{x})\dd \V{x}  \right]
    \:,
\end{equation}
with $\rho_\star$ solution of
\begin{equation}
    \label{eq:SaddleInitCorr}
    \frac{\delta F}{\delta \rho(\V{x})} = \kappa(\V{x})
    \:.
\end{equation}
Using the expression~\eqref{eq:DistrInitDens}, this gives
\begin{equation}
    \int_{\bar\rho}^{\rho_\star(\V{x})} \frac{2 D(r)}{\sigma(r)} \dd r
    = \kappa(\V{x})
    \:.
\end{equation}
At leading order in $\kappa$, 
one then finds
\begin{equation}
    \label{eq:SolInitCorrelLeadingOrder}
    \rho_\star(\V{x}) = \bar\rho + \kappa(\V{x}) \frac{\sigma(\bar\rho)}{2 D(\bar\rho)}
    + O(\kappa^2)
    \:.
\end{equation}
Additionally, substituting the saddle-point equation~\eqref{eq:SaddleInitCorr} into~\eqref{eq:GenCorrelInitSol}, we obtain that
\begin{equation}
    \frac{\delta G[\kappa]}{\delta \kappa(\V{x})}
    = \Lambda^d \rho_\star(\V{x})
    \:.
\end{equation}
Thus, from the solution~\eqref{eq:SolInitCorrelLeadingOrder} we obtain the correlations~\eqref{eq:CorrelInitDistr}
\begin{equation}
    \label{eq:ResultInitCorrel}
    \moy{\rho(\V{x},0) \rho(\V{y},0)}_c \underset{\Lambda \to \infty}{\simeq} \Lambda^{-d}  \frac{\sigma(\bar\rho)}{2 D(\bar\rho)}
    \delta(\V{x}-\V{y})
    \:.
\end{equation}
We recovered here the well-known form of the equilibrium correlations~\cite{Derrida:2025a}, which are local and whose amplitude is fully determined by $D$ and $\sigma$. Note that the rescaling parameter $\Lambda$ appears explicitly in these correlations. This is however artificial, since if we write the correlations in terms of the microscopic density $\rho_0$~\eqref{eq:RescalingDensity}, we obtain
\begin{equation}
    \label{eq:ResultInitCorrel2}
    \moy{\rho_0(\V{x},0) \rho_0(\V{y},0)}_c = \frac{\sigma(\bar\rho)}{2 D(\bar\rho)}
    \delta(\V{x}-\V{y})
    \:,
\end{equation}
which no longer depends on the arbitrary scale $\Lambda$, but requires that $||\V{x} - \V{y}|| \to \infty$.

\subsection{Dynamical two-point correlations}

We now turn to the determination of the dynamical two-point correlations, by relying on the evolution equations~(\ref{eq:ConsEqDens},\ref{eq:jflucHydro}). Since in the large-$\Lambda$ limit the noise is small, we can expand the equations around the stationary solution,
\begin{gather}
    \label{eq:ExpDensForCorrel}
    \rho(\V{x},t) = \bar\rho + \Lambda^{-d/2} \rho_1(\V{x},t) + O(\Lambda^{-d})
    \:,\\
    \label{eq:ExpCurrentForCorrel}
    \V{j}(\V{x},t) =  \Lambda^{-d/2} \V{j}_1(\V{x},t) + O(\Lambda^{-d})
    \:.
\end{gather}
The equations of fluctuating hydrodynamics~(\ref{eq:ConsEqDens},\ref{eq:jflucHydro}) thus reduce to the linear equations
\begin{equation}
    \label{eq:RhoJfirstorder}
    \partial_t \rho_1 + \V{\nabla} \cdot \V{j}_1 = 0
    \:,
    \quad
    \V{j}_1 = - D(\bar\rho)\V{\nabla} \rho_1 - \sqrt{\sigma(\bar\rho)} \: \V{\eta}
    \:,
\end{equation}
which can be solved explicitly. For the computation of the two-point correlations, the simplification~\eqref{eq:RhoJfirstorder} at leading order in $\Lambda^{-d/2}$ is equivalent to the saddle-point calculations performed for $\Lambda \to \infty$ in the action formalism above~\cite{Sadhu:2016}. The determination of higher-order correlation functions from this method would require expanding~(\ref{eq:ExpDensForCorrel},\ref{eq:RhoJfirstorder}) to higher orders in $\Lambda^{-d/2}$. These computations typically become quickly cumbersome, but since we are only interested in the two-point correlations, a direct manipulation of the equations~(\ref{eq:ConsEqDens},\ref{eq:jflucHydro}) is more straightforward than the full saddle-point computation.
Introducing the heat kernel
\begin{equation}
    \label{eq:HeatKernel}
    K(\V{x},t| \V{x}' , t') =
    \Theta(t-t')
    \frac{\e^{- \frac{(\V{x}-\V{x}')^2}{4 D(\bar\rho) (t-t')}}}
    {(4 \pi D(\bar\rho)(t-t'))^{d/2}}
    \:,
\end{equation}
we can express $\rho_1(\V{x},t)$ as
\begin{multline}
    \rho_1(\V{x},t) = \sqrt{\sigma(\bar\rho)}
    \int \dd^d \V{z} \int_0^t \dd \tau \: K(\V{x},t| \V{z} , \tau)  \: \V{\nabla} \cdot \V{\eta}(\V{z},\tau)
    \\
    + \int \dd^d \V{z} \:  K(\V{x},t| \V{z} ,0 ) \rho_1(\V{z},0)
    \label{eq:NoiseDensCorrel}
    \:.
\end{multline}
Performing an integration by parts in the first term yields a more convenient representation:
\begin{multline}
    \rho_1(\V{x},t) = -\sqrt{\sigma(\bar\rho)}
    \int \dd^d \V{z} \int_0^t \dd \tau \: \V{\eta}(\V{z},\tau)  \cdot \V{\nabla}_{\V{z}}  K(\V{x},t| \V{z} , \tau) 
    \\
    + \int \dd^d \V{z} \:  K(\V{x},t| \V{z} ,0 ) \rho_1(\V{z},0)
    \:.
\end{multline}
This expression will be our starting point to compute various correlation functions. For instance, 
\begin{align}
    &\moy{\rho_1(\V{x},t) \V{\eta}(\V{x}',t')}_c
    \nonumber
    \\
    &=- \sqrt{\sigma(\bar\rho)} \int \dd^d \V{z} \int_0^t \dd \tau 
    \moy{ \eta^a(\V{z},\tau) \V{\eta}(\V{x}',t') }_c
    \nabla^a_{\V{z}} \cdot  K(\V{x},t| \V{z} , \tau)
    \nonumber
    \\
    &= - \sqrt{\sigma(\bar\rho)} \Theta(t-t') \V{\nabla}_{\V{x}'} K(\V{x},t| \V{x}' , t') 
    \label{eq:CorrelCurrentNoise}
    \:,
\end{align}
where we have used the noise correlator~\eqref{eq:DefNoiseCorrel}, and used Einstein's convention to sum over the repeated index $a$ in the second line.

The computation of the density-density correlations is a bit more involved:
\begin{multline}
    \moy{\rho_1(\V{x},t) \rho_1(\V{x}',t')}_c = 
    \\
    \sigma(\bar\rho) \int \dd^d \V{z} \int_0^{\min(t,t')} \dd \tau \:
     \V{\nabla}_{\V{z}} K(\V{x},t| \V{z} , \tau) \cdot 
     \V{\nabla}_{\V{z}} K(\V{x'},t'| \V{z} , \tau)
    \\
    +\frac{\sigma(\bar\rho)}{2D(\bar\rho)} \int \dd^d \V{z} \: K(\V{x},t| \V{z},0)
    K(\V{x}',t'| \V{z},0)
    \:,
\end{multline}
where we have used the noise correlations~\eqref{eq:DefNoiseCorrel} and the initial density correlations~\eqref{eq:ResultInitCorrel}. Using that $K$ satisfies the diffusion equation
\begin{equation}
    D(\bar\rho) \Delta_z K(\V{x},t| \V{z},\tau) = -\partial_\tau K(\V{x},t| \V{z},\tau)
    \:,
\end{equation}
we obtain after an integration by parts
\begin{multline}
    \moy{\rho_1(\V{x},t) \rho_1(\V{x}',t')}_c = 
    \\
    \frac{\sigma(\bar\rho)}{2D(\bar\rho)} \int \dd^d \V{z} \int_0^{\min(t,t')} \dd \tau \:
     \partial_\tau \left[ K(\V{x},t| \V{z} , \tau) 
      K(\V{x'},t'| \V{z} , \tau) \right]
    \\
    +\frac{\sigma(\bar\rho)}{2D(\bar\rho)} \int \dd^d \V{z} \: K(\V{x},t| \V{z},0)
    K(\V{x}',t'| \V{z},0)
    \:.
\end{multline}
Performing the integral over $\tau$, we finally get
\begin{multline}
    \label{eq:TwoPtCorrelDens}
    \moy{\rho_1(\V{x},t) \rho_1(\V{x}',t')}_c = 
    \frac{\sigma(\bar\rho)}{2D(\bar\rho)} \Big[ 
    \Theta(t-t') K(\V{x},t| \V{x}' , t')
    \\
    + \Theta(t'-t) K(\V{x'},t'| \V{x} , t) \Big]
    \:,
\end{multline}
with here the convention that $\Theta(0) = \frac{1}{2}$. Note that for $t=t'$ we recover the local equilibrium correlations computed at $t=t'=0$~\eqref{eq:ResultInitCorrel}, as expected. However, for $t \neq t'$ the density presents long-range spatial correlations. The result~\eqref{eq:TwoPtCorrelDens} is for now written in terms of the first order $\rho_1$ of the actual macroscopic density $\rho$. Using the relation~\eqref{eq:ExpDensForCorrel} this becomes,
\begin{multline}
    \label{eq:TwoPtCorrelDens1}
    \moy{\rho(\V{x},t) \rho(\V{x}',t')}_c 
    \underset{\Lambda \to \infty}{\simeq}
    \Lambda^{-d}
    \frac{\sigma(\bar\rho)}{2D(\bar\rho)} \Big[ 
    \Theta(t-t') K(\V{x},t| \V{x}' , t')
    \\
    + \Theta(t'-t) K(\V{x'},t'| \V{x} , t) \Big]
    \:.
\end{multline}
As noted above for the equilibrium correlations~\eqref{eq:ResultInitCorrel}, this result depends on the rescaling parameter $\Lambda$. However, if we write the result in terms of the microscopic density $\rho_0$~\eqref{eq:RescalingDensity} this parameters disappear,
\begin{multline}
    \label{eq:TwoPtCorrelDens2}
    \moy{\rho_0(\V{x},t) \rho_0(\V{x}',t')}_c 
    \simeq
    \frac{\sigma(\bar\rho)}{2D(\bar\rho)} \Big[ 
    \Theta(t-t') K(\V{x},t| \V{x}' , t')
    \\
    + \Theta(t'-t) K(\V{x'},t'| \V{x} , t) \Big]
    \:,
\end{multline}
where $||\V{x} - \V{x}'||$ must be large. Note that these density correlations are Gaussian due to the fact that, at leading order in $\Lambda^{-d/2}$, the density obeys a linear diffusion equation~\eqref{eq:RhoJfirstorder}, which results from the weak noise present in the macroscopic equations~(\ref{eq:ConsEqDens},\ref{eq:jflucHydro}).

Finally, using the results~(\ref{eq:CorrelCurrentNoise},\ref{eq:TwoPtCorrelDens}) and the expression of the current~\eqref{eq:RhoJfirstorder}, we similarly obtain
\begin{align}
    \label{eq:CurDenCorrel}
    \Lambda^d \moy{\rho(\V{x},t) \V{j}(\V{x}',t')}_c
    & \underset{\Lambda \to \infty}{\simeq}
    \frac{\sigma(\bar\rho)}{2}
    \V{\nabla}_{\V{x}'} \Big[
    \Theta(t-t') K(\V{x},t| \V{x}' , t')
    \nonumber \\
   & - \Theta(t'-t) K(\V{x}',t'| \V{x}, t)
    \Big]
    \:,
\end{align}
and
\begin{align}
    &\Lambda^d \moy{j^a(\V{x},t) j^b(\V{x}',t')}_c 
    \underset{\Lambda \to \infty}{\simeq}
    \sigma(\bar\rho) \delta_{a,b} \delta(\V{x}-\V{x'}) \delta(t-t')
    \nonumber \\
    &- \frac{D(\bar\rho) \sigma(\bar\rho)}{2} 
    \nabla^a_{\V{x}}\nabla^b_{\V{x}'}
    \Big[ 
    \Theta(t-t') K(\V{x},t| \V{x}' , t')
    \nonumber \\
    &\qquad \qquad\qquad \qquad \quad + \Theta(t'-t) K(\V{x'},t'| \V{x} , t) \Big]
    \:.
\end{align}
Note that all these correlations are nonlocal for $t \neq t'$.

\subsection{Application: integrated current-density correlations}

As an application of the previous results, we consider the integrated current through a given point in space, which we choose without loss of generality to be the origin,
\begin{equation}
    \V{Q}_t \equiv \int_0^t \V{j}_0(\V{0},\tau)\dd \tau
    = \Lambda \int_{0}^{t/\Lambda^2} \V{j}(\V{0},\tau) \dd \tau
    \:,
\end{equation}
with the microscopic current $\V{j}_0$ related to the macroscopic current by $\V{j}(\V{x},t)= \Lambda \V{j}_0(\Lambda \V{x},\Lambda^2 t)$. The correlations between $\V{Q}_t$ and the microscopic density take the form
\begin{equation}
    \moy{\V{Q}_{\Lambda^2 t} \: \rho_0(\Lambda \V{x}, \Lambda^2 t) }_c
    = \Lambda \int_0^{t} \moy{ \rho(\V{x},t) \V{j}(\V{0},\tau) }_c \dd \tau
    \:.
\end{equation}
This quantity generalises to arbitrary dimensions the correlation profile~\eqref{eq:QtRho} discussed above.
Using the expression of the current-density correlation function~\eqref{eq:CurDenCorrel}, we have
\begin{multline}
    \moy{\V{Q}_{\Lambda^2 t} \: \rho_0(\Lambda \V{x}, \Lambda^2 t) }_c
    \underset{\Lambda \to \infty}{\simeq}
    \\
    \Lambda^{1-d} \frac{\sigma(\bar\rho)}{2}
    \V{\nabla}_{\V{x}'} \int_0^{t} K(\V{x},t | \V{x}', \tau) \dd \tau \Bigg|_{\V{x}'=0}
    \:.
\end{multline}
Computing the temporal integral by using the explicit form of the kernel $K$~\eqref{eq:HeatKernel} yields
\begin{multline}
    \moy{\V{Q}_{\Lambda^2 t} \: \rho_0(\Lambda \V{x}, \Lambda^2 t) }_c
    \underset{\Lambda \to \infty}{\simeq}
    \\
    \Lambda^{1-d} \frac{\sigma(\bar\rho)}{4 \pi^{d/2} D(\bar\rho)}
    \frac{\V{x}}{|| \V{x} ||^d}
    \Gamma \left( \frac{d}{2}, \frac{\V{x}^2}{4 D(\bar\rho) t} \right)
    \:,
\end{multline}
with $\Gamma(a,z)$ the incomplete Gamma function. Equivalently, we have, for large $\V{x}$ and $t$,
\begin{equation}
    \label{eq:CorrelIntegCurrDens}
    \moy{\V{Q}_{t} \: \rho_0(\V{x}, t) }_c
    \simeq
    \frac{\sigma(\bar\rho)}{4 \pi^{d/2} D(\bar\rho)}
    \frac{\V{x}}{|| \V{x} ||^d}
    \Gamma \left( \frac{d}{2}, \frac{\V{x}^2}{4 D(\bar\rho) t} \right)
    \:.
\end{equation}
For $d=1$, we recover the result~\eqref{eq:QtRho} discussed above, leading to a nonstationary correlation profile.
For $d\geq 2$, for a given large $\V{x}$, we get in the limit $t \to \infty$,
\begin{equation}
    \label{eq:CorrelIntegCurrDensAsympt}
    \moy{\V{Q}_{t} \: \rho_0(\V{x}, t) }_c
    \simeq
    \frac{\sigma(\bar\rho) \Gamma \left( \frac{d}{2} \right)}{4 \pi^{d/2} D(\bar\rho)}
    \frac{\V{x}}{|| \V{x} ||^d}
    \:.
\end{equation}
This shows that these correlations are stationary at long time, and display a universal power-law decay as $|| \V{x} ||^{1-d}$ at large distances. The exponent does not depend on the microscopic details of the model, but only on the spatial dimension.
Conversely, the prefactor of the decay is system dependent, through the ratio $\sigma(\bar\rho)/D(\bar\rho)$.
In the case of the 
SEP, corresponding to $D(\rho) = 1$ and $\sigma(\rho) = 2 \rho(1-\rho)$, the quantity~\eqref{eq:CorrelIntegCurrDens} has been computed exactly from the microscopic dynamics~\cite{Berlioz:2024} --- at large distances, it matches with the result~\eqref{eq:CorrelIntegCurrDensAsympt}. This shows that the MFT formalism is able to capture the large-scale properties of interacting particle systems, in any dimension. In particular, using the expressions of the transport coefficients given in Section~\ref{sec:TrCoefs}, the MFT can be applied to study the large-scale dynamical behaviour of interacting Brownian particles beyond the one-dimensional case.

\section{Conclusion}

By combining the expression of the collective diffusion coefficient of pairwise interacting Brownian particles~\eqref{eq:DiffCoef} with the macroscopic fluctuation theory (MFT)~\cite{Bertini:2015}, we have accessed the large-scale dynamical properties of this system. In particular, we have computed explicitly the diffusion coefficient for various emblematic interaction potentials. Using these expressions into either known results from MFT or newly derived expressions in the case of infinite channels or higher-dimensional systems, we have exactly determined the long-time behaviour of different observables, such as the integrated current through a section of the system. Our results open the way to the application of the MFT framework to fully characterise the large-scale behaviour of interacting Brownian particles.
The treatment of non-pairwise interactions, such as hydrodynamic interactions, within the MFT framework 
remains an open problem.

\begin{acknowledgements}
DV thanks Pascal Viot for his help with numerical resources.
AG acknowledges the financial support of the Emergence program at Sorbonne Université, Paris.
\end{acknowledgements}

\appendix

\section{The thermal de Broglie wavelength}
\label{App:DeBroglie}

In this Appendix, we show that, although the thermal de Broglie wavelength $\ell_0 = \sqrt{\beta h^2/(2 \pi m)}$ appears explicitly in the expression of the partition function~\eqref{eq:DefPartFct}, it is actually not involved in the pressure $P(\rho)$, and thus neither in the diffusion coefficient~\eqref{eq:DiffCoef}.  Note that the full partition function~\eqref{eq:DefFullPartFct} involves an integration over the momentum degrees of freedom, which are irrelevant here since we consider an overdamped dynamics. This integration is at the origin of the appearance of the parameter $\ell_0$ in the normalisation of the partition function~\eqref{eq:DefPartFct}.

We first compute the grand canonical partition function
\begin{align}
    \label{eq:GrandPartition}
    &\mathcal{Z}(\beta, \varphi)
    \equiv \sum_{N=0}^\infty \varphi^N Z_N(\beta)
    \\
    &= \sum_{N=0}^\infty \frac{1}{N!} \left(\frac{\varphi}{\ell_0^d} \right)^N
    \int_0^L \dd^d \V{x}_1 \cdots \dd^d \V{x}_N \: 
    \e^{- \frac{\beta}{2} \sum_{i \neq j} V(\V{x}_i - \V{x}_j)}
    \:,
    \nonumber
\end{align}
where $\varphi = \e^{\beta \mu}$ is the fugacity and $\mu$ the chemical potential. We introduce the grand potential density $\phi_G$, defined by
\begin{equation}
    \beta \phi_G(\beta,\varphi) V \underset{V \to \infty}{\simeq} - \ln \mathcal{Z}(\beta, \varphi)
    \:.
\end{equation}
The free energy density~\eqref{eq:DefFreeEnerDens} is obtained from the grand potential by a Legendre transform,
\begin{equation}
    f(\rho;\beta) = \min_{\varphi} \left[ \phi_G(\beta,\varphi) + \frac{\rho}{\beta} \ln \varphi \right]
    \:.
\end{equation}
In practice, it is computed as
\begin{equation}
    f(\rho;\beta) = \phi_G(\beta,\varphi_\star) + \frac{\rho}{\beta} \ln \varphi_\star
    \:,
\end{equation}
where $\varphi^\star(\rho;\beta)$ is the solution of
\begin{equation}
    \label{eq:EqLegendreFfromGC}
    \partial_\varphi \phi_G + \frac{\rho}{\beta \varphi} = 0
    \:.
\end{equation}
The pressure is then defined from the free energy via Eq.~\eqref{eq:LinkPf}, which explicitly gives
\begin{align}
    P(\rho) &= \rho \partial_\rho f(\rho;\beta) - f(\rho;\beta)
    \nonumber
    \\
    &= \rho \left[ \partial_\rho \varphi_\star  \left( \partial_\varphi \phi_G + \frac{\rho}{\beta \varphi_\star}
    \right) 
    + \frac{\ln \varphi_\star}{\beta}
    \right]
    - \phi_G - \frac{\rho}{\beta} \ln \varphi_\star
    \nonumber
    \\
    &= - \phi_G( \beta, \varphi_\star(\rho;\beta) )
    \label{eq:PressureFromGrandPot}
    \:,
\end{align}
by using~\eqref{eq:EqLegendreFfromGC}. We thus recovered the well-known property that the pressure is the opposite of the grand potential~\cite{Hill:1986}. Importantly, due to the definition~\eqref{eq:GrandPartition}, the thermal de Broglie wavelength $\ell_0$ enters $\phi_G$ only through the combination $\varphi/\ell_0^d$,
\begin{equation}
    \phi_G(\beta,\varphi) \equiv \tilde{\phi}_G \left( \beta, \tilde{\varphi} = \frac{\varphi}{\ell_0^d} \right)
    \:.
\end{equation}
Consequently, the equation~\eqref{eq:EqLegendreFfromGC} for $\varphi^\star$ becomes
\begin{equation}
    \label{eq:LegendreNew}
    \partial_{\tilde{\varphi}} \tilde\phi_G + \frac{\rho}{\beta \tilde\varphi} = 0
    \:,
\end{equation}
and the pressure~\eqref{eq:PressureFromGrandPot} reads
\begin{equation}
    \label{eq:PfromGPotNew}
    P(\rho) = - \tilde{\phi}_G(\beta, \tilde{\varphi}_\star)
    \:.
\end{equation}
Since neither Eq.~\eqref{eq:LegendreNew} nor Eq.~\eqref{eq:PfromGPotNew} involve the thermal de Broglie wavelength $\ell_0$, the pressure expressed in terms of the density $P(\rho)$ does not depend on $\ell_0$. Therefore, the diffusion coefficient~\eqref{eq:DiffCoef} depends only on the microscopic parameters present in the evolution equations~\eqref{eq:EqBrownianPart}, and not on $\ell_0$, as it should. This is a general property, which we have verified explicitly on the different examples considered in Section~\ref{sec:TrCoefs}. Since the pressure is also given by~\eqref{eq:RelDpFpp}, then $f''(\rho)$ also does not depend on $\ell_0$.

\section{Numerical computation of the transport coefficients}
\label{app:NumDSigma}

In one dimension, the transport coefficients can be computed by using the method described in~\cite{Kollmann:2003}. From the result~\eqref{eq:TwoPtCorrelDens2}, we can show that the method actually works in arbitrary dimension.
Indeed, by computing the Fourier transform of the two-point correlation function~\eqref{eq:TwoPtCorrelDens2}, we obtain the dynamical structure factor
\begin{align}
    \label{eq:DefStrFacMicro}
    S(\V{k},t) &\equiv \frac{1}{N}
    \moy{ \sum_{i, j = 1}^N \e^{\I \V{k} \cdot (\V{x}_i(t) - \V{x}_j(0))}}\textbf{}
    \\
    &= \int \dd^d \V{x} \moy{\rho_0(\V{x},t) \rho_0(\V{0},0)}_c
    \nonumber
    \\
    & \underset{\V{k} \to 0}{\simeq}
    \frac{\sigma(\bar\rho)}{2\bar\rho D(\bar\rho)} \e^{- \V{k}^2 D(\bar\rho) t}
    \label{eq:ExprStrFact}
    \:.
\end{align}
In one dimension, we recover the result of~\cite{Kollmann:2003}, which remains valid in any dimension.
One can thus measure in a numerical simulation the structure factor $S(\V{k},t)$ by using the definition~\eqref{eq:DefStrFacMicro}. Then, using~\eqref{eq:ExprStrFact}, the diffusion coefficient $D(\rho)$ is obtained from the small-$\V{k}$ behaviour of
\begin{equation}
    - \ln \frac{S(\V{k},t)}{S(\V{k},0)} \underset{\V{k} \to 0}{\simeq} \V{k}^2 D(\bar\rho) t
    \:,
\end{equation}
while the mobility $\sigma(\rho)$ is deduced from $S(\V{0},0) = \sigma(\bar\rho)/(2 \bar\rho D(\bar\rho))$. 

\section{Numerical resolution of the HNC equation}
\label{app:SolHNC}

The hypernetted-chain equation~\eqref{eq:HNC} is solved numerically by discretising the integral. To reduce the number of equations to solve, we use the rotational symmetry of the pair correlation function $g(\V{r}) = g(r)$. The set of nonlinear equations is solved numerically using Newton's method. To avoid the singularities of the potential, we introduce a cutoff $V_{\mathrm{max}}$ and replace the true potential by
\begin{equation}
    \tilde{V}_0(x) =
    \left\lbrace
    \begin{array}{ll}
         V_0(x) & \text{if } V_0(x) < V_{\mathrm{max}} \;,  \\[0.1cm]
         V_{\mathrm{max}} &  \text{if } V_0(x) > V_{\mathrm{max}}  \:.
    \end{array}
    \right.
\end{equation}
We typically choose $V_{\mathrm{max}} \in [20,50]$, and use $N = 400$ points. To ensure stability of Newton's method, we first solve the equations at a low density $\rho = 0.01$ using as an initial guess $g(r) = \e^{- \beta V_0(r)}$. We then use the obtained solution as an initial guess to solve the equations at density $\rho + \delta \rho$, and iterate until we reach the desired density.

The pair correlation obtained from this method is then numerically integrated using the relation~\eqref{eq:Pfromgr}. Using the values computed for different densities, the diffusion coefficient can be obtained using~\eqref{eq:DiffCoef}. The result of this procedure is shown in Fig.~\ref{fig:D-first3} for different interaction potentials $V_0$.


\begin{thebibliography}{95}%
\makeatletter
\providecommand \@ifxundefined [1]{%
 \@ifx{#1\undefined}
}%
\providecommand \@ifnum [1]{%
 \ifnum #1\expandafter \@firstoftwo
 \else \expandafter \@secondoftwo
 \fi
}%
\providecommand \@ifx [1]{%
 \ifx #1\expandafter \@firstoftwo
 \else \expandafter \@secondoftwo
 \fi
}%
\providecommand \natexlab [1]{#1}%
\providecommand \enquote  [1]{``#1''}%
\providecommand \bibnamefont  [1]{#1}%
\providecommand \bibfnamefont [1]{#1}%
\providecommand \citenamefont [1]{#1}%
\providecommand \href@noop [0]{\@secondoftwo}%
\providecommand \href [0]{\begingroup \@sanitize@url \@href}%
\providecommand \@href[1]{\@@startlink{#1}\@@href}%
\providecommand \@@href[1]{\endgroup#1\@@endlink}%
\providecommand \@sanitize@url [0]{\catcode `\\12\catcode `\$12\catcode
  `\&12\catcode `\#12\catcode `\^12\catcode `\_12\catcode `\%12\relax}%
\providecommand \@@startlink[1]{}%
\providecommand \@@endlink[0]{}%
\providecommand \url  [0]{\begingroup\@sanitize@url \@url }%
\providecommand \@url [1]{\endgroup\@href {#1}{\urlprefix }}%
\providecommand \urlprefix  [0]{URL }%
\providecommand \Eprint [0]{\href }%
\providecommand \doibase [0]{http://dx.doi.org/}%
\providecommand \selectlanguage [0]{\@gobble}%
\providecommand \bibinfo  [0]{\@secondoftwo}%
\providecommand \bibfield  [0]{\@secondoftwo}%
\providecommand \translation [1]{[#1]}%
\providecommand \BibitemOpen [0]{}%
\providecommand \bibitemStop [0]{}%
\providecommand \bibitemNoStop [0]{.\EOS\space}%
\providecommand \EOS [0]{\spacefactor3000\relax}%
\providecommand \BibitemShut  [1]{\csname bibitem#1\endcsname}%
\let\auto@bib@innerbib\@empty
\bibitem [{\citenamefont {Spohn}(1991)}]{Spohn:1991}%
  \BibitemOpen
  \bibfield  {author} {\bibinfo {author} {\bibfnamefont {H.}~\bibnamefont
  {Spohn}},\ }\href@noop {} {\emph {\bibinfo {title} {Large scale dynamics of
  interacting particles}}}\ (\bibinfo  {publisher} {Springer Berlin,
  Heidelberg},\ \bibinfo {year} {1991})\BibitemShut {NoStop}%
\bibitem [{\citenamefont {Evans}\ and\ \citenamefont
  {Hanney}(2005)}]{Evans:2005}%
  \BibitemOpen
  \bibfield  {author} {\bibinfo {author} {\bibfnamefont {M.~R.}\ \bibnamefont
  {Evans}}\ and\ \bibinfo {author} {\bibfnamefont {T.}~\bibnamefont {Hanney}},\
  }\href {\doibase 10.1088/0305-4470/38/19/r01} {\bibfield  {journal} {\bibinfo
   {journal} {J. Phys. A}\ }\textbf {\bibinfo {volume} {38}},\ \bibinfo {pages}
  {R195} (\bibinfo {year} {2005})}\BibitemShut {NoStop}%
\bibitem [{\citenamefont {Derrida}(2007)}]{Derrida:2007}%
  \BibitemOpen
  \bibfield  {author} {\bibinfo {author} {\bibfnamefont {B.}~\bibnamefont
  {Derrida}},\ }\href {\doibase 10.1088/1742-5468/2007/07/p07023} {\bibfield
  {journal} {\bibinfo  {journal} {J. Stat. Mech.}\ }\textbf {\bibinfo {volume}
  {2007}},\ \bibinfo {pages} {P07023} (\bibinfo {year} {2007})}\BibitemShut
  {NoStop}%
\bibitem [{\citenamefont {Campa}\ \emph {et~al.}(2009)\citenamefont {Campa},
  \citenamefont {Dauxois},\ and\ \citenamefont {Ruffo}}]{Campa:2009}%
  \BibitemOpen
  \bibfield  {author} {\bibinfo {author} {\bibfnamefont {A.}~\bibnamefont
  {Campa}}, \bibinfo {author} {\bibfnamefont {T.}~\bibnamefont {Dauxois}}, \
  and\ \bibinfo {author} {\bibfnamefont {S.}~\bibnamefont {Ruffo}},\ }\href
  {\doibase https://doi.org/10.1016/j.physrep.2009.07.001} {\bibfield
  {journal} {\bibinfo  {journal} {Phys. Rep.}\ }\textbf {\bibinfo {volume}
  {480}},\ \bibinfo {pages} {57} (\bibinfo {year} {2009})}\BibitemShut
  {NoStop}%
\bibitem [{\citenamefont {Chou}\ \emph {et~al.}(2011)\citenamefont {Chou},
  \citenamefont {Mallick},\ and\ \citenamefont {Zia}}]{Chou:2011}%
  \BibitemOpen
  \bibfield  {author} {\bibinfo {author} {\bibfnamefont {T.}~\bibnamefont
  {Chou}}, \bibinfo {author} {\bibfnamefont {K.}~\bibnamefont {Mallick}}, \
  and\ \bibinfo {author} {\bibfnamefont {R.~K.}\ \bibnamefont {Zia}},\ }\href
  {\doibase 10.1088/0034-4885/74/11/116601} {\bibfield  {journal} {\bibinfo
  {journal} {Rep. Prog. Phys.}\ }\textbf {\bibinfo {volume} {74}},\ \bibinfo
  {pages} {116601} (\bibinfo {year} {2011})}\BibitemShut {NoStop}%
\bibitem [{\citenamefont {Bertini}\ \emph {et~al.}(2015)\citenamefont
  {Bertini}, \citenamefont {De~Sole}, \citenamefont {Gabrielli}, \citenamefont
  {Jona-Lasinio},\ and\ \citenamefont {Landim}}]{Bertini:2015}%
  \BibitemOpen
  \bibfield  {author} {\bibinfo {author} {\bibfnamefont {L.}~\bibnamefont
  {Bertini}}, \bibinfo {author} {\bibfnamefont {A.}~\bibnamefont {De~Sole}},
  \bibinfo {author} {\bibfnamefont {D.}~\bibnamefont {Gabrielli}}, \bibinfo
  {author} {\bibfnamefont {G.}~\bibnamefont {Jona-Lasinio}}, \ and\ \bibinfo
  {author} {\bibfnamefont {C.}~\bibnamefont {Landim}},\ }\href {\doibase
  10.1103/RevModPhys.87.593} {\bibfield  {journal} {\bibinfo  {journal} {Rev.
  Mod. Phys.}\ }\textbf {\bibinfo {volume} {87}},\ \bibinfo {pages} {593}
  (\bibinfo {year} {2015})}\BibitemShut {NoStop}%
\bibitem [{\citenamefont {Derrida}(2025)}]{Derrida:2025a}%
  \BibitemOpen
  \bibfield  {author} {\bibinfo {author} {\bibfnamefont {B.}~\bibnamefont
  {Derrida}},\ }\href {\doibase 10.21468/SciPostPhysLectNotes.106} {\bibfield
  {journal} {\bibinfo  {journal} {SciPost Phys. Lect. Notes}\ ,\ \bibinfo
  {pages} {106}} (\bibinfo {year} {2025})}\BibitemShut {NoStop}%
\bibitem [{\citenamefont {Derrida}\ and\ \citenamefont
  {Gerschenfeld}(2009{\natexlab{a}})}]{Derrida:2009}%
  \BibitemOpen
  \bibfield  {author} {\bibinfo {author} {\bibfnamefont {B.}~\bibnamefont
  {Derrida}}\ and\ \bibinfo {author} {\bibfnamefont {A.}~\bibnamefont
  {Gerschenfeld}},\ }\href {\doibase 10.1007/s10955-009-9772-7} {\bibfield
  {journal} {\bibinfo  {journal} {J. Stat. Phys.}\ }\textbf {\bibinfo {volume}
  {136}},\ \bibinfo {pages} {1} (\bibinfo {year}
  {2009}{\natexlab{a}})}\BibitemShut {NoStop}%
\bibitem [{\citenamefont {Mallick}(2015)}]{Mallick:2015}%
  \BibitemOpen
  \bibfield  {author} {\bibinfo {author} {\bibfnamefont {K.}~\bibnamefont
  {Mallick}},\ }\href {\doibase 10.1016/j.physa.2014.07.046} {\bibfield
  {journal} {\bibinfo  {journal} {Physica A}\ }\textbf {\bibinfo {volume}
  {418}},\ \bibinfo {pages} {17} (\bibinfo {year} {2015})},\ \bibinfo {note}
  {proceedings of the 13th International Summer School on Fundamental Problems
  in Statistical Physics}\BibitemShut {NoStop}%
\bibitem [{\citenamefont {Mallick}\ \emph {et~al.}(2022)\citenamefont
  {Mallick}, \citenamefont {Moriya},\ and\ \citenamefont
  {Sasamoto}}]{Mallick:2022}%
  \BibitemOpen
  \bibfield  {author} {\bibinfo {author} {\bibfnamefont {K.}~\bibnamefont
  {Mallick}}, \bibinfo {author} {\bibfnamefont {H.}~\bibnamefont {Moriya}}, \
  and\ \bibinfo {author} {\bibfnamefont {T.}~\bibnamefont {Sasamoto}},\ }\href
  {\doibase 10.1103/PhysRevLett.129.040601} {\bibfield  {journal} {\bibinfo
  {journal} {Phys. Rev. Lett.}\ }\textbf {\bibinfo {volume} {129}},\ \bibinfo
  {pages} {040601} (\bibinfo {year} {2022})}\BibitemShut {NoStop}%
\bibitem [{\citenamefont {Rouse}(1953)}]{Rouse:1953}%
  \BibitemOpen
  \bibfield  {author} {\bibinfo {author} {\bibfnamefont {J.}~\bibnamefont
  {Rouse}, \bibfnamefont {Prince~E.}},\ }\href {\doibase 10.1063/1.1699180}
  {\bibfield  {journal} {\bibinfo  {journal} {J. Chem. Phys.}\ }\textbf
  {\bibinfo {volume} {21}},\ \bibinfo {pages} {1272} (\bibinfo {year}
  {1953})}\BibitemShut {NoStop}%
\bibitem [{\citenamefont {Dufrêche}\ \emph {et~al.}(2005)\citenamefont
  {Dufrêche}, \citenamefont {Bernard},\ and\ \citenamefont
  {Turq}}]{Dufreche:2005}%
  \BibitemOpen
  \bibfield  {author} {\bibinfo {author} {\bibfnamefont {J.-F.}\ \bibnamefont
  {Dufrêche}}, \bibinfo {author} {\bibfnamefont {O.}~\bibnamefont {Bernard}},
  \ and\ \bibinfo {author} {\bibfnamefont {P.}~\bibnamefont {Turq}},\ }\href
  {\doibase 10.1016/j.molliq.2004.07.036} {\bibfield  {journal} {\bibinfo
  {journal} {J. Mol. Liq.}\ }\textbf {\bibinfo {volume} {118}},\ \bibinfo
  {pages} {189} (\bibinfo {year} {2005})},\ \bibinfo {note} {contributions to
  the 28th International Conference on Solution Chemistry}\BibitemShut
  {NoStop}%
\bibitem [{\citenamefont {Liu}\ \emph {et~al.}(2015)\citenamefont {Liu},
  \citenamefont {Huang},\ and\ \citenamefont {Suo}}]{Liu:2015}%
  \BibitemOpen
  \bibfield  {author} {\bibinfo {author} {\bibfnamefont {Q.}~\bibnamefont
  {Liu}}, \bibinfo {author} {\bibfnamefont {S.}~\bibnamefont {Huang}}, \ and\
  \bibinfo {author} {\bibfnamefont {Z.}~\bibnamefont {Suo}},\ }\href {\doibase
  10.1103/PhysRevLett.114.224301} {\bibfield  {journal} {\bibinfo  {journal}
  {Phys. Rev. Lett.}\ }\textbf {\bibinfo {volume} {114}},\ \bibinfo {pages}
  {224301} (\bibinfo {year} {2015})}\BibitemShut {NoStop}%
\bibitem [{\citenamefont {Dandekar}\ \emph {et~al.}(2023)\citenamefont
  {Dandekar}, \citenamefont {Krapivsky},\ and\ \citenamefont
  {Mallick}}]{Dandekar:2023}%
  \BibitemOpen
  \bibfield  {author} {\bibinfo {author} {\bibfnamefont {R.}~\bibnamefont
  {Dandekar}}, \bibinfo {author} {\bibfnamefont {P.~L.}\ \bibnamefont
  {Krapivsky}}, \ and\ \bibinfo {author} {\bibfnamefont {K.}~\bibnamefont
  {Mallick}},\ }\href {\doibase 10.1103/PhysRevE.107.044129} {\bibfield
  {journal} {\bibinfo  {journal} {Phys. Rev. E}\ }\textbf {\bibinfo {volume}
  {107}},\ \bibinfo {pages} {044129} (\bibinfo {year} {2023})}\BibitemShut
  {NoStop}%
\bibitem [{\citenamefont {Touzo}\ \emph {et~al.}(2024)\citenamefont {Touzo},
  \citenamefont {Le~Doussal},\ and\ \citenamefont {Schehr}}]{Touzo:2024}%
  \BibitemOpen
  \bibfield  {author} {\bibinfo {author} {\bibfnamefont {L.}~\bibnamefont
  {Touzo}}, \bibinfo {author} {\bibfnamefont {P.}~\bibnamefont {Le~Doussal}}, \
  and\ \bibinfo {author} {\bibfnamefont {G.}~\bibnamefont {Schehr}},\ }\href
  {\doibase 10.1103/PhysRevE.109.014136} {\bibfield  {journal} {\bibinfo
  {journal} {Phys. Rev. E}\ }\textbf {\bibinfo {volume} {109}},\ \bibinfo
  {pages} {014136} (\bibinfo {year} {2024})}\BibitemShut {NoStop}%
\bibitem [{\citenamefont {Spohn}(1987)}]{Spohn:1987}%
  \BibitemOpen
  \bibfield  {author} {\bibinfo {author} {\bibfnamefont {H.}~\bibnamefont
  {Spohn}},\ }\href {\doibase 10.1007/BF01206151} {\bibfield  {journal}
  {\bibinfo  {journal} {J. Stat. Phys.}\ }\textbf {\bibinfo {volume} {47}},\
  \bibinfo {pages} {669} (\bibinfo {year} {1987})}\BibitemShut {NoStop}%
\bibitem [{\citenamefont {Touzo}\ \emph {et~al.}(2025)\citenamefont {Touzo},
  \citenamefont {Le~Doussal},\ and\ \citenamefont {Schehr}}]{Touzo:2024a}%
  \BibitemOpen
  \bibfield  {author} {\bibinfo {author} {\bibfnamefont {L.}~\bibnamefont
  {Touzo}}, \bibinfo {author} {\bibfnamefont {P.}~\bibnamefont {Le~Doussal}}, \
  and\ \bibinfo {author} {\bibfnamefont {G.}~\bibnamefont {Schehr}},\ }\href
  {\doibase 10.1007/s10955-025-03452-7} {\bibfield  {journal} {\bibinfo
  {journal} {J. Stat. Phys.}\ }\textbf {\bibinfo {volume} {192}},\ \bibinfo
  {pages} {79} (\bibinfo {year} {2025})}\BibitemShut {NoStop}%
\bibitem [{\citenamefont {Dandekar}\ \emph {et~al.}(2024)\citenamefont
  {Dandekar}, \citenamefont {Krapivsky},\ and\ \citenamefont
  {Mallick}}]{Dandekar:2024}%
  \BibitemOpen
  \bibfield  {author} {\bibinfo {author} {\bibfnamefont {R.}~\bibnamefont
  {Dandekar}}, \bibinfo {author} {\bibfnamefont {P.~L.}\ \bibnamefont
  {Krapivsky}}, \ and\ \bibinfo {author} {\bibfnamefont {K.}~\bibnamefont
  {Mallick}},\ }\href {\doibase 10.1103/PhysRevE.110.064153} {\bibfield
  {journal} {\bibinfo  {journal} {Phys. Rev. E}\ }\textbf {\bibinfo {volume}
  {110}},\ \bibinfo {pages} {064153} (\bibinfo {year} {2024})}\BibitemShut
  {NoStop}%
\bibitem [{\citenamefont {Kawasaki}(1994)}]{Kawasaki:1994}%
  \BibitemOpen
  \bibfield  {author} {\bibinfo {author} {\bibfnamefont {K.}~\bibnamefont
  {Kawasaki}},\ }\href {\doibase 10.1016/0378-4371(94)90533-9} {\bibfield
  {journal} {\bibinfo  {journal} {Physica A}\ }\textbf {\bibinfo {volume}
  {208}},\ \bibinfo {pages} {35} (\bibinfo {year} {1994})}\BibitemShut
  {NoStop}%
\bibitem [{\citenamefont {Dean}(1996)}]{Dean:1996}%
  \BibitemOpen
  \bibfield  {author} {\bibinfo {author} {\bibfnamefont {D.~S.}\ \bibnamefont
  {Dean}},\ }\href {\doibase 10.1088/0305-4470/29/24/001} {\bibfield  {journal}
  {\bibinfo  {journal} {J. Phys. A}\ }\textbf {\bibinfo {volume} {29}},\
  \bibinfo {pages} {L613} (\bibinfo {year} {1996})}\BibitemShut {NoStop}%
\bibitem [{\citenamefont {Illien}(2025)}]{Illien:2024}%
  \BibitemOpen
  \bibfield  {author} {\bibinfo {author} {\bibfnamefont {P.}~\bibnamefont
  {Illien}},\ }\href {\doibase 10.1088/1361-6633/adee2e} {\bibfield  {journal}
  {\bibinfo  {journal} {Rep. Prog. Phys.}\ }\textbf {\bibinfo {volume} {88}},\
  \bibinfo {pages} {086601} (\bibinfo {year} {2025})}\BibitemShut {NoStop}%
\bibitem [{\citenamefont {Démery}\ \emph {et~al.}(2014)\citenamefont
  {Démery}, \citenamefont {Bénichou},\ and\ \citenamefont
  {Jacquin}}]{Demery:2014}%
  \BibitemOpen
  \bibfield  {author} {\bibinfo {author} {\bibfnamefont {V.}~\bibnamefont
  {Démery}}, \bibinfo {author} {\bibfnamefont {O.}~\bibnamefont {Bénichou}},
  \ and\ \bibinfo {author} {\bibfnamefont {H.}~\bibnamefont {Jacquin}},\ }\href
  {\doibase 10.1088/1367-2630/16/5/053032} {\bibfield  {journal} {\bibinfo
  {journal} {New J. Phys.}\ }\textbf {\bibinfo {volume} {16}},\ \bibinfo
  {pages} {053032} (\bibinfo {year} {2014})}\BibitemShut {NoStop}%
\bibitem [{\citenamefont {Dean}\ and\ \citenamefont
  {Podgornik}(2014)}]{Dean:2014}%
  \BibitemOpen
  \bibfield  {author} {\bibinfo {author} {\bibfnamefont {D.~S.}\ \bibnamefont
  {Dean}}\ and\ \bibinfo {author} {\bibfnamefont {R.}~\bibnamefont
  {Podgornik}},\ }\href {\doibase 10.1103/PhysRevE.89.032117} {\bibfield
  {journal} {\bibinfo  {journal} {Phys. Rev. E}\ }\textbf {\bibinfo {volume}
  {89}},\ \bibinfo {pages} {032117} (\bibinfo {year} {2014})}\BibitemShut
  {NoStop}%
\bibitem [{\citenamefont {Démery}\ and\ \citenamefont
  {Dean}(2016)}]{Demery:2016}%
  \BibitemOpen
  \bibfield  {author} {\bibinfo {author} {\bibfnamefont {V.}~\bibnamefont
  {Démery}}\ and\ \bibinfo {author} {\bibfnamefont {D.~S.}\ \bibnamefont
  {Dean}},\ }\href {\doibase 10.1088/1742-5468/2016/02/023106} {\bibfield
  {journal} {\bibinfo  {journal} {J. Stat. Mech.}\ }\textbf {\bibinfo {volume}
  {2016}},\ \bibinfo {pages} {023106} (\bibinfo {year} {2016})}\BibitemShut
  {NoStop}%
\bibitem [{\citenamefont {Poncet}\ \emph {et~al.}(2017)\citenamefont {Poncet},
  \citenamefont {B\'enichou}, \citenamefont {D\'emery},\ and\ \citenamefont
  {Oshanin}}]{Poncet:2017}%
  \BibitemOpen
  \bibfield  {author} {\bibinfo {author} {\bibfnamefont {A.}~\bibnamefont
  {Poncet}}, \bibinfo {author} {\bibfnamefont {O.}~\bibnamefont {B\'enichou}},
  \bibinfo {author} {\bibfnamefont {V.}~\bibnamefont {D\'emery}}, \ and\
  \bibinfo {author} {\bibfnamefont {G.}~\bibnamefont {Oshanin}},\ }\href
  {\doibase 10.1103/PhysRevLett.118.118002} {\bibfield  {journal} {\bibinfo
  {journal} {Phys. Rev. Lett.}\ }\textbf {\bibinfo {volume} {118}},\ \bibinfo
  {pages} {118002} (\bibinfo {year} {2017})}\BibitemShut {NoStop}%
\bibitem [{\citenamefont {Kr\"{u}ger}\ and\ \citenamefont
  {Dean}(2017)}]{Kruger:2017}%
  \BibitemOpen
  \bibfield  {author} {\bibinfo {author} {\bibfnamefont {M.}~\bibnamefont
  {Kr\"{u}ger}}\ and\ \bibinfo {author} {\bibfnamefont {D.~S.}\ \bibnamefont
  {Dean}},\ }\href {http://dx.doi.org/10.1063/1.4979659} {\bibfield  {journal}
  {\bibinfo  {journal} {J. Chem. Phys.}\ }\textbf {\bibinfo {volume} {146}}
  (\bibinfo {year} {2017})}\BibitemShut {NoStop}%
\bibitem [{\citenamefont {Kr\"{u}ger}\ \emph {et~al.}(2018)\citenamefont
  {Kr\"{u}ger}, \citenamefont {Solon}, \citenamefont {Démery}, \citenamefont
  {Rohwer},\ and\ \citenamefont {Dean}}]{Kruger:2018}%
  \BibitemOpen
  \bibfield  {author} {\bibinfo {author} {\bibfnamefont {M.}~\bibnamefont
  {Kr\"{u}ger}}, \bibinfo {author} {\bibfnamefont {A.}~\bibnamefont {Solon}},
  \bibinfo {author} {\bibfnamefont {V.}~\bibnamefont {Démery}}, \bibinfo
  {author} {\bibfnamefont {C.~M.}\ \bibnamefont {Rohwer}}, \ and\ \bibinfo
  {author} {\bibfnamefont {D.~S.}\ \bibnamefont {Dean}},\ }\href
  {http://dx.doi.org/10.1063/1.5019424} {\bibfield  {journal} {\bibinfo
  {journal} {J. Chem. Phys.}\ }\textbf {\bibinfo {volume} {148}} (\bibinfo
  {year} {2018})}\BibitemShut {NoStop}%
\bibitem [{\citenamefont {Mahdisoltani}\ and\ \citenamefont
  {Golestanian}(2021)}]{Mahdisoltani:2021}%
  \BibitemOpen
  \bibfield  {author} {\bibinfo {author} {\bibfnamefont {S.}~\bibnamefont
  {Mahdisoltani}}\ and\ \bibinfo {author} {\bibfnamefont {R.}~\bibnamefont
  {Golestanian}},\ }\href {\doibase 10.1103/PhysRevLett.126.158002} {\bibfield
  {journal} {\bibinfo  {journal} {Phys. Rev. Lett.}\ }\textbf {\bibinfo
  {volume} {126}},\ \bibinfo {pages} {158002} (\bibinfo {year}
  {2021})}\BibitemShut {NoStop}%
\bibitem [{\citenamefont {Venturelli}\ \emph {et~al.}(2025)\citenamefont
  {Venturelli}, \citenamefont {Illien}, \citenamefont {Grabsch},\ and\
  \citenamefont {B\'enichou}}]{Venturelli:2024}%
  \BibitemOpen
  \bibfield  {author} {\bibinfo {author} {\bibfnamefont {D.}~\bibnamefont
  {Venturelli}}, \bibinfo {author} {\bibfnamefont {P.}~\bibnamefont {Illien}},
  \bibinfo {author} {\bibfnamefont {A.}~\bibnamefont {Grabsch}}, \ and\
  \bibinfo {author} {\bibfnamefont {O.}~\bibnamefont {B\'enichou}},\ }\href
  {\doibase 10.1103/55qy-sflc} {\bibfield  {journal} {\bibinfo  {journal}
  {Phys. Rev. Lett.}\ }\textbf {\bibinfo {volume} {135}},\ \bibinfo {pages}
  {127101} (\bibinfo {year} {2025})}\BibitemShut {NoStop}%
\bibitem [{\citenamefont {Muzzeddu}\ \emph {et~al.}(2025)\citenamefont
  {Muzzeddu}, \citenamefont {Kalz}, \citenamefont {Gambassi}, \citenamefont
  {Sharma},\ and\ \citenamefont {Metzler}}]{Muzzeddu:2024}%
  \BibitemOpen
  \bibfield  {author} {\bibinfo {author} {\bibfnamefont {P.~L.}\ \bibnamefont
  {Muzzeddu}}, \bibinfo {author} {\bibfnamefont {E.}~\bibnamefont {Kalz}},
  \bibinfo {author} {\bibfnamefont {A.}~\bibnamefont {Gambassi}}, \bibinfo
  {author} {\bibfnamefont {A.}~\bibnamefont {Sharma}}, \ and\ \bibinfo {author}
  {\bibfnamefont {R.}~\bibnamefont {Metzler}},\ }\href {\doibase
  10.1088/1367-2630/adbdea} {\bibfield  {journal} {\bibinfo  {journal} {New J.
  Phys.}\ }\textbf {\bibinfo {volume} {27}},\ \bibinfo {pages} {033025}
  (\bibinfo {year} {2025})}\BibitemShut {NoStop}%
\bibitem [{\citenamefont {Spohn}(1983)}]{Spohn:1983}%
  \BibitemOpen
  \bibfield  {author} {\bibinfo {author} {\bibfnamefont {H.}~\bibnamefont
  {Spohn}},\ }\href {\doibase 10.1088/0305-4470/16/18/029} {\bibfield
  {journal} {\bibinfo  {journal} {J. Phys. A}\ }\textbf {\bibinfo {volume}
  {16}},\ \bibinfo {pages} {4275} (\bibinfo {year} {1983})}\BibitemShut
  {NoStop}%
\bibitem [{\citenamefont {Derrida}\ and\ \citenamefont
  {Gerschenfeld}(2009{\natexlab{b}})}]{Derrida:2009a}%
  \BibitemOpen
  \bibfield  {author} {\bibinfo {author} {\bibfnamefont {B.}~\bibnamefont
  {Derrida}}\ and\ \bibinfo {author} {\bibfnamefont {A.}~\bibnamefont
  {Gerschenfeld}},\ }\href {\doibase 10.1007/s10955-009-9830-1} {\bibfield
  {journal} {\bibinfo  {journal} {J. Stat. Phys.}\ }\textbf {\bibinfo {volume}
  {137}},\ \bibinfo {pages} {978} (\bibinfo {year}
  {2009}{\natexlab{b}})}\BibitemShut {NoStop}%
\bibitem [{\citenamefont {Krapivsky}\ and\ \citenamefont
  {Meerson}(2012)}]{Krapivsky:2012}%
  \BibitemOpen
  \bibfield  {author} {\bibinfo {author} {\bibfnamefont {P.}~\bibnamefont
  {Krapivsky}}\ and\ \bibinfo {author} {\bibfnamefont {B.}~\bibnamefont
  {Meerson}},\ }\href {\doibase 10.1103/PhysRevE.86.031106} {\bibfield
  {journal} {\bibinfo  {journal} {Phys. Rev. E}\ }\textbf {\bibinfo {volume}
  {86}},\ \bibinfo {pages} {031106} (\bibinfo {year} {2012})}\BibitemShut
  {NoStop}%
\bibitem [{\citenamefont {Krapivsky}\ \emph {et~al.}(2014)\citenamefont
  {Krapivsky}, \citenamefont {Mallick},\ and\ \citenamefont
  {Sadhu}}]{Krapivsky:2014}%
  \BibitemOpen
  \bibfield  {author} {\bibinfo {author} {\bibfnamefont {P.~L.}\ \bibnamefont
  {Krapivsky}}, \bibinfo {author} {\bibfnamefont {K.}~\bibnamefont {Mallick}},
  \ and\ \bibinfo {author} {\bibfnamefont {T.}~\bibnamefont {Sadhu}},\ }\href
  {\doibase 10.1103/PhysRevLett.113.078101} {\bibfield  {journal} {\bibinfo
  {journal} {Phys. Rev. Lett.}\ }\textbf {\bibinfo {volume} {113}},\ \bibinfo
  {pages} {078101} (\bibinfo {year} {2014})}\BibitemShut {NoStop}%
\bibitem [{\citenamefont {Krapivsky}\ \emph
  {et~al.}(2015{\natexlab{a}})\citenamefont {Krapivsky}, \citenamefont
  {Mallick},\ and\ \citenamefont {Sadhu}}]{Krapivsky:2015}%
  \BibitemOpen
  \bibfield  {author} {\bibinfo {author} {\bibfnamefont {P.~L.}\ \bibnamefont
  {Krapivsky}}, \bibinfo {author} {\bibfnamefont {K.}~\bibnamefont {Mallick}},
  \ and\ \bibinfo {author} {\bibfnamefont {T.}~\bibnamefont {Sadhu}},\ }\href
  {\doibase 10.1088/1742-5468/2015/09/P09007} {\bibfield  {journal} {\bibinfo
  {journal} {J. Stat. Mech.}\ }\textbf {\bibinfo {volume} {2015}},\ \bibinfo
  {pages} {P09007} (\bibinfo {year} {2015}{\natexlab{a}})}\BibitemShut
  {NoStop}%
\bibitem [{\citenamefont {Krapivsky}\ \emph
  {et~al.}(2015{\natexlab{b}})\citenamefont {Krapivsky}, \citenamefont
  {Mallick},\ and\ \citenamefont {Sadhu}}]{Krapivsky:2015a}%
  \BibitemOpen
  \bibfield  {author} {\bibinfo {author} {\bibfnamefont {P.~L.}\ \bibnamefont
  {Krapivsky}}, \bibinfo {author} {\bibfnamefont {K.}~\bibnamefont {Mallick}},
  \ and\ \bibinfo {author} {\bibfnamefont {T.}~\bibnamefont {Sadhu}},\ }\href
  {\doibase 10.1007/s10955-015-1291-0} {\bibfield  {journal} {\bibinfo
  {journal} {J. Stat. Phys.}\ }\textbf {\bibinfo {volume} {160}},\ \bibinfo
  {pages} {885} (\bibinfo {year} {2015}{\natexlab{b}})}\BibitemShut {NoStop}%
\bibitem [{\citenamefont {Sadhu}\ and\ \citenamefont
  {Derrida}(2015)}]{Sadhu:2015}%
  \BibitemOpen
  \bibfield  {author} {\bibinfo {author} {\bibfnamefont {T.}~\bibnamefont
  {Sadhu}}\ and\ \bibinfo {author} {\bibfnamefont {B.}~\bibnamefont
  {Derrida}},\ }\href {\doibase 10.1088/1742-5468/2015/09/p09008} {\bibfield
  {journal} {\bibinfo  {journal} {J. Stat. Mech.}\ }\textbf {\bibinfo {volume}
  {2015}},\ \bibinfo {pages} {P09008} (\bibinfo {year} {2015})}\BibitemShut
  {NoStop}%
\bibitem [{\citenamefont {Poncet}\ \emph {et~al.}(2021)\citenamefont {Poncet},
  \citenamefont {Grabsch}, \citenamefont {Illien},\ and\ \citenamefont
  {B\'enichou}}]{Poncet:2021}%
  \BibitemOpen
  \bibfield  {author} {\bibinfo {author} {\bibfnamefont {A.}~\bibnamefont
  {Poncet}}, \bibinfo {author} {\bibfnamefont {A.}~\bibnamefont {Grabsch}},
  \bibinfo {author} {\bibfnamefont {P.}~\bibnamefont {Illien}}, \ and\ \bibinfo
  {author} {\bibfnamefont {O.}~\bibnamefont {B\'enichou}},\ }\href {\doibase
  10.1103/PhysRevLett.127.220601} {\bibfield  {journal} {\bibinfo  {journal}
  {Phys. Rev. Lett.}\ }\textbf {\bibinfo {volume} {127}},\ \bibinfo {pages}
  {220601} (\bibinfo {year} {2021})}\BibitemShut {NoStop}%
\bibitem [{\citenamefont {Grabsch}\ \emph {et~al.}(2022)\citenamefont
  {Grabsch}, \citenamefont {Poncet}, \citenamefont {Rizkallah}, \citenamefont
  {Illien},\ and\ \citenamefont {B{\'e}nichou}}]{Grabsch:2022}%
  \BibitemOpen
  \bibfield  {author} {\bibinfo {author} {\bibfnamefont {A.}~\bibnamefont
  {Grabsch}}, \bibinfo {author} {\bibfnamefont {A.}~\bibnamefont {Poncet}},
  \bibinfo {author} {\bibfnamefont {P.}~\bibnamefont {Rizkallah}}, \bibinfo
  {author} {\bibfnamefont {P.}~\bibnamefont {Illien}}, \ and\ \bibinfo {author}
  {\bibfnamefont {O.}~\bibnamefont {B{\'e}nichou}},\ }\href {\doibase
  10.1126/sciadv.abm5043} {\bibfield  {journal} {\bibinfo  {journal} {Sci.
  Adv.}\ }\textbf {\bibinfo {volume} {8}},\ \bibinfo {pages} {eabm5043}
  (\bibinfo {year} {2022})}\BibitemShut {NoStop}%
\bibitem [{\citenamefont {Bettelheim}\ \emph
  {et~al.}(2022{\natexlab{a}})\citenamefont {Bettelheim}, \citenamefont
  {Smith},\ and\ \citenamefont {Meerson}}]{Bettelheim:2022}%
  \BibitemOpen
  \bibfield  {author} {\bibinfo {author} {\bibfnamefont {E.}~\bibnamefont
  {Bettelheim}}, \bibinfo {author} {\bibfnamefont {N.~R.}\ \bibnamefont
  {Smith}}, \ and\ \bibinfo {author} {\bibfnamefont {B.}~\bibnamefont
  {Meerson}},\ }\href {\doibase 10.1103/PhysRevLett.128.130602} {\bibfield
  {journal} {\bibinfo  {journal} {Phys. Rev. Lett.}\ }\textbf {\bibinfo
  {volume} {128}},\ \bibinfo {pages} {130602} (\bibinfo {year}
  {2022}{\natexlab{a}})}\BibitemShut {NoStop}%
\bibitem [{\citenamefont {Krajenbrink}\ and\ \citenamefont
  {Le~Doussal}(2023)}]{Krajenbrink:2022}%
  \BibitemOpen
  \bibfield  {author} {\bibinfo {author} {\bibfnamefont {A.}~\bibnamefont
  {Krajenbrink}}\ and\ \bibinfo {author} {\bibfnamefont {P.}~\bibnamefont
  {Le~Doussal}},\ }\href {\doibase 10.1103/PhysRevE.107.014137} {\bibfield
  {journal} {\bibinfo  {journal} {Phys. Rev. E}\ }\textbf {\bibinfo {volume}
  {107}},\ \bibinfo {pages} {014137} (\bibinfo {year} {2023})}\BibitemShut
  {NoStop}%
\bibitem [{\citenamefont {Rizkallah}\ \emph {et~al.}(2023)\citenamefont
  {Rizkallah}, \citenamefont {Grabsch}, \citenamefont {Illien},\ and\
  \citenamefont {Bénichou}}]{Rizkallah:2022}%
  \BibitemOpen
  \bibfield  {author} {\bibinfo {author} {\bibfnamefont {P.}~\bibnamefont
  {Rizkallah}}, \bibinfo {author} {\bibfnamefont {A.}~\bibnamefont {Grabsch}},
  \bibinfo {author} {\bibfnamefont {P.}~\bibnamefont {Illien}}, \ and\ \bibinfo
  {author} {\bibfnamefont {O.}~\bibnamefont {Bénichou}},\ }\href {\doibase
  10.1088/1742-5468/aca8fb} {\bibfield  {journal} {\bibinfo  {journal} {J.
  Stat. Mech.}\ }\textbf {\bibinfo {volume} {2023}},\ \bibinfo {pages} {013202}
  (\bibinfo {year} {2023})}\BibitemShut {NoStop}%
\bibitem [{\citenamefont {Grabsch}\ \emph {et~al.}(2023)\citenamefont
  {Grabsch}, \citenamefont {Rizkallah}, \citenamefont {Poncet}, \citenamefont
  {Illien},\ and\ \citenamefont {B\'enichou}}]{Grabsch:2023}%
  \BibitemOpen
  \bibfield  {author} {\bibinfo {author} {\bibfnamefont {A.}~\bibnamefont
  {Grabsch}}, \bibinfo {author} {\bibfnamefont {P.}~\bibnamefont {Rizkallah}},
  \bibinfo {author} {\bibfnamefont {A.}~\bibnamefont {Poncet}}, \bibinfo
  {author} {\bibfnamefont {P.}~\bibnamefont {Illien}}, \ and\ \bibinfo {author}
  {\bibfnamefont {O.}~\bibnamefont {B\'enichou}},\ }\href {\doibase
  10.1103/PhysRevE.107.044131} {\bibfield  {journal} {\bibinfo  {journal}
  {Phys. Rev. E}\ }\textbf {\bibinfo {volume} {107}},\ \bibinfo {pages}
  {044131} (\bibinfo {year} {2023})}\BibitemShut {NoStop}%
\bibitem [{\citenamefont {Grabsch}\ \emph
  {et~al.}(2024{\natexlab{a}})\citenamefont {Grabsch}, \citenamefont
  {Rizkallah},\ and\ \citenamefont {Bénichou}}]{Grabsch:2024}%
  \BibitemOpen
  \bibfield  {author} {\bibinfo {author} {\bibfnamefont {A.}~\bibnamefont
  {Grabsch}}, \bibinfo {author} {\bibfnamefont {P.}~\bibnamefont {Rizkallah}},
  \ and\ \bibinfo {author} {\bibfnamefont {O.}~\bibnamefont {Bénichou}},\
  }\href {\doibase 10.21468/SciPostPhys.16.1.016} {\bibfield  {journal}
  {\bibinfo  {journal} {SciPost Phys.}\ }\textbf {\bibinfo {volume} {16}},\
  \bibinfo {pages} {016} (\bibinfo {year} {2024}{\natexlab{a}})}\BibitemShut
  {NoStop}%
\bibitem [{\citenamefont {Grabsch}\ and\ \citenamefont
  {B\'enichou}(2024)}]{Grabsch:2024b}%
  \BibitemOpen
  \bibfield  {author} {\bibinfo {author} {\bibfnamefont {A.}~\bibnamefont
  {Grabsch}}\ and\ \bibinfo {author} {\bibfnamefont {O.}~\bibnamefont
  {B\'enichou}},\ }\href {\doibase 10.1103/PhysRevLett.132.217101} {\bibfield
  {journal} {\bibinfo  {journal} {Phys. Rev. Lett.}\ }\textbf {\bibinfo
  {volume} {132}},\ \bibinfo {pages} {217101} (\bibinfo {year}
  {2024})}\BibitemShut {NoStop}%
\bibitem [{\citenamefont {Bodineau}\ and\ \citenamefont
  {Derrida}(2025)}]{Bodineau:2025}%
  \BibitemOpen
  \bibfield  {author} {\bibinfo {author} {\bibfnamefont {T.}~\bibnamefont
  {Bodineau}}\ and\ \bibinfo {author} {\bibfnamefont {B.}~\bibnamefont
  {Derrida}},\ }\href {\doibase 10.1007/s10955-025-03439-4} {\bibfield
  {journal} {\bibinfo  {journal} {J. Stat. Phys.}\ }\textbf {\bibinfo {volume}
  {192}},\ \bibinfo {pages} {1} (\bibinfo {year} {2025})}\BibitemShut {NoStop}%
\bibitem [{\citenamefont {Saha}\ and\ \citenamefont {Sadhu}(2025)}]{Saha:2025}%
  \BibitemOpen
  \bibfield  {author} {\bibinfo {author} {\bibfnamefont {S.}~\bibnamefont
  {Saha}}\ and\ \bibinfo {author} {\bibfnamefont {T.}~\bibnamefont {Sadhu}},\
  }\href {\doibase 10.48550/arXiv.2501.03164} {\bibfield  {journal} {\bibinfo
  {journal} {arXiv:2501.03164}\ } (\bibinfo {year} {2025}),\
  10.48550/arXiv.2501.03164}\BibitemShut {NoStop}%
\bibitem [{\citenamefont {Grabsch}\ \emph {et~al.}(2025)\citenamefont
  {Grabsch}, \citenamefont {Venturelli},\ and\ \citenamefont
  {B\'enichou}}]{Grabsch:2025b}%
  \BibitemOpen
  \bibfield  {author} {\bibinfo {author} {\bibfnamefont {A.}~\bibnamefont
  {Grabsch}}, \bibinfo {author} {\bibfnamefont {D.}~\bibnamefont {Venturelli}},
  \ and\ \bibinfo {author} {\bibfnamefont {O.}~\bibnamefont {B\'enichou}},\
  }\href {\doibase 10.1103/gwdh-3vqm} {\bibfield  {journal} {\bibinfo
  {journal} {Phys. Rev. Lett.}\ }\textbf {\bibinfo {volume} {135}},\ \bibinfo
  {pages} {137102} (\bibinfo {year} {2025})}\BibitemShut {NoStop}%
\bibitem [{\citenamefont {Baxter}(1968)}]{Baxter:1968}%
  \BibitemOpen
  \bibfield  {author} {\bibinfo {author} {\bibfnamefont {R.~J.}\ \bibnamefont
  {Baxter}},\ }\href {\doibase 10.1063/1.1670482} {\bibfield  {journal}
  {\bibinfo  {journal} {J. Chem. Phys.}\ }\textbf {\bibinfo {volume} {49}},\
  \bibinfo {pages} {2770} (\bibinfo {year} {1968})}\BibitemShut {NoStop}%
\bibitem [{\citenamefont {Percus}(1982)}]{Percus:1982}%
  \BibitemOpen
  \bibfield  {author} {\bibinfo {author} {\bibfnamefont {J.}~\bibnamefont
  {Percus}},\ }\href {\doibase doi.org/10.1007/BF01011623} {\bibfield
  {journal} {\bibinfo  {journal} {J. Stat. Phys.}\ }\textbf {\bibinfo {volume}
  {28}},\ \bibinfo {pages} {67} (\bibinfo {year} {1982})}\BibitemShut {NoStop}%
\bibitem [{\citenamefont {Bravo~Yuste}\ \emph {et~al.}(2025)\citenamefont
  {Bravo~Yuste}, \citenamefont {Baumgaertner},\ and\ \citenamefont
  {Abad}}]{Yuste:2025}%
  \BibitemOpen
  \bibfield  {author} {\bibinfo {author} {\bibfnamefont {S.}~\bibnamefont
  {Bravo~Yuste}}, \bibinfo {author} {\bibfnamefont {A.}~\bibnamefont
  {Baumgaertner}}, \ and\ \bibinfo {author} {\bibfnamefont {E.}~\bibnamefont
  {Abad}},\ }\href {\doibase 10.1063/5.0271037} {\bibfield  {journal} {\bibinfo
   {journal} {J. Chem. Phys.}\ }\textbf {\bibinfo {volume} {162}},\ \bibinfo
  {pages} {224114} (\bibinfo {year} {2025})}\BibitemShut {NoStop}%
\bibitem [{\citenamefont {Le~Bon}\ \emph {et~al.}(2025)\citenamefont {Le~Bon},
  \citenamefont {Carof},\ and\ \citenamefont {Illien}}]{Illien:2025}%
  \BibitemOpen
  \bibfield  {author} {\bibinfo {author} {\bibfnamefont {L.}~\bibnamefont
  {Le~Bon}}, \bibinfo {author} {\bibfnamefont {A.}~\bibnamefont {Carof}}, \
  and\ \bibinfo {author} {\bibfnamefont {P.}~\bibnamefont {Illien}},\ }\href
  {\doibase 10.1103/5mjd-m46h} {\bibfield  {journal} {\bibinfo  {journal}
  {Phys. Rev. E}\ }\textbf {\bibinfo {volume} {112}},\ \bibinfo {pages}
  {034127} (\bibinfo {year} {2025})}\BibitemShut {NoStop}%
\bibitem [{\citenamefont {Van~den Broeck}\ \emph {et~al.}(1981)\citenamefont
  {Van~den Broeck}, \citenamefont {Lostak},\ and\ \citenamefont
  {Lekkerkerker}}]{Lekkerkerker:1981}%
  \BibitemOpen
  \bibfield  {author} {\bibinfo {author} {\bibfnamefont {C.}~\bibnamefont
  {Van~den Broeck}}, \bibinfo {author} {\bibfnamefont {F.}~\bibnamefont
  {Lostak}}, \ and\ \bibinfo {author} {\bibfnamefont {H.~N.~W.}\ \bibnamefont
  {Lekkerkerker}},\ }\href {\doibase 10.1063/1.441244} {\bibfield  {journal}
  {\bibinfo  {journal} {J. Chem. Phys.}\ }\textbf {\bibinfo {volume} {74}},\
  \bibinfo {pages} {2006} (\bibinfo {year} {1981})}\BibitemShut {NoStop}%
\bibitem [{\citenamefont {Cichocki}\ and\ \citenamefont
  {Felderhof}(1991)}]{Cichocki:1991}%
  \BibitemOpen
  \bibfield  {author} {\bibinfo {author} {\bibfnamefont {B.}~\bibnamefont
  {Cichocki}}\ and\ \bibinfo {author} {\bibfnamefont {B.~U.}\ \bibnamefont
  {Felderhof}},\ }\href {\doibase 10.1063/1.460319} {\bibfield  {journal}
  {\bibinfo  {journal} {J. Chem. Phys.}\ }\textbf {\bibinfo {volume} {94}},\
  \bibinfo {pages} {556} (\bibinfo {year} {1991})}\BibitemShut {NoStop}%
\bibitem [{\citenamefont {Butt{\`a}}\ and\ \citenamefont
  {Lebowitz}(1999)}]{Butta:1999}%
  \BibitemOpen
  \bibfield  {author} {\bibinfo {author} {\bibfnamefont {P.}~\bibnamefont
  {Butt{\`a}}}\ and\ \bibinfo {author} {\bibfnamefont {J.~L.}\ \bibnamefont
  {Lebowitz}},\ }\href {\doibase 10.1023/A:1004512807858} {\bibfield  {journal}
  {\bibinfo  {journal} {J. Stat. Phys.}\ }\textbf {\bibinfo {volume} {94}},\
  \bibinfo {pages} {653} (\bibinfo {year} {1999})}\BibitemShut {NoStop}%
\bibitem [{\citenamefont {Felderhof}(2009)}]{Felderhof:2009}%
  \BibitemOpen
  \bibfield  {author} {\bibinfo {author} {\bibfnamefont {B.~U.}\ \bibnamefont
  {Felderhof}},\ }\href {\doibase 10.1063/1.3204469} {\bibfield  {journal}
  {\bibinfo  {journal} {J. Chem. Phys.}\ }\textbf {\bibinfo {volume} {131}},\
  \bibinfo {pages} {064504} (\bibinfo {year} {2009})}\BibitemShut {NoStop}%
\bibitem [{\citenamefont {Hill}(1986)}]{Hill:1986}%
  \BibitemOpen
  \bibfield  {author} {\bibinfo {author} {\bibfnamefont {T.~L.}\ \bibnamefont
  {Hill}},\ }\href@noop {} {\emph {\bibinfo {title} {An introduction to
  statistical thermodynamics}}}\ (\bibinfo  {publisher} {Courier Corporation},\
  \bibinfo {year} {1986})\BibitemShut {NoStop}%
\bibitem [{\citenamefont {Arita}\ \emph {et~al.}(2017)\citenamefont {Arita},
  \citenamefont {Krapivsky},\ and\ \citenamefont {Mallick}}]{Arita:2017}%
  \BibitemOpen
  \bibfield  {author} {\bibinfo {author} {\bibfnamefont {C.}~\bibnamefont
  {Arita}}, \bibinfo {author} {\bibfnamefont {P.~L.}\ \bibnamefont
  {Krapivsky}}, \ and\ \bibinfo {author} {\bibfnamefont {K.}~\bibnamefont
  {Mallick}},\ }\href {\doibase 10.1103/PhysRevE.95.032121} {\bibfield
  {journal} {\bibinfo  {journal} {Phys. Rev. E}\ }\textbf {\bibinfo {volume}
  {95}},\ \bibinfo {pages} {032121} (\bibinfo {year} {2017})}\BibitemShut
  {NoStop}%
\bibitem [{\citenamefont {Arita}\ \emph {et~al.}(2018)\citenamefont {Arita},
  \citenamefont {Krapivsky},\ and\ \citenamefont {Mallick}}]{Arita:2018}%
  \BibitemOpen
  \bibfield  {author} {\bibinfo {author} {\bibfnamefont {C.}~\bibnamefont
  {Arita}}, \bibinfo {author} {\bibfnamefont {P.~L.}\ \bibnamefont
  {Krapivsky}}, \ and\ \bibinfo {author} {\bibfnamefont {K.}~\bibnamefont
  {Mallick}},\ }\href {\doibase 10.1088/1751-8121/aaac89} {\bibfield  {journal}
  {\bibinfo  {journal} {J. Phys. A}\ }\textbf {\bibinfo {volume} {51}},\
  \bibinfo {pages} {125002} (\bibinfo {year} {2018})}\BibitemShut {NoStop}%
\bibitem [{Note1()}]{Note1}%
  \BibitemOpen
  \bibinfo {note} {Note that the same argument also holds for an underdamped
  dynamics.}\BibitemShut {Stop}%
\bibitem [{\citenamefont {Calogero}(1975)}]{Calogero:1975}%
  \BibitemOpen
  \bibfield  {author} {\bibinfo {author} {\bibfnamefont {F.}~\bibnamefont
  {Calogero}},\ }\href {\doibase 10.1007/BF02790495} {\bibfield  {journal}
  {\bibinfo  {journal} {Lett. Nuovo Cimento}\ }\textbf {\bibinfo {volume}
  {13}},\ \bibinfo {pages} {411} (\bibinfo {year} {1975})}\BibitemShut
  {NoStop}%
\bibitem [{\citenamefont {Moser}(1975)}]{Moser:1975}%
  \BibitemOpen
  \bibfield  {author} {\bibinfo {author} {\bibfnamefont {J.}~\bibnamefont
  {Moser}},\ }\href {\doibase https://doi.org/10.1016/0001-8708(75)90151-6}
  {\bibfield  {journal} {\bibinfo  {journal} {Adv. Math.}\ }\textbf {\bibinfo
  {volume} {16}},\ \bibinfo {pages} {197} (\bibinfo {year} {1975})}\BibitemShut
  {NoStop}%
\bibitem [{\citenamefont {Polychronakos}(2006)}]{Polychronakos:2006}%
  \BibitemOpen
  \bibfield  {author} {\bibinfo {author} {\bibfnamefont {A.~P.}\ \bibnamefont
  {Polychronakos}},\ }\href {\doibase 10.1088/0305-4470/39/41/S07} {\bibfield
  {journal} {\bibinfo  {journal} {J. Phys. A}\ }\textbf {\bibinfo {volume}
  {39}},\ \bibinfo {pages} {12793} (\bibinfo {year} {2006})}\BibitemShut
  {NoStop}%
\bibitem [{\citenamefont {Lewin}(2022)}]{Lewin:2022}%
  \BibitemOpen
  \bibfield  {author} {\bibinfo {author} {\bibfnamefont {M.}~\bibnamefont
  {Lewin}},\ }\href {\doibase 10.1063/5.0086835} {\bibfield  {journal}
  {\bibinfo  {journal} {J. Math. Phys.}\ }\textbf {\bibinfo {volume} {63}},\
  \bibinfo {pages} {061101} (\bibinfo {year} {2022})}\BibitemShut {NoStop}%
\bibitem [{\citenamefont {Choquard}(2000)}]{Choquard:2000}%
  \BibitemOpen
  \bibfield  {author} {\bibinfo {author} {\bibfnamefont {P.}~\bibnamefont
  {Choquard}},\ }\enquote {\bibinfo {title} {Classical and quantum partition
  functions of the {C}alogero---{M}oser---{S}utherland model},}\ in\ \href
  {\doibase 10.1007/978-1-4612-1206-5_8} {\emph {\bibinfo {booktitle}
  {Calogero---{M}oser--- {S}utherland Models}}},\ \bibinfo {editor} {edited by\
  \bibinfo {editor} {\bibfnamefont {J.~F.}\ \bibnamefont {van Diejen}}\ and\
  \bibinfo {editor} {\bibfnamefont {L.}~\bibnamefont {Vinet}}}\ (\bibinfo
  {publisher} {Springer New York},\ \bibinfo {address} {New York, NY},\
  \bibinfo {year} {2000})\ pp.\ \bibinfo {pages} {117--125}\BibitemShut
  {NoStop}%
\bibitem [{\citenamefont {Santos}(2016)}]{Santos:2016}%
  \BibitemOpen
  \bibfield  {author} {\bibinfo {author} {\bibfnamefont {A.}~\bibnamefont
  {Santos}},\ }\href {\doibase 10.1007/978-3-319-29668-5} {\bibfield  {journal}
  {\bibinfo  {journal} {Lect. Notes Phys.}\ }\textbf {\bibinfo {volume}
  {923}},\ \bibinfo {pages} {064601} (\bibinfo {year} {2016})}\BibitemShut
  {NoStop}%
\bibitem [{\citenamefont {Wei}\ \emph {et~al.}(2000)\citenamefont {Wei},
  \citenamefont {Bechinger},\ and\ \citenamefont {Leiderer}}]{Wei:2000}%
  \BibitemOpen
  \bibfield  {author} {\bibinfo {author} {\bibfnamefont {Q.-H.}\ \bibnamefont
  {Wei}}, \bibinfo {author} {\bibfnamefont {C.}~\bibnamefont {Bechinger}}, \
  and\ \bibinfo {author} {\bibfnamefont {P.}~\bibnamefont {Leiderer}},\ }\href
  {\doibase 10.1126/science.287.5453.625} {\bibfield  {journal} {\bibinfo
  {journal} {Science}\ }\textbf {\bibinfo {volume} {287}},\ \bibinfo {pages}
  {625} (\bibinfo {year} {2000})}\BibitemShut {NoStop}%
\bibitem [{\citenamefont {Lin}\ \emph {et~al.}(2005)\citenamefont {Lin},
  \citenamefont {Meron}, \citenamefont {Cui}, \citenamefont {Rice},\ and\
  \citenamefont {Diamant}}]{Lin:2005}%
  \BibitemOpen
  \bibfield  {author} {\bibinfo {author} {\bibfnamefont {B.}~\bibnamefont
  {Lin}}, \bibinfo {author} {\bibfnamefont {M.}~\bibnamefont {Meron}}, \bibinfo
  {author} {\bibfnamefont {B.}~\bibnamefont {Cui}}, \bibinfo {author}
  {\bibfnamefont {S.~A.}\ \bibnamefont {Rice}}, \ and\ \bibinfo {author}
  {\bibfnamefont {H.}~\bibnamefont {Diamant}},\ }\href {\doibase
  10.1103/physrevlett.94.216001} {\bibfield  {journal} {\bibinfo  {journal}
  {Phys. Rev. Lett.}\ }\textbf {\bibinfo {volume} {94}},\ \bibinfo {pages}
  {216001} (\bibinfo {year} {2005})}\BibitemShut {NoStop}%
\bibitem [{\citenamefont {Schweers}\ \emph {et~al.}(2023)\citenamefont
  {Schweers}, \citenamefont {Antonov}, \citenamefont {Ryabov},\ and\
  \citenamefont {Maass}}]{Schweers:2023}%
  \BibitemOpen
  \bibfield  {author} {\bibinfo {author} {\bibfnamefont {S.}~\bibnamefont
  {Schweers}}, \bibinfo {author} {\bibfnamefont {A.~P.}\ \bibnamefont
  {Antonov}}, \bibinfo {author} {\bibfnamefont {A.}~\bibnamefont {Ryabov}}, \
  and\ \bibinfo {author} {\bibfnamefont {P.}~\bibnamefont {Maass}},\ }\href
  {\doibase 10.1103/PhysRevE.107.L042102} {\bibfield  {journal} {\bibinfo
  {journal} {Phys. Rev. E}\ }\textbf {\bibinfo {volume} {107}},\ \bibinfo
  {pages} {L042102} (\bibinfo {year} {2023})}\BibitemShut {NoStop}%
\bibitem [{\citenamefont {Yuste}\ and\ \citenamefont
  {Santos}(1993)}]{Yuste:1993}%
  \BibitemOpen
  \bibfield  {author} {\bibinfo {author} {\bibfnamefont {S.~B.}\ \bibnamefont
  {Yuste}}\ and\ \bibinfo {author} {\bibfnamefont {A.}~\bibnamefont {Santos}},\
  }\href {\doibase 10.1007/BF01048029} {\bibfield  {journal} {\bibinfo
  {journal} {J. Stat. Phys.}\ }\textbf {\bibinfo {volume} {72}},\ \bibinfo
  {pages} {703} (\bibinfo {year} {1993})}\BibitemShut {NoStop}%
\bibitem [{\citenamefont {Hansen}\ and\ \citenamefont
  {McDonald}(2005)}]{Hansen:2005}%
  \BibitemOpen
  \bibfield  {author} {\bibinfo {author} {\bibfnamefont {J.-P.}\ \bibnamefont
  {Hansen}}\ and\ \bibinfo {author} {\bibfnamefont {I.~R.}\ \bibnamefont
  {McDonald}},\ }\href@noop {} {\emph {\bibinfo {title} {Theory of simple
  liquids}}},\ \bibinfo {edition} {3rd}\ ed.\ (\bibinfo  {publisher} {Academic
  press},\ \bibinfo {year} {2005})\BibitemShut {NoStop}%
\bibitem [{\citenamefont {Riesz}(1988)}]{Riesz:1988}%
  \BibitemOpen
  \bibfield  {author} {\bibinfo {author} {\bibfnamefont {M.}~\bibnamefont
  {Riesz}},\ }in\ \href {\doibase 10.1007/978-3-642-37535-4_38} {\emph
  {\bibinfo {booktitle} {Collected Papers}}}\ (\bibinfo  {publisher}
  {Springer},\ \bibinfo {year} {1988})\ pp.\ \bibinfo {pages}
  {482--526}\BibitemShut {NoStop}%
\bibitem [{\citenamefont {Berlioz}\ \emph
  {et~al.}(2025{\natexlab{a}})\citenamefont {Berlioz}, \citenamefont
  {B\'enichou},\ and\ \citenamefont {Grabsch}}]{Berlioz:2025}%
  \BibitemOpen
  \bibfield  {author} {\bibinfo {author} {\bibfnamefont {T.}~\bibnamefont
  {Berlioz}}, \bibinfo {author} {\bibfnamefont {O.}~\bibnamefont {B\'enichou}},
  \ and\ \bibinfo {author} {\bibfnamefont {A.}~\bibnamefont {Grabsch}},\ }\href
  {\doibase 10.1103/4j5q-j4ht} {\bibfield  {journal} {\bibinfo  {journal}
  {Phys. Rev. Lett.}\ }\textbf {\bibinfo {volume} {134}},\ \bibinfo {pages}
  {247101} (\bibinfo {year} {2025}{\natexlab{a}})}\BibitemShut {NoStop}%
\bibitem [{\citenamefont {Harris}(1965)}]{Harris:1965}%
  \BibitemOpen
  \bibfield  {author} {\bibinfo {author} {\bibfnamefont {T.~E.}\ \bibnamefont
  {Harris}},\ }\href {\doibase 10.2307/3212197} {\bibfield  {journal} {\bibinfo
   {journal} {J. Appl. Probab.}\ }\textbf {\bibinfo {volume} {2}},\ \bibinfo
  {pages} {323} (\bibinfo {year} {1965})}\BibitemShut {NoStop}%
\bibitem [{\citenamefont {Kollmann}(2003)}]{Kollmann:2003}%
  \BibitemOpen
  \bibfield  {author} {\bibinfo {author} {\bibfnamefont {M.}~\bibnamefont
  {Kollmann}},\ }\href {\doibase 10.1103/PhysRevLett.90.180602} {\bibfield
  {journal} {\bibinfo  {journal} {Phys. Rev. Lett.}\ }\textbf {\bibinfo
  {volume} {90}},\ \bibinfo {pages} {180602} (\bibinfo {year}
  {2003})}\BibitemShut {NoStop}%
\bibitem [{\citenamefont {Hegde}\ \emph {et~al.}(2014)\citenamefont {Hegde},
  \citenamefont {Sabhapandit},\ and\ \citenamefont {Dhar}}]{Hegde:2014}%
  \BibitemOpen
  \bibfield  {author} {\bibinfo {author} {\bibfnamefont {C.}~\bibnamefont
  {Hegde}}, \bibinfo {author} {\bibfnamefont {S.}~\bibnamefont {Sabhapandit}},
  \ and\ \bibinfo {author} {\bibfnamefont {A.}~\bibnamefont {Dhar}},\ }\href
  {\doibase 10.1103/PhysRevLett.113.120601} {\bibfield  {journal} {\bibinfo
  {journal} {Phys. Rev. Lett.}\ }\textbf {\bibinfo {volume} {113}},\ \bibinfo
  {pages} {120601} (\bibinfo {year} {2014})}\BibitemShut {NoStop}%
\bibitem [{\citenamefont {Illien}\ \emph {et~al.}(2013)\citenamefont {Illien},
  \citenamefont {B\'enichou}, \citenamefont {Mej\'{\i}a-Monasterio},
  \citenamefont {Oshanin},\ and\ \citenamefont {Voituriez}}]{Illien:2013}%
  \BibitemOpen
  \bibfield  {author} {\bibinfo {author} {\bibfnamefont {P.}~\bibnamefont
  {Illien}}, \bibinfo {author} {\bibfnamefont {O.}~\bibnamefont {B\'enichou}},
  \bibinfo {author} {\bibfnamefont {C.}~\bibnamefont {Mej\'{\i}a-Monasterio}},
  \bibinfo {author} {\bibfnamefont {G.}~\bibnamefont {Oshanin}}, \ and\
  \bibinfo {author} {\bibfnamefont {R.}~\bibnamefont {Voituriez}},\ }\href
  {\doibase 10.1103/PhysRevLett.111.038102} {\bibfield  {journal} {\bibinfo
  {journal} {Phys. Rev. Lett.}\ }\textbf {\bibinfo {volume} {111}},\ \bibinfo
  {pages} {038102} (\bibinfo {year} {2013})}\BibitemShut {NoStop}%
\bibitem [{\citenamefont {Imamura}\ \emph {et~al.}(2017)\citenamefont
  {Imamura}, \citenamefont {Mallick},\ and\ \citenamefont
  {Sasamoto}}]{Imamura:2017}%
  \BibitemOpen
  \bibfield  {author} {\bibinfo {author} {\bibfnamefont {T.}~\bibnamefont
  {Imamura}}, \bibinfo {author} {\bibfnamefont {K.}~\bibnamefont {Mallick}}, \
  and\ \bibinfo {author} {\bibfnamefont {T.}~\bibnamefont {Sasamoto}},\ }\href
  {\doibase 10.1103/PhysRevLett.118.160601} {\bibfield  {journal} {\bibinfo
  {journal} {Phys. Rev. Lett.}\ }\textbf {\bibinfo {volume} {118}},\ \bibinfo
  {pages} {160601} (\bibinfo {year} {2017})}\BibitemShut {NoStop}%
\bibitem [{\citenamefont {Imamura}\ \emph {et~al.}(2021)\citenamefont
  {Imamura}, \citenamefont {Mallick},\ and\ \citenamefont
  {Sasamoto}}]{Imamura:2021}%
  \BibitemOpen
  \bibfield  {author} {\bibinfo {author} {\bibfnamefont {T.}~\bibnamefont
  {Imamura}}, \bibinfo {author} {\bibfnamefont {K.}~\bibnamefont {Mallick}}, \
  and\ \bibinfo {author} {\bibfnamefont {T.}~\bibnamefont {Sasamoto}},\ }\href
  {\doibase 10.1007/s00220-021-03954-x} {\bibfield  {journal} {\bibinfo
  {journal} {Commun. Math. Phys.}\ }\textbf {\bibinfo {volume} {384}},\
  \bibinfo {pages} {1409} (\bibinfo {year} {2021})}\BibitemShut {NoStop}%
\bibitem [{Note2()}]{Note2}%
  \BibitemOpen
  \bibinfo {note} {There is a typo in the formula given in Ref.~\cite
  {Berlioz:2025}. The correct expression is~\protect \eqref
  {eq:CovXtYt}.}\BibitemShut {Stop}%
\bibitem [{\citenamefont {Berlioz}\ \emph
  {et~al.}(2025{\natexlab{b}})\citenamefont {Berlioz}, \citenamefont
  {B{\'e}nichou},\ and\ \citenamefont {Grabsch}}]{Berlioz:2025b}%
  \BibitemOpen
  \bibfield  {author} {\bibinfo {author} {\bibfnamefont {T.}~\bibnamefont
  {Berlioz}}, \bibinfo {author} {\bibfnamefont {O.}~\bibnamefont
  {B{\'e}nichou}}, \ and\ \bibinfo {author} {\bibfnamefont {A.}~\bibnamefont
  {Grabsch}},\ }\href {\doibase 10.48550/arXiv.2509.12017} {\bibfield
  {journal} {\bibinfo  {journal} {arXiv:2509.12017}\ } (\bibinfo {year}
  {2025}{\natexlab{b}}),\ 10.48550/arXiv.2509.12017}\BibitemShut {NoStop}%
\bibitem [{\citenamefont {Bodineau}\ and\ \citenamefont
  {Derrida}(2004)}]{Bodineau:2004}%
  \BibitemOpen
  \bibfield  {author} {\bibinfo {author} {\bibfnamefont {T.}~\bibnamefont
  {Bodineau}}\ and\ \bibinfo {author} {\bibfnamefont {B.}~\bibnamefont
  {Derrida}},\ }\href {\doibase 10.1103/PhysRevLett.92.180601} {\bibfield
  {journal} {\bibinfo  {journal} {Phys. Rev. Lett.}\ }\textbf {\bibinfo
  {volume} {92}},\ \bibinfo {pages} {180601} (\bibinfo {year}
  {2004})}\BibitemShut {NoStop}%
\bibitem [{\citenamefont {Bertini}\ \emph {et~al.}(2005)\citenamefont
  {Bertini}, \citenamefont {De~Sole}, \citenamefont {Gabrielli}, \citenamefont
  {Jona-Lasinio},\ and\ \citenamefont {Landim}}]{Bertini:2005a}%
  \BibitemOpen
  \bibfield  {author} {\bibinfo {author} {\bibfnamefont {L.}~\bibnamefont
  {Bertini}}, \bibinfo {author} {\bibfnamefont {A.}~\bibnamefont {De~Sole}},
  \bibinfo {author} {\bibfnamefont {D.}~\bibnamefont {Gabrielli}}, \bibinfo
  {author} {\bibfnamefont {G.}~\bibnamefont {Jona-Lasinio}}, \ and\ \bibinfo
  {author} {\bibfnamefont {C.}~\bibnamefont {Landim}},\ }\href {\doibase
  10.1103/PhysRevLett.94.030601} {\bibfield  {journal} {\bibinfo  {journal}
  {Phys. Rev. Lett.}\ }\textbf {\bibinfo {volume} {94}},\ \bibinfo {pages}
  {030601} (\bibinfo {year} {2005})}\BibitemShut {NoStop}%
\bibitem [{\citenamefont {Bodineau}\ and\ \citenamefont
  {Derrida}(2005)}]{Bodineau:2005}%
  \BibitemOpen
  \bibfield  {author} {\bibinfo {author} {\bibfnamefont {T.}~\bibnamefont
  {Bodineau}}\ and\ \bibinfo {author} {\bibfnamefont {B.}~\bibnamefont
  {Derrida}},\ }\href {\doibase 10.1103/PhysRevE.72.066110} {\bibfield
  {journal} {\bibinfo  {journal} {Phys. Rev. E}\ }\textbf {\bibinfo {volume}
  {72}},\ \bibinfo {pages} {066110} (\bibinfo {year} {2005})}\BibitemShut
  {NoStop}%
\bibitem [{\citenamefont {Sadhu}\ and\ \citenamefont
  {Derrida}(2016)}]{Sadhu:2016}%
  \BibitemOpen
  \bibfield  {author} {\bibinfo {author} {\bibfnamefont {T.}~\bibnamefont
  {Sadhu}}\ and\ \bibinfo {author} {\bibfnamefont {B.}~\bibnamefont
  {Derrida}},\ }\href {\doibase 10.1088/1742-5468/2016/11/113202} {\bibfield
  {journal} {\bibinfo  {journal} {J Stat. Mech.}\ }\textbf {\bibinfo {volume}
  {2016}},\ \bibinfo {pages} {113202} (\bibinfo {year} {2016})}\BibitemShut
  {NoStop}%
\bibitem [{\citenamefont {Bettelheim}\ \emph
  {et~al.}(2022{\natexlab{b}})\citenamefont {Bettelheim}, \citenamefont
  {Smith},\ and\ \citenamefont {Meerson}}]{Bettelheim:2022a}%
  \BibitemOpen
  \bibfield  {author} {\bibinfo {author} {\bibfnamefont {E.}~\bibnamefont
  {Bettelheim}}, \bibinfo {author} {\bibfnamefont {N.~R.}\ \bibnamefont
  {Smith}}, \ and\ \bibinfo {author} {\bibfnamefont {B.}~\bibnamefont
  {Meerson}},\ }\href {\doibase 10.1088/1742-5468/ac8a4d} {\bibfield  {journal}
  {\bibinfo  {journal} {J. Stat. Mech.}\ }\textbf {\bibinfo {volume} {2022}},\
  \bibinfo {pages} {093103} (\bibinfo {year} {2022}{\natexlab{b}})}\BibitemShut
  {NoStop}%
\bibitem [{\citenamefont {Saha}\ and\ \citenamefont {Sadhu}(2023)}]{Saha:2023}%
  \BibitemOpen
  \bibfield  {author} {\bibinfo {author} {\bibfnamefont {S.}~\bibnamefont
  {Saha}}\ and\ \bibinfo {author} {\bibfnamefont {T.}~\bibnamefont {Sadhu}},\
  }\href {\doibase 10.1088/1742-5468/ace3b2} {\bibfield  {journal} {\bibinfo
  {journal} {J. Stat. Mech.}\ }\textbf {\bibinfo {volume} {2023}},\ \bibinfo
  {pages} {073207} (\bibinfo {year} {2023})}\BibitemShut {NoStop}%
\bibitem [{\citenamefont {Grabsch}\ \emph
  {et~al.}(2024{\natexlab{b}})\citenamefont {Grabsch}, \citenamefont {Moriya},
  \citenamefont {Mallick}, \citenamefont {Sasamoto},\ and\ \citenamefont
  {B\'enichou}}]{Grabsch:2024d}%
  \BibitemOpen
  \bibfield  {author} {\bibinfo {author} {\bibfnamefont {A.}~\bibnamefont
  {Grabsch}}, \bibinfo {author} {\bibfnamefont {H.}~\bibnamefont {Moriya}},
  \bibinfo {author} {\bibfnamefont {K.}~\bibnamefont {Mallick}}, \bibinfo
  {author} {\bibfnamefont {T.}~\bibnamefont {Sasamoto}}, \ and\ \bibinfo
  {author} {\bibfnamefont {O.}~\bibnamefont {B\'enichou}},\ }\href {\doibase
  10.1103/PhysRevLett.133.117102} {\bibfield  {journal} {\bibinfo  {journal}
  {Phys. Rev. Lett.}\ }\textbf {\bibinfo {volume} {133}},\ \bibinfo {pages}
  {117102} (\bibinfo {year} {2024}{\natexlab{b}})}\BibitemShut {NoStop}%
\bibitem [{\citenamefont {Sharma}\ \emph {et~al.}(2024)\citenamefont {Sharma},
  \citenamefont {Saha}, \citenamefont {Jangid},\ and\ \citenamefont
  {Sadhu}}]{Sharma:2024}%
  \BibitemOpen
  \bibfield  {author} {\bibinfo {author} {\bibfnamefont {K.}~\bibnamefont
  {Sharma}}, \bibinfo {author} {\bibfnamefont {S.}~\bibnamefont {Saha}},
  \bibinfo {author} {\bibfnamefont {S.}~\bibnamefont {Jangid}}, \ and\ \bibinfo
  {author} {\bibfnamefont {T.}~\bibnamefont {Sadhu}},\ }\href {\doibase
  10.48550/arXiv.2405.00654} {\bibfield  {journal} {\bibinfo  {journal}
  {arXiv:2405.00654}\ } (\bibinfo {year} {2024}),\
  10.48550/arXiv.2405.00654}\BibitemShut {NoStop}%
\bibitem [{\citenamefont {P\'erez-Espigares}\ \emph {et~al.}(2016)\citenamefont
  {P\'erez-Espigares}, \citenamefont {Garrido},\ and\ \citenamefont
  {Hurtado}}]{Espigares:2016}%
  \BibitemOpen
  \bibfield  {author} {\bibinfo {author} {\bibfnamefont {C.}~\bibnamefont
  {P\'erez-Espigares}}, \bibinfo {author} {\bibfnamefont {P.~L.}\ \bibnamefont
  {Garrido}}, \ and\ \bibinfo {author} {\bibfnamefont {P.~I.}\ \bibnamefont
  {Hurtado}},\ }\href {\doibase 10.1103/PhysRevE.93.040103} {\bibfield
  {journal} {\bibinfo  {journal} {Phys. Rev. E}\ }\textbf {\bibinfo {volume}
  {93}},\ \bibinfo {pages} {040103} (\bibinfo {year} {2016})}\BibitemShut
  {NoStop}%
\bibitem [{\citenamefont {Tiz\'on-Escamilla}\ \emph
  {et~al.}(2017{\natexlab{a}})\citenamefont {Tiz\'on-Escamilla}, \citenamefont
  {Hurtado},\ and\ \citenamefont {Garrido}}]{Escamilla:2017}%
  \BibitemOpen
  \bibfield  {author} {\bibinfo {author} {\bibfnamefont {N.}~\bibnamefont
  {Tiz\'on-Escamilla}}, \bibinfo {author} {\bibfnamefont {P.~I.}\ \bibnamefont
  {Hurtado}}, \ and\ \bibinfo {author} {\bibfnamefont {P.~L.}\ \bibnamefont
  {Garrido}},\ }\href {\doibase 10.1103/PhysRevE.95.032119} {\bibfield
  {journal} {\bibinfo  {journal} {Phys. Rev. E}\ }\textbf {\bibinfo {volume}
  {95}},\ \bibinfo {pages} {032119} (\bibinfo {year}
  {2017}{\natexlab{a}})}\BibitemShut {NoStop}%
\bibitem [{\citenamefont {Tiz\'on-Escamilla}\ \emph
  {et~al.}(2017{\natexlab{b}})\citenamefont {Tiz\'on-Escamilla}, \citenamefont
  {P\'erez-Espigares}, \citenamefont {Garrido},\ and\ \citenamefont
  {Hurtado}}]{Escamilla:2017a}%
  \BibitemOpen
  \bibfield  {author} {\bibinfo {author} {\bibfnamefont {N.}~\bibnamefont
  {Tiz\'on-Escamilla}}, \bibinfo {author} {\bibfnamefont {C.}~\bibnamefont
  {P\'erez-Espigares}}, \bibinfo {author} {\bibfnamefont {P.~L.}\ \bibnamefont
  {Garrido}}, \ and\ \bibinfo {author} {\bibfnamefont {P.~I.}\ \bibnamefont
  {Hurtado}},\ }\href {\doibase 10.1103/PhysRevLett.119.090602} {\bibfield
  {journal} {\bibinfo  {journal} {Phys. Rev. Lett.}\ }\textbf {\bibinfo
  {volume} {119}},\ \bibinfo {pages} {090602} (\bibinfo {year}
  {2017}{\natexlab{b}})}\BibitemShut {NoStop}%
\bibitem [{Note3()}]{Note3}%
  \BibitemOpen
  \bibinfo {note} {See for instance Ref.~\cite {Saha:2023}, where reflective or
  absorbing boundary conditions for the one-dimensional SEP in a semi-infinite
  line are recovered as limiting situations of a system coupled to a particle
  reservoir.}\BibitemShut {Stop}%
\bibitem [{\citenamefont {Bodineau}\ \emph {et~al.}(2008)\citenamefont
  {Bodineau}, \citenamefont {Derrida},\ and\ \citenamefont
  {Lebowitz}}]{Bodineau:2008}%
  \BibitemOpen
  \bibfield  {author} {\bibinfo {author} {\bibfnamefont {T.}~\bibnamefont
  {Bodineau}}, \bibinfo {author} {\bibfnamefont {B.}~\bibnamefont {Derrida}}, \
  and\ \bibinfo {author} {\bibfnamefont {J.~L.}\ \bibnamefont {Lebowitz}},\
  }\href {\doibase 10.1007/s10955-008-9518-y} {\bibfield  {journal} {\bibinfo
  {journal} {J. Stat. Phys.}\ }\textbf {\bibinfo {volume} {131}},\ \bibinfo
  {pages} {821} (\bibinfo {year} {2008})}\BibitemShut {NoStop}%
\bibitem [{\citenamefont {Berlioz}\ \emph {et~al.}(2024)\citenamefont
  {Berlioz}, \citenamefont {Venturelli}, \citenamefont {Grabsch},\ and\
  \citenamefont {Bénichou}}]{Berlioz:2024}%
  \BibitemOpen
  \bibfield  {author} {\bibinfo {author} {\bibfnamefont {T.}~\bibnamefont
  {Berlioz}}, \bibinfo {author} {\bibfnamefont {D.}~\bibnamefont {Venturelli}},
  \bibinfo {author} {\bibfnamefont {A.}~\bibnamefont {Grabsch}}, \ and\
  \bibinfo {author} {\bibfnamefont {O.}~\bibnamefont {Bénichou}},\ }\href
  {\doibase 10.1088/1742-5468/ad874a} {\bibfield  {journal} {\bibinfo
  {journal} {J. Stat. Mech.}\ }\textbf {\bibinfo {volume} {2024}},\ \bibinfo
  {pages} {113208} (\bibinfo {year} {2024})}\BibitemShut {NoStop}%
\end{thebibliography}

%

\end{document}